\documentclass[prd,preprint,tightenlines,floatfix,showpacs,preprintnumbers,nofootinbib,eqsecnum,superscriptaddress]{revtex4}

 \usepackage[dvips,final]{graphicx}
  \usepackage{amssymb}
   \usepackage{amsmath}
    \usepackage{amsfonts}
     \usepackage{epsfig}
      \usepackage{bm}

\begin{document}

\title{
Exclusive \mbox{\boldmath $\textit{\textbf{pp}} \to \textit{\textbf{pp}} \pi^{+}\pi^{-}$} reaction:\\
from the threshold to LHC}

\author{P.~Lebiedowicz}
\email{piotr.lebiedowicz@ifj.edu.pl}
\affiliation{Institute of Nuclear Physics PAN, PL-31-342 Cracow, Poland}
 
\author{A.~Szczurek}
\email{antoni.szczurek@ifj.edu.pl}
\affiliation{University of Rzesz\'ow, PL-35-959 Rzesz\'ow, Poland}
\affiliation{Institute of Nuclear Physics PAN, PL-31-342 Cracow, Poland} 

\begin{abstract}
We evaluate differential distributions for the four-body
$p p \to p p \pi^+ \pi^-$ reaction which constitutes 
a irreducible background to three-body processes $p p \to p p M$,
where $M$ are a broad resonances in the $\pi^+ \pi^-$ channel, e.g. 
$M=\sigma, \rho^{0}, f_{0}(980), f_{2}(1275), f_{0}(1500)$.
We include both double-diffractive contribution
(both pomeron and reggeon exchanges) as well as the pion-pion 
rescattering contribution.
The first process dominates at higher energies and small 
pion-pion invariant masses while the second becomes important at 
lower energies and higher pion-pion invariant masses.
The amplitude(s) is(are) calculated in the Regge approach. 
We compare our results with measured cross sections 
for the ISR experiments at CERN.
We make predictions for future experiments at PANDA, RHIC, Tevatron and LHC energies.
Differential distributions in effective two-pion mass, 
pion rapidities and transverse momenta of pions are presented.
The two-dimensional distribution in $(y_{\pi^+}, y_{\pi^-})$
is particularly interesting.
The higher the incident energy, the higher preference
for the same-hemisphere emission of pions.
The processes considered constitute a sizeable contribution to
the total nucleon-nucleon cross section as well as to pion
inclusive cross section.
\end{abstract}


\maketitle

\section{Introduction}

Diffractive processes although very difficult from 
the point of view of perturbative QCD are very attractive from 
the general point of view of the reaction mechanism.
There are several classes of diffractive-type processes \cite{AG81} 
in high-energy nucleon-nucleon collisions such as:

$\bullet$ (a) elastic scattering,

$\bullet$ (b) single-diffractive excitation of one of the nucleons,

$\bullet$ (c) double-diffractive excitation of both participating nucleons,

$\bullet$ (d) central (double)-diffractive production of a simple final state.

The energy dependence of the first three types of the reaction 
was measured and can be nicely described \cite{K79}
in a somewhat academic two-state (but fullfilling unitarity)
Good-Walker model \cite{Good-Walker}. 
The last case was not studied in too much detail either experimentally
or theoretically. At not too high energies the dominant diffractive 
final state is the $\pi^+ \pi^-$ and $\pi^0 \pi^0$ continuum.
The multi-pion and $KK$ continuum is expected to be smaller.

There is recently a growing interest in understanding
exclusive three-body reactions $p p \to p p M$ at high energies,
where the meson (resonance) $M$ is produced in the central
rapidity region. Many of the resonances decay into $\pi\pi$
and/or $KK$ channels. The representative examples are:
$M=\sigma, \rho^{0}, f_{0}(980), \phi, f_{2}(1275), 
f_{0}(1500), \chi_{c}(0^{+})$. It is clear that these
resonances are seen (or will be seen) "on" the background
of a $\pi\pi$ or $KK$ continuum
\footnote{In general, the resonance and continuum contributions 
may interfere. This may produce even a dip.
A good example is $f_{0}(980)$ production (see \cite{LSK09,Alde97}).}.
Therefore a good understanding of the continuum
seems indispensable. In the present analysis we concentrate on the 
$\pi^+ \pi^-$ channel. Similar analysis can be done
for $\pi^0 \pi^0$ exclusive production.

At larger energies two-pomeron exchange mechanism
dominates in central production 
(see \cite{AG81} and references therein). 
In calculating the amplitude related to double
diffractive mechanism for $p p \to p p \pi^+ \pi^-$
we follow the general rules of Pumplin and Henyey \cite{PH76} 
(for early rough estimates see also Ref.\cite{AKLR75}).

At lower energies subleading reggeons must be
included in addition to the pomeron exchanges.
We include a new mechanism relevant 
at lower energies (FAIR, J-PARC) relying on the exchange of two pion. 
We shall call this mechanism pion-pion rescattering for brevity.

We discuss interplay of all the mechanisms
in a quite rich four-body phase space.

\section{The $\pi N$ elastic cross section}

At low energies, the total cross sections for $\pi^+ p$ and $\pi^- p$ 
show a significant energy-dependent asymmetry defined as:
\begin{equation}
A_{tot}^{\pi p}(W) \equiv
\frac{|\sigma_{tot}^{\pi^+ p}(W) - \sigma_{tot}^{\pi^- p}(W)|}
     { \sigma_{tot}^{\pi^+ p}(W) + \sigma_{tot}^{\pi^- p}(W) } \; .
\label{A_tot_pip}
\end{equation}
The total cross section tests, via optical theorem, only imaginary 
part of the scattering amplitude.
In our case of the $2 \to 4$ reactions 
\footnote{$2 \to 4$ reaction denotes a type of the reaction
with two initial and four final particles.}
we should use rather full scattering amplitude.
In contrast to the total cross section the elastic scattering cross 
sections for $\pi^+ p$ and $\pi^- p$ show
at low energies rather small asymmetry defined as:
\begin{equation}
A_{el}^{\pi p}(W) \equiv
\frac{|\sigma_{el}^{\pi^+ p}(W) - \sigma_{el}^{\pi^- p}(W)|}
     { \sigma_{el}^{\pi^+ p}(W) + \sigma_{el}^{\pi^- p}(W) } \; .
\label{A_el_pip}
\end{equation}
A reliably model should explain such details of the interaction.

Therefore to fix parameters of our double-diffractive model 
we consider first elastic pion-proton scattering. 
The amplitude for the elastic scattering of pions on nucleons
is written in the simplified Regge-like form:
\begin{eqnarray}
{\cal M}_{\pi^{\pm} p \to \pi^{\pm} p}(s,t)
&=& \mathrm{i} \; s \; C_{I\!\!P}\left( \frac{s}{s_0} \right)^{\alpha_{I\!\!P}(t)-1} 
\exp \left( \frac{B^{I\!\!P}_{\pi N}}{2} t \right)
\nonumber \\
&+& (a_{f} + \mathrm{i}) \; s \; C_{f} \left( \frac{s}{s_0} \right)^{\alpha_{I\!\!R}(t)-1}
\exp \left( \frac{B^{I\!\!R}_{\pi N}}{2} t \right) 
\nonumber \\
&\pm& (a_{\rho} - \mathrm{i})\; s \; C_{\rho} \left( \frac{s}{s_0} \right)^{\alpha_{I\!\!R}(t)-1}
\exp\left( \frac{B^{I\!\!R}_{\pi N}}{2} t \right)
\; ,
\label{piN_amplitude}
\end{eqnarray}
where $a_{f}$ = -0.860895 and $a_{\rho}$ = -1.16158.
The strength parameters $C_{I\!\!P}$, $C_f$, $C_{\rho}$ are taken from
the Donnachie-Landshoff model \cite{DL92} for total cross section:
\begin{eqnarray}
C_{I\!\!P}=13.63 \; mb,\qquad
C_f=31.79 \; mb,\qquad
C_{\rho}=4.23 \; mb \; .
\label{strength_parameters}
\end{eqnarray}
This means that our effective phenomenological model describes 
the available total cross sections.
The pomeron and reggeon trajectories 
determined from elastic and total cross sections
are given in the form
($\alpha_{i}(t)=\alpha_{i}(0)+\alpha_{i}^{'}t$):
\begin{eqnarray}
\alpha_{I\!\!P}(t) = 1.088 + 0.25 t ,\qquad
\alpha_{I\!\!R}(t) = 0.5475 + 0.93 t \; .
\label{trajectories}
\end{eqnarray}
The values of the intercept $\alpha_{I\!\!P}(0)$ and $\alpha_{I\!\!R}(0)$ are also
taken from the Donnachie-Landshoff model \cite{DL92} for consistency.
The effective slope parameter can be written as
\begin{equation}
B_{eff} \equiv B_{\pi N}(W_{\pi N}) = B_0 + 2 \alpha'_{i}
 \ln \left( \frac{s}{s_0} \right) \; .
\label{slope_W}
\end{equation}
We take $\alpha'_{i}$ = 0.25/0.93 for pomeron and reggeon exchanges, respectively.
The slope parameter $B_{\pi N}$, taken the same for the pomeron and reggeons,
must be fitted to the data. From the fit to the data \cite{dsig_dt} 
we find $B_0$ = 5.5 GeV$^{-2}$. 
The effective slope observed in $t$-distributions is of course 
much larger ($B_{eff}$ = 7-10 GeV$^{-2}$ for $P_{lab}$ = 3-200 GeV \cite{dsig_dt}).

The differential elastic cross section is expressed with the help
of the scattering amplitude as:
\begin{equation}
\frac{d\sigma_{el}}{dt}=\frac{1}{16\pi s^{2}}|{\cal M}(s,t)|^{2}\; .
\label{dsigma_dt_elastic}
\end{equation}
%

\begin{figure}[!h]    %
\includegraphics[width=0.45\textwidth]{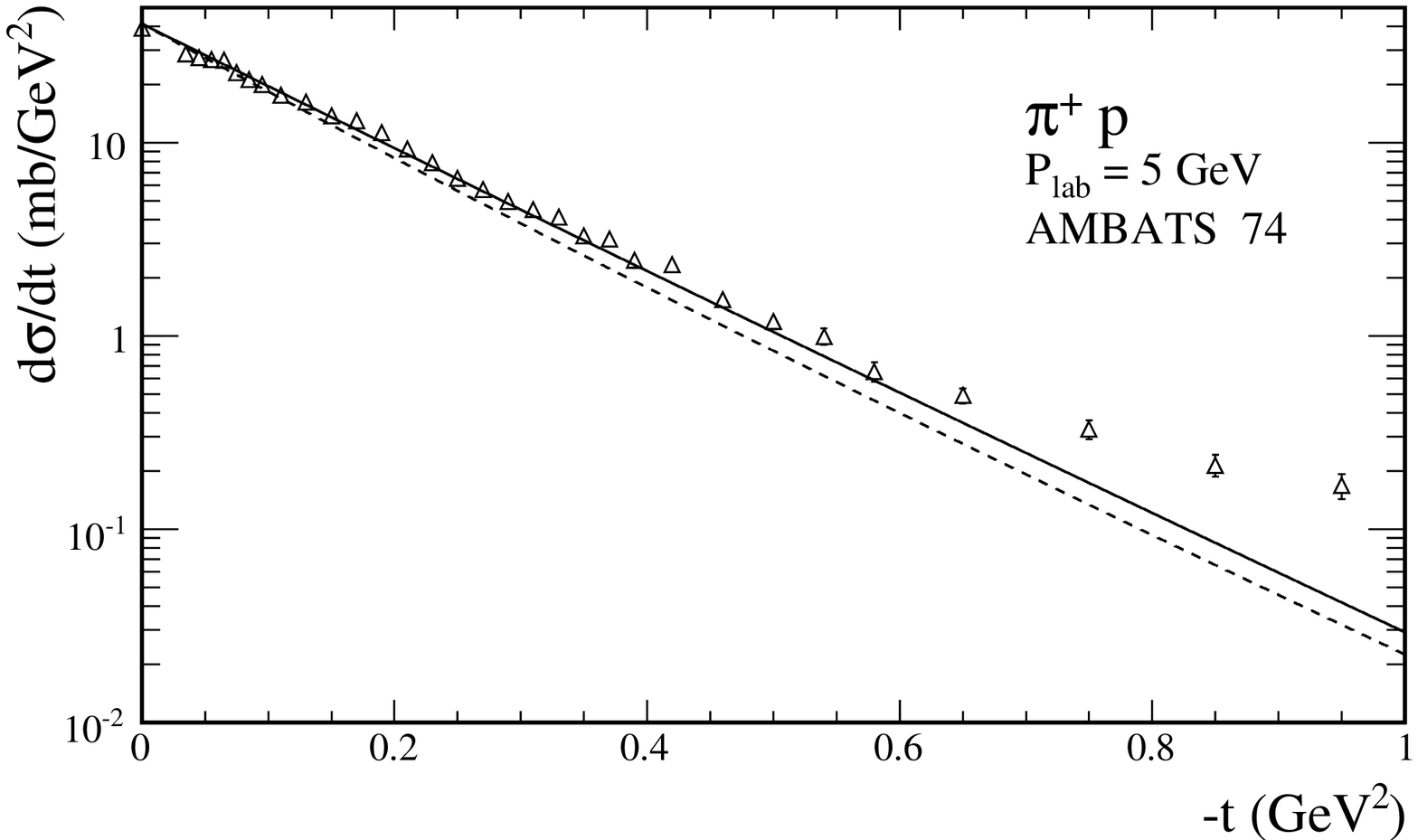}\qquad
\includegraphics[width=0.45\textwidth]{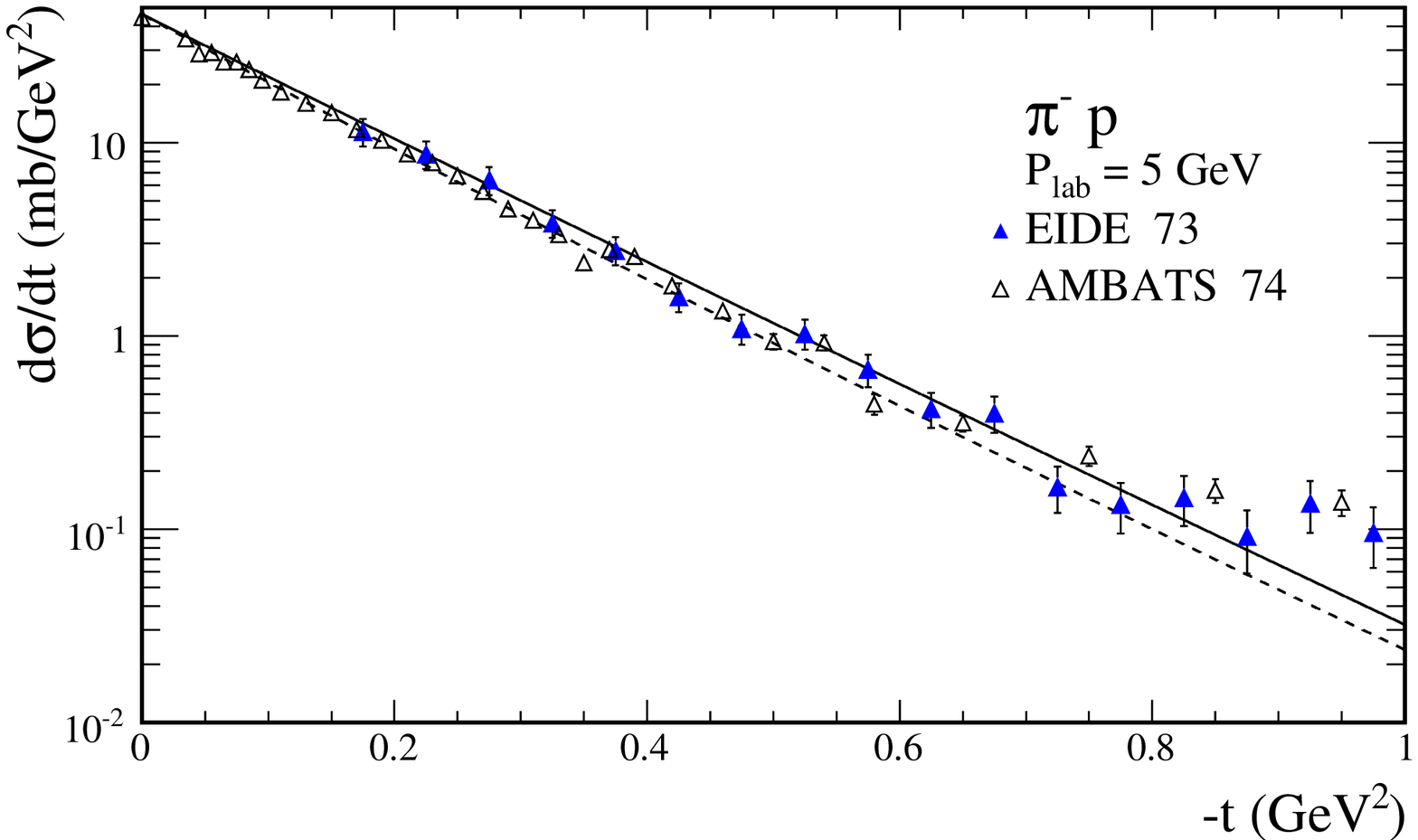}
\includegraphics[width=0.45\textwidth]{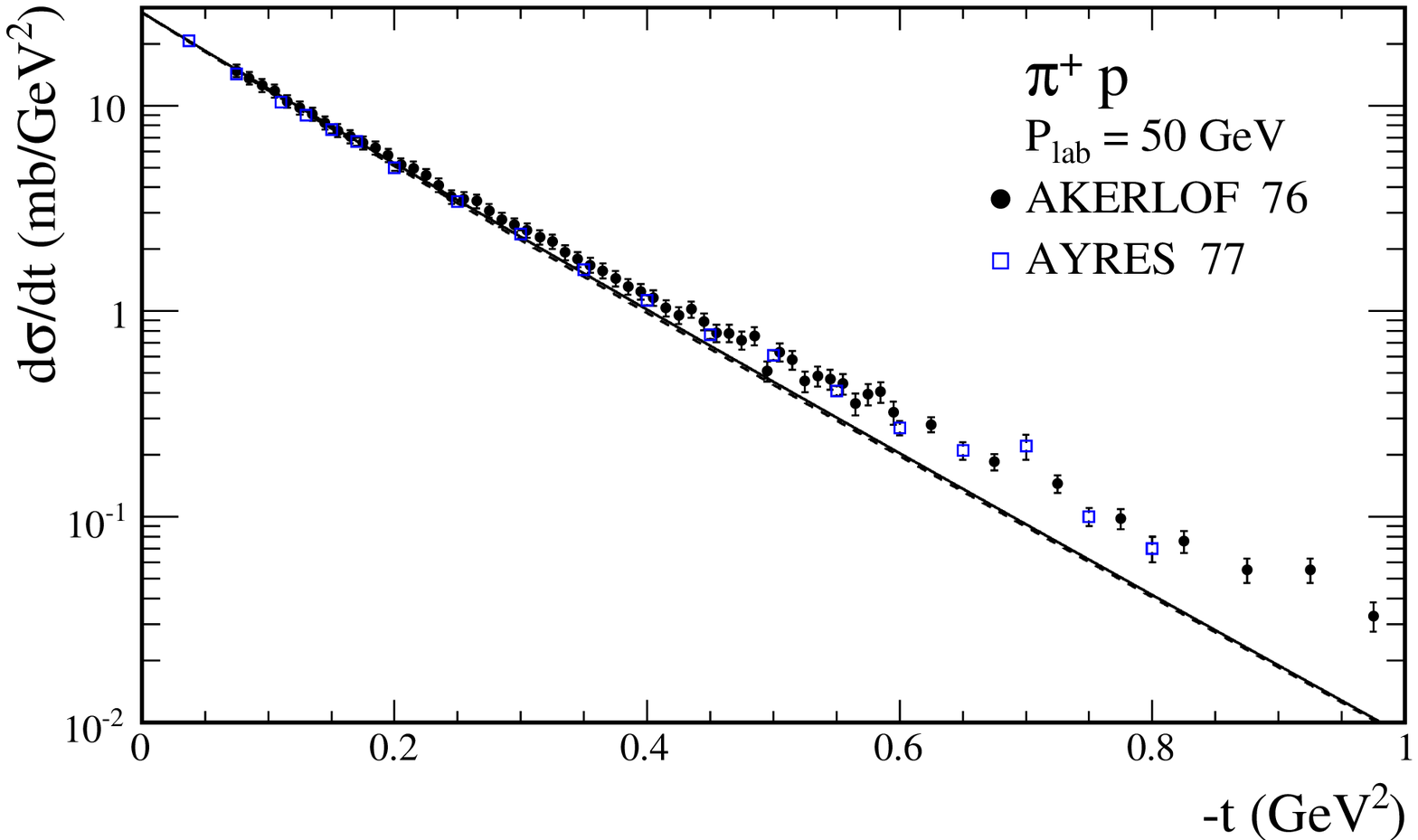}\qquad
\includegraphics[width=0.45\textwidth]{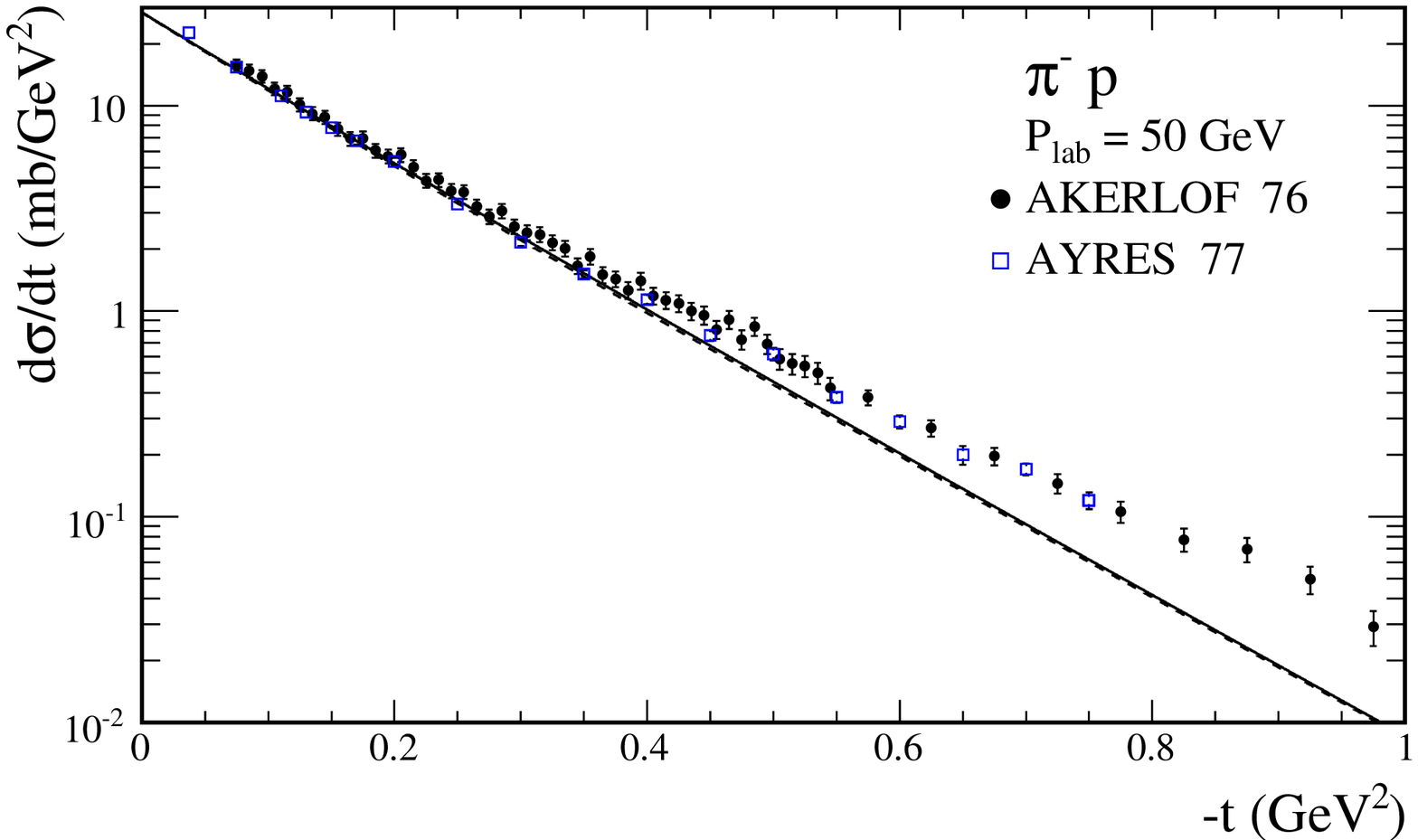}
\includegraphics[width=0.45\textwidth]{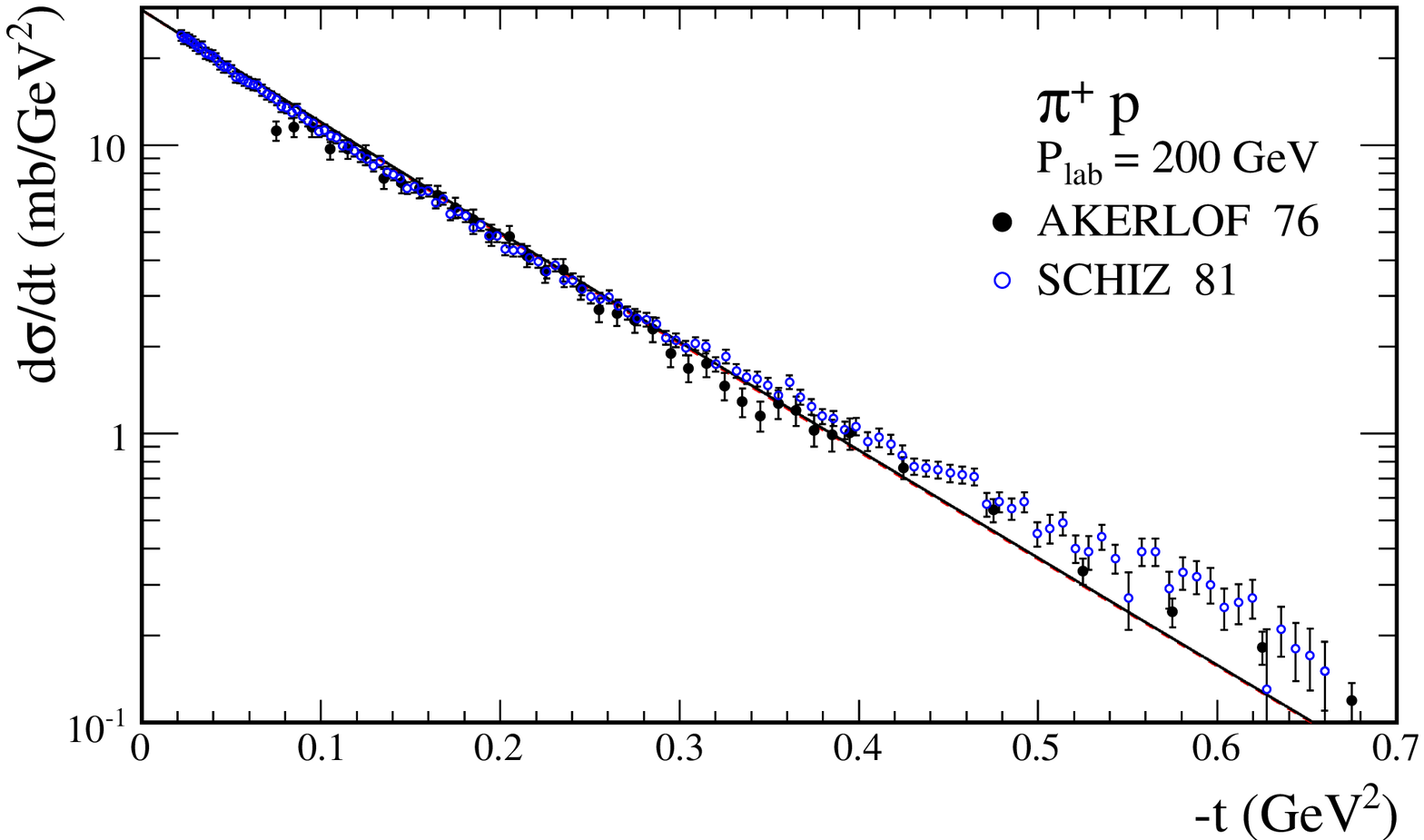}\qquad
\includegraphics[width=0.45\textwidth]{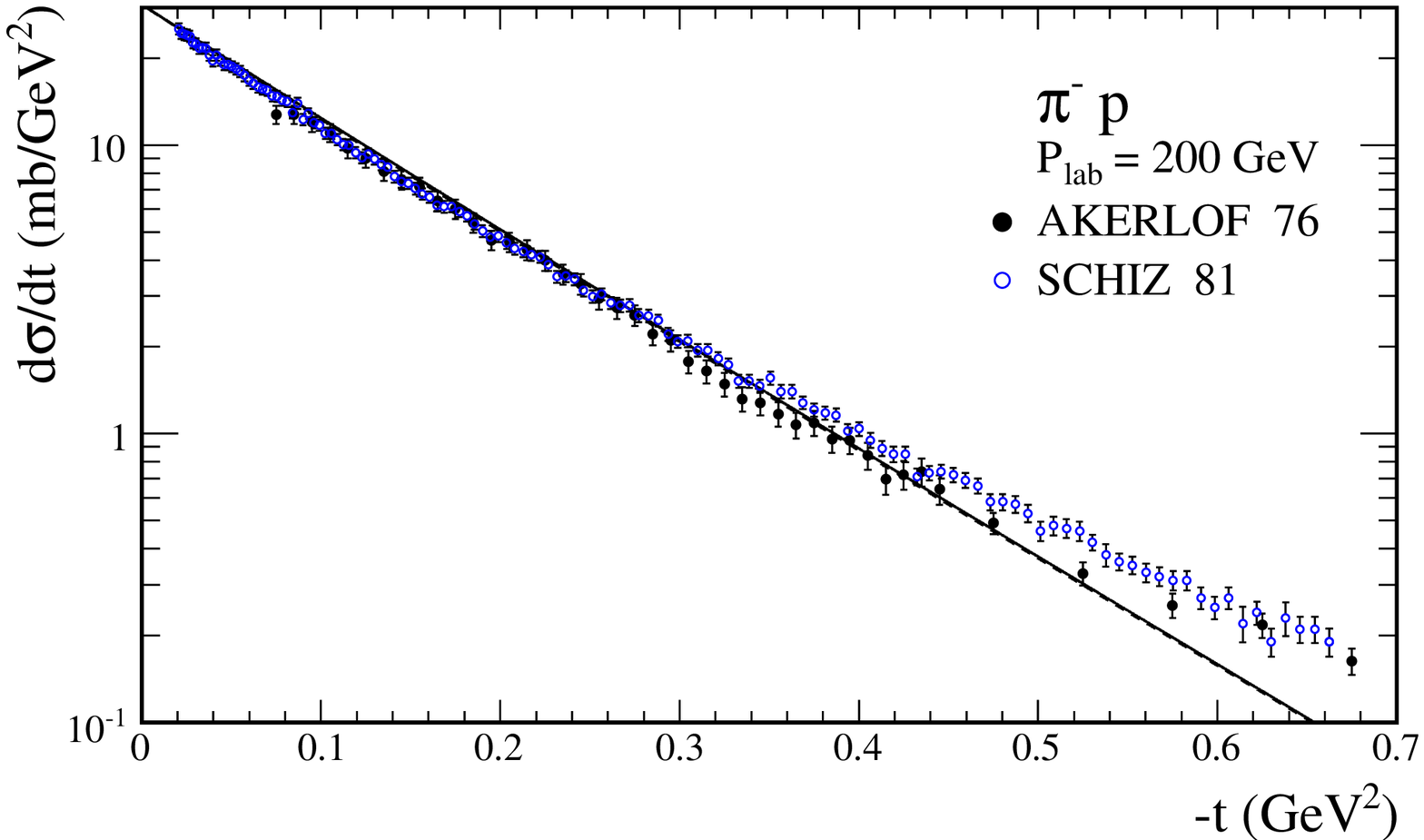}
   \caption{\label{fig:dsig_dt_piN_piN}
   \small 
Differential distributions for $\pi^+ p$ (left) and $\pi^- p$ (right)
elastic scattering for different energies calculated with the amplitude
(\ref{piN_amplitude}) and parameters as given in the text.
In this calculation the slope parameter was taken as 
$B_{\pi N}$ = 5.5 GeV$^{-2}$ (dashed line). 
A fit to the world $\pi N$ elastic scattering data suggest that
the pomeron and reggeon slopes may be slightly different.
The solid line shows such a result.
The details are explained when discussing Fig.\ref{fig:piN_elastic}.
The experimental data are taken from Ref.\cite{dsig_dt}.
}
\end{figure}

The differential distributions $d\sigma_{el}/dt$
for both $\pi^+ p$ and $\pi^- p$ elastic scattering 
for three incident-beam momenta of $P_{lab}=$ 5 GeV, 
$P_{lab}=$ 50 GeV and $P_{lab}=$ 200 GeV are shown 
in Fig.\ref{fig:dsig_dt_piN_piN}.
Under a detailed inspection one can observe that the local 
slope parameter
\begin{equation}
B_{eff}(t) \equiv \frac{d}{dt} \ln \left( \frac{d\sigma_{el}}{dt} \right)
\label{effective_slope}
\end{equation}
is $t$-dependent and is slightly larger for $\pi^- p$ than for $\pi^+ p$.
Such an effect was observed experimentally in Ref.\cite{dsig_dt}.
The local slope decreases with increasing $t$.
A rather good description of experimental $d\sigma_{el}/dt$ is achieved.

Our one-parameter ($B_{\pi N}$) model here is consistent with the simple
Donnachie-Landshoff model for total cross section \cite{DL92}.
A more refined model should include absorption effects due to 
pion-nucleon rescatterings. The analysis of absorption effects clearly goes beyond 
the scope of the present paper. 
Our model sufficiently well describes the $\pi N$ data 
and includes absorption effects in an effective way. This has advantage for 
the $p p \to p p \pi \pi$ reaction discussed in the present paper where the 
$\pi N$ absorption effects do not need to be included explicitly. This considerably
simplifies the calculation for the 2$\to$4 reaction and actually this makes 
the calculation of the 2$\to$4 reaction feasible.

\begin{figure}[!h]   
\includegraphics[width=0.45\textwidth]{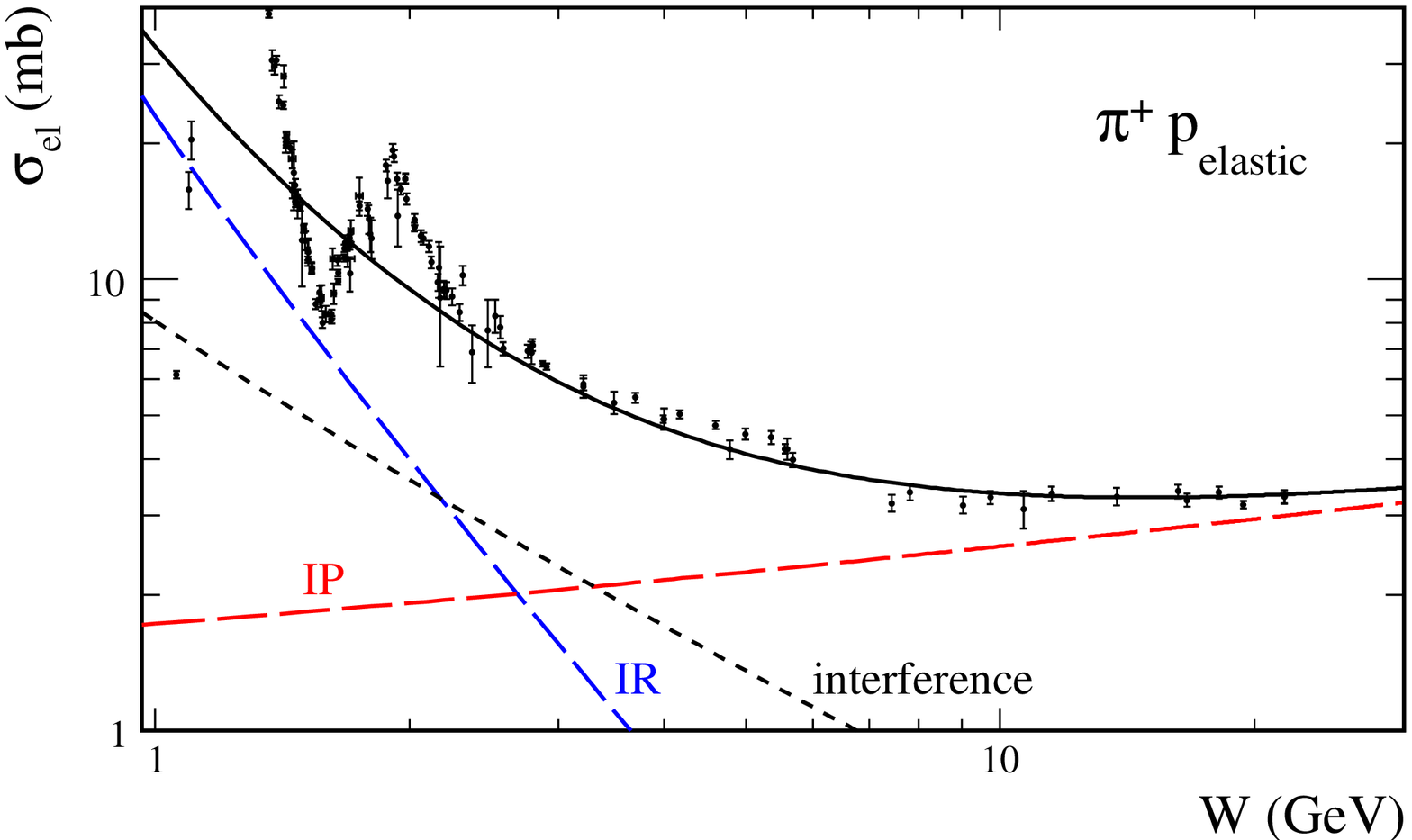}\qquad
\includegraphics[width=0.45\textwidth]{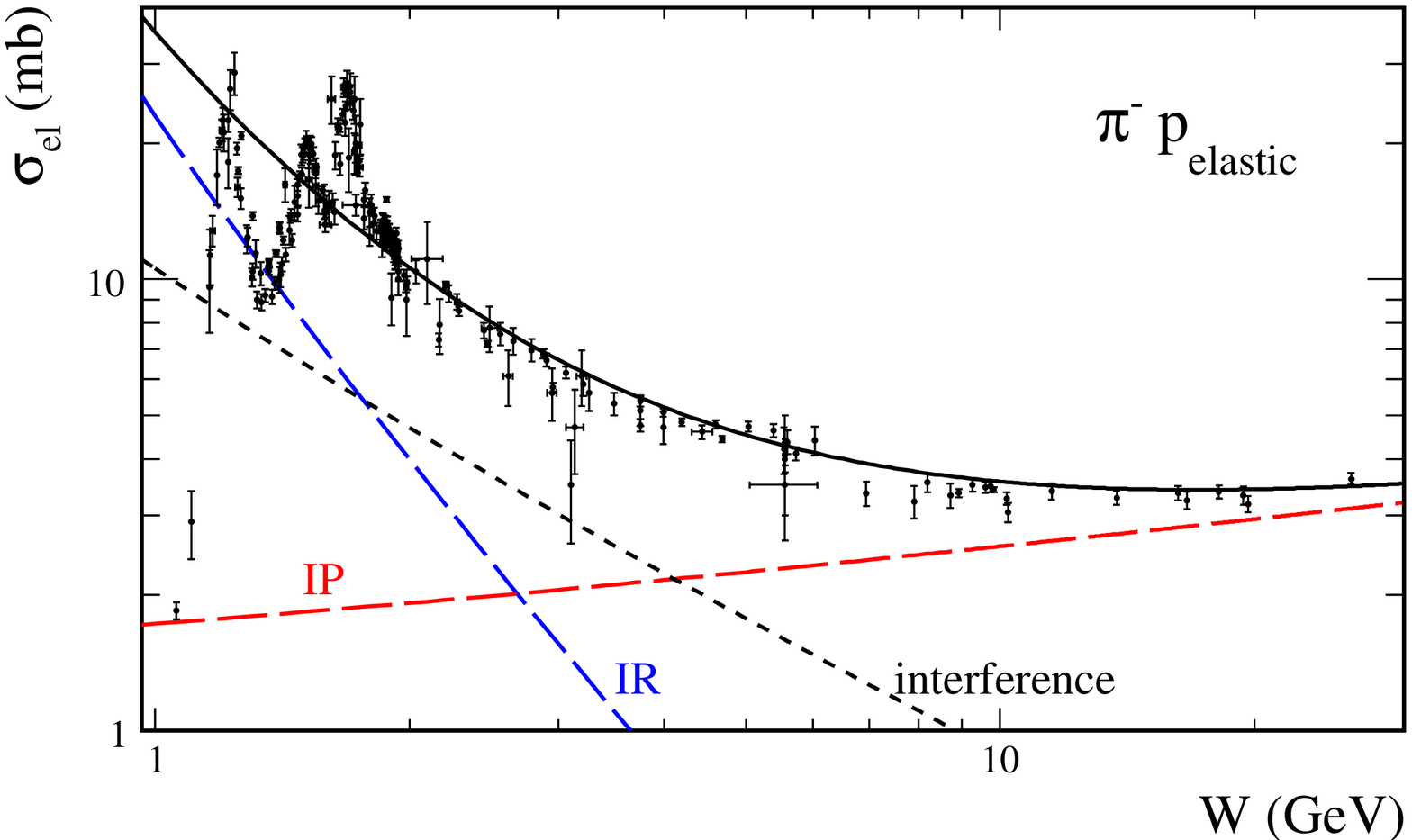}
   \caption{\label{fig:piN_elastic}
   \small 
The integrated cross section for $\pi N$ elastic scattering.
The experimental data are taken from Ref.\cite{Amsler_PDG08}.
}
\end{figure}

Before we shall go to the $p p \to p p \pi^+ \pi^-$ reaction,
we have to discuss the parameters of the $\pi N$ interaction.
The strenght parameters of the pomeron and reggeon couplings
are taken from the Donnachie-Landshoff analysis of the total
cross section in several hadronic reactions \cite{DL92}
as discussed above.
The only free parameters -- the slope parameters, are adjusted to 
the elastic $\pi^+ p$ and $\pi^- p$ scattering.
With $B_{I\!\!P}$ = 5.5 GeV$^{-2}$ and $B_{I\!\!R}$ = 4 GeV$^{-2}$
we nicely describe the existing experimental data for $\pi p$ scattering 
as can be seen from Fig.\ref{fig:piN_elastic} (solid lines). 
The long-dashed lines show pomeron ($I\!\!P$) and reggeon ($I\!\!R$) 
contributions and the short-dashed lines their interference term.
In the Regge approach, high energy cross section is dominated 
by pomeron exchange. The reggeon exchange dominates in the resonance
region. There is a region of energies where
the interference term dominates. This is very different than
for the total cross section where the cross section is just a sum
of the pomeron and reggeon terms. We get a nice description
of the data for $\sqrt{s} >$ 2.5 GeV. The region below contains
resonances and is therefore very difficult for modeling.

Having fixed the parameters we can proceed to our four-body 
$p p \to p p \pi^+ \pi^-$ reaction.

\section{Central double diffractive contribution}

\begin{figure}[!h]   
\includegraphics[width=0.32\textwidth]{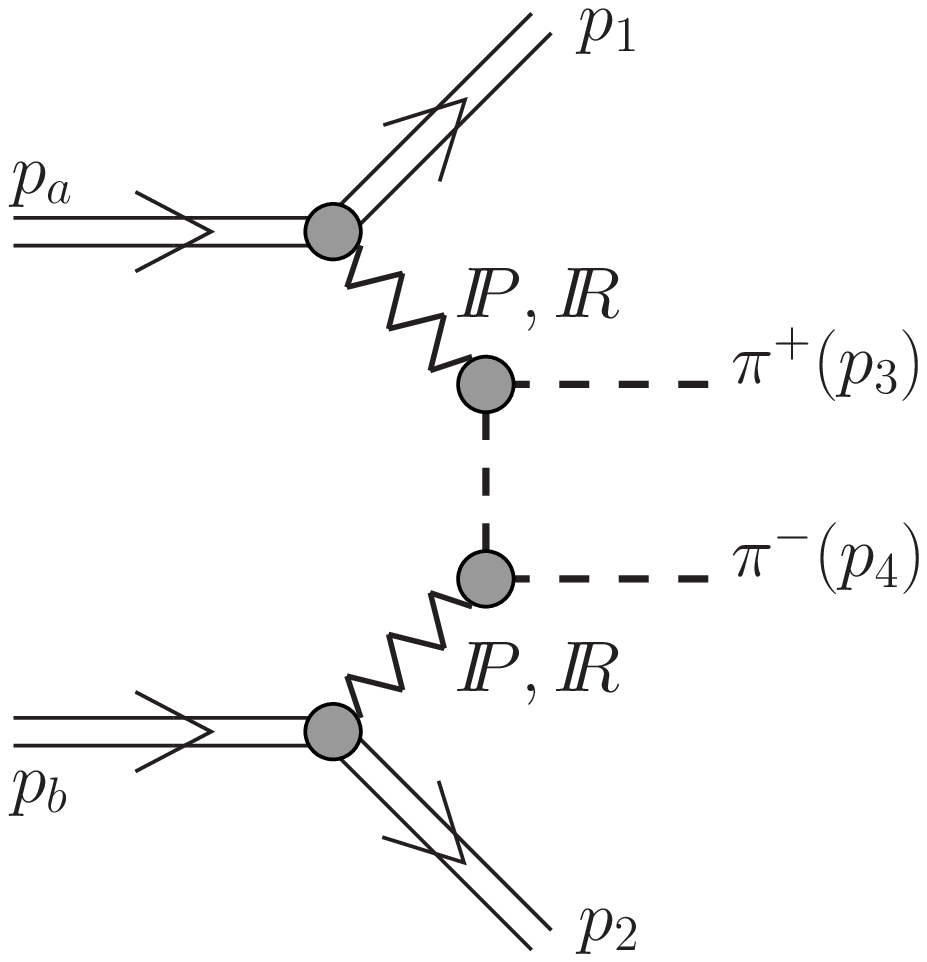}
\includegraphics[width=0.32\textwidth]{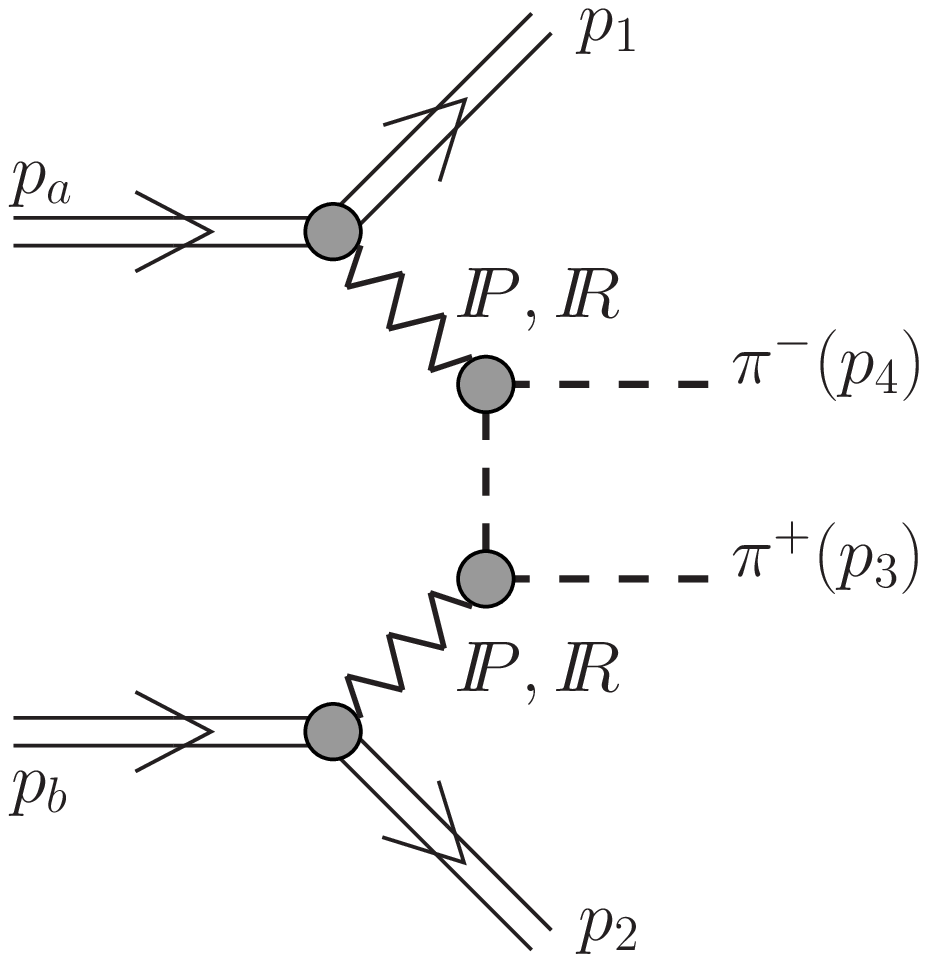}
   \caption{\label{fig:central_double_diffraction_diagrams}
   \small 
A sketch of the dominant mechanisms of exclusive production
of $\pi^{+}\pi^{-}$ pairs at high energies.
}
\end{figure}

The general situation is sketched 
in Fig.\ref{fig:central_double_diffraction_diagrams}.
The corresponding amplitude for the $pp\to pp\pi^{+}\pi^{-}$ process 
(with four-momenta $p_{a}+p_{b}\rightarrow p_{1}+p_{2}+p_{3}+p_{4}$)
can be written as
\begin{eqnarray} \nonumber
\mathcal{M}^{pp\to pp\pi\pi}&=&
M_{13}(t_1,s_{13})F(t_{a})\frac{1}{t_{a}-m_{\pi}^{2}}F(t_{a})M_{24}(t_2,s_{24})\\
&+&M_{14}(t_1,s_{14})F(t_{b})\frac{1}{t_{b}-m_{\pi}^{2}}F(t_{b})M_{23}(t_2,s_{23})\; ,
\label{Regge_amplitude_DD}
\end{eqnarray}
where $M_{ik}$ denotes "interaction" between nucleon $i=1$
(forward nucleon) or $i=2$ (backward nucleon) 
and one of the two pions $k=3$ ($\pi^{+}$), $k=4$ ($\pi^{-}$). 
In the Regge phenomenology they can be written as
\begin{eqnarray}
M_{ik}(t_i,s_{ik}) &=&  
\mathrm{i} \; s_{ik} \; C_{I\!\!P} \left(\frac{s_{ik}}{s_0}\right)^{\alpha_{I\!\!P}(t_{i})-1} \;
\exp\left( {\frac{B_{I\!\!P}}{2} \; t_i}\right) 
\nonumber \\
&+& (a_{f} + \mathrm{i}) \; s_{ik} \; C_{f} \left(\frac{s_{ik}}{s_0}\right)^{\alpha_{I\!\!R}(t_{i})-1} \;
\exp\left( {\frac{B_{I\!\!R}}{2} \; t_i}\right) 
\nonumber \\
&\pm& (a_{\rho} - \mathrm{i})\; s_{ik} \; C_{\rho} \left(\frac{s_{ik}}{s_0}\right)^{\alpha_{I\!\!R}(t_{i})-1} \;
\exp\left( {\frac{B_{I\!\!R}}{2} \; t_i}\right) 
\; .
\label{piN_amplitude}
\end{eqnarray}

Above $s_{ik}=W_{ik}^{2}$, where $W_{ik}$ is the center-of-mass
energy in the ($i,k$) subsystems.
The third term is with the sign plus if $k=3$
and with the sign minus if $k=4$.
The normalization constants 
($C_{I\!\!P}$, $C_f$, $C_{\rho}$)
can be estimated from
the fit to the total $\pi N$ cross section (\ref{strength_parameters}).
The values of the Regge trajectories (\ref{trajectories}) are also
taken from the Donnachie-Landshoff model \cite{DL92}.
The first term describes exchange of the leading
(pomeron) trajectory while the next terms describe
the subleading reggeon exchanges.
At high $\pi N$ subsystem energies $W_{ik}>20$ GeV
only the pomeron exchange survive.
%
%

%

The extra form factors $F(t_{a})$ and $F(t_{b})$ "correct" 
for off-shellness of the intermediate pions 
in the middle of the diagrams shown in 
Fig.\ref{fig:central_double_diffraction_diagrams}.
In the following they are parametrized as
\begin{equation} 
F(t_{1,2})=\exp\left(\frac{t_{1,2}-m_{\pi}^{2}}{\Lambda^{2}_{off, E}}\right) \;,
\label{off-shell_form_factors}
\end{equation}
i.e. normalized to unity on the pion-mass-shell.
In the following for brevity we shall use notation $t_{1,2}$ 
which means $t_1$ or $t_2$.
In general, the parameter $\Lambda_{off, E}$ is not known 
but in principle could be fitted to the
(normalized) experimental data. 
From our general experience in hadronic physics 
we expect $\Lambda_{off, E}\sim$ 1 GeV.
How to extract $\Lambda_{off, E}$ will be discussed in the result section.

The parametrization \cite{DL92} can be used only for 
$W_{ik}>2-3$ GeV. Bellow $W_{ik}=2$ GeV resonances in $\pi N$
subsystems are present. In principle, their contribution
could and should be included explicitly
\footnote{The higher the center-of-mass energy 
the smaller the relative resonance contribution.}
\footnote{In the standard terminology the resonances belong
to single-diffractive contribution to be distinguished
from double-diffractive contribution discussed here.}.
The amplitude (\ref{Regge_amplitude_DD}) with (\ref{piN_amplitude})
is used to calculate the corresponding cross section
including limitations of the four-body phase-space.
To exclude resonance regions we shall "correct" the parametrization (\ref{Regge_amplitude_DD}) with (\ref{piN_amplitude})
by multiplying by a purely phenomenological smooth cut-off correction factor:
\begin{eqnarray} 
f_{cont}^{\pi N}(W_{ik})=\frac{\exp \left( \frac{W-W_{0}}{a}\right)}{1+\exp \left( \frac{W-W_{0}}{a}\right)} \; .
\label{Correction_factor}
\end{eqnarray}
The parameter $W_{0}$ gives the position of the cut and 
parameter $a$ describes how sharp is the cut off. 
The first parameter can have a significant influence
on the results. We shall take $W_{0}=2-3$ GeV
and $a=0.1-0.5$ GeV. For large energies
$f_{cont}^{\pi N}(W_{ik})\approx$ 1 and close to
kinematical threshold $W_{ik}=m_{\pi}+M_{N}:$
$f_{cont}^{\pi N}(W_{ik})\approx$ 0.

\section{Pion-pion rescattering}

\begin{figure}[!h]    %
\includegraphics[width=0.27\textwidth]{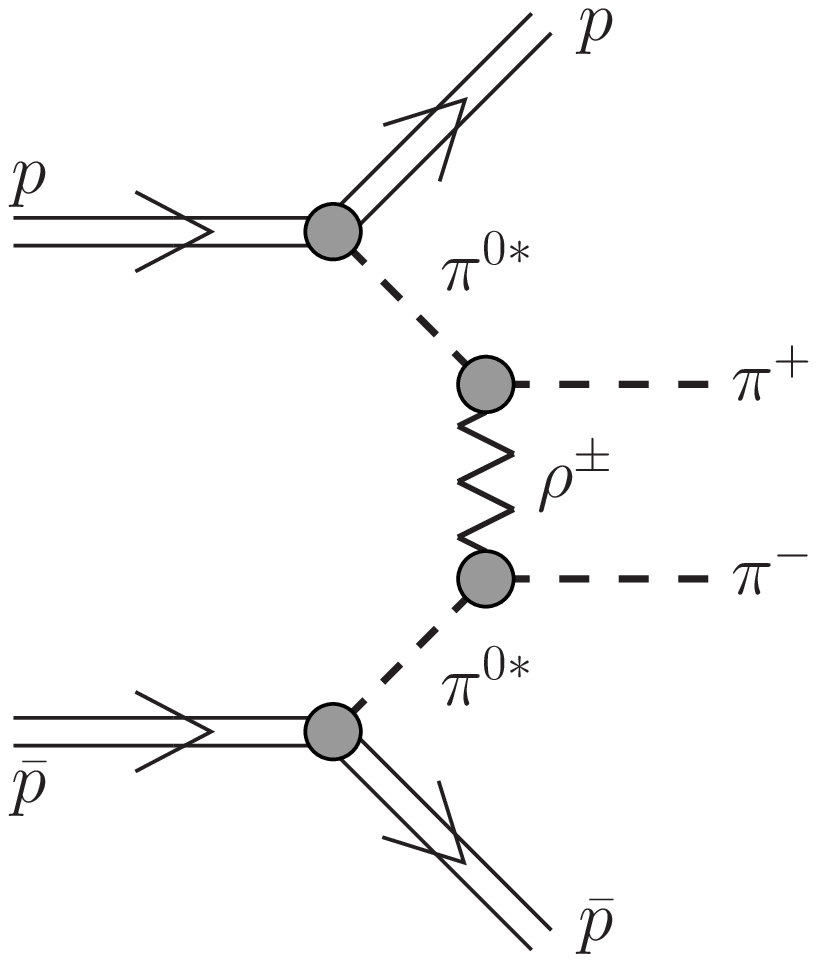}
\includegraphics[width=0.27\textwidth]{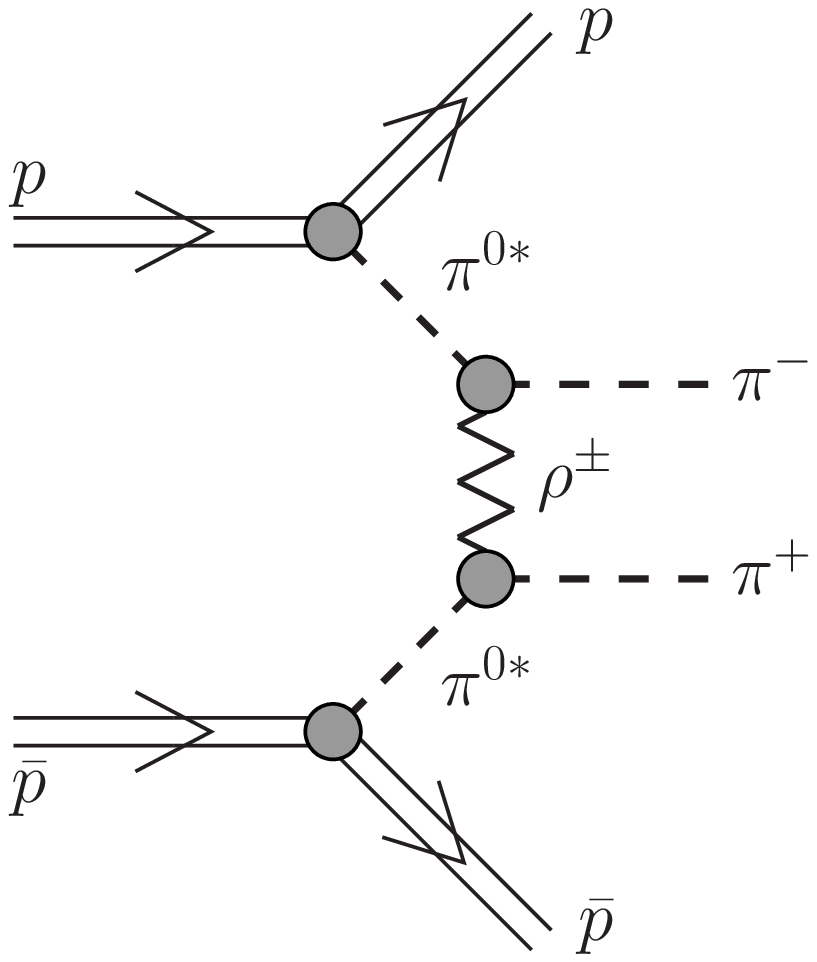}
   \caption{\label{fig:pion-pion_rescattering_diagrams}
   \small 
A sketch of the high-energy pion-pion rescattering mechanisms.
}
\end{figure}

For the $p p \to p p \pi^+ \pi^-$ or
$p \bar{p} \to p \bar{p} \pi^+ \pi^-$ reactions
there is another type of semi-diffractive contribution shown
in Fig.\ref{fig:pion-pion_rescattering_diagrams}. 

Similarly as for the $p \bar{p} \to N \bar{N} f_{0}(1500)$ reaction
(see \cite{SL09})
the amplitude squared - averaged over the initial and summed
over the final state - for these processes can be written as:
%
%
%
\begin{eqnarray}
\overline{|{\cal M}|^2}=\dfrac{1}{4}&
\left[ \left( E_a + m \right)\left( E_1+ m \right)
\left(\dfrac{{\bf p}_a^2}{(E_a + m)^2} + \dfrac{{\bf p}_1^2}{(E_1 + m)^2} -
\dfrac{ 2 {\bf p}_a \cdot {\bf p}_1}{(E_a + m)(E_1 + m)}\right)\right] \times 2 \nonumber \\
  \times &
\dfrac{g_{\pi NN}^2}{(t_1 - m_{\pi}^2)^2} F_{\pi NN}^2(t_1)
\; \times \; |{\cal M}_{\pi^{0*}\pi^{0*}\to\pi^+\pi^-}(s_{34},t_{0};t_{1},t_{2})|^2 
\; \times \; \dfrac{g_{\pi NN}^2}{(t_2 - m_{\pi}^2)^2} F_{\pi NN}^2(t_2) \nonumber \\
  \times &
\left[ \left( E_b + m \right)\left( E_2+ m \right)
\left(\dfrac{{\bf p}_b^2}{(E_b + m)^2} + \dfrac{{\bf p}_2^2}{(E_2 + m)^2} -
\dfrac{ 2 {\bf p}_b \cdot {\bf p}_2}{(E_b + m)(E_2 + m)}\right)\right] \times 2 \;. \nonumber \\
\label{pion-pion_amplitude}
\end{eqnarray}
%
%
In the formula above $m$ is the mass of the nucleon,
$E_a, E_b$ and $E_1, E_2$ are energies of initial and outgoing nucleons,
${\bf p}_a, {\bf p}_b$ and ${\bf p}_1, {\bf p}_2$ are corresponding
three-momenta and $m_{\pi}$ is the pion mass.
The factor $g_{\pi NN}$ is the familiar pion nucleon coupling constant 
and is relatively well known \cite{ELT02} ($\frac{g_{\pi NN}^2}{4 \pi}$ = 13.5 -- 14.6).
In our calculations the coupling constants are taken as
$g^2_{\pi N N}/4\pi$ = 13.5.

At high-energies the pion-pion scattering amplitude 
of the subprocess $\pi^{0*}\pi^{0*}\rightarrow\pi^+\pi^-$ 
with virtual initial pions can be written, 
similarly as for ${\pi}N$ scattering:
\begin{eqnarray}
{\cal M}_{\pi^{0*} \pi^{0*} \to \pi^{+} \pi^{-}} (s_{34},t_{0};t_{1},t_{2})&=&
(a_{\rho} - \mathrm{i}) \; s_{34}\left( \frac{s_{34}}{s_{0}}\right)^{\alpha_{I\!\!R}(t_{0})-1} \exp \left( {\frac{B_{\pi\pi}}{2}t_{0}}\right) F_{\pi^{0*}}(t_{1})F_{\pi^{0*}}(t_{2})\; .\nonumber
\\
\label{Regge_amplitude}
\end{eqnarray}
We have parametrized the $t_0$ dependence in the exponential form.
The slope parameter is not well know, however it may be expected
to be $B_{\pi\pi}$ $\thicksim$ 4-6 GeV$^{-2}$.
In the formula above $F_{\pi^*}(t_{1,2})$ are extra correction 
factors due to off-shellness of initial pions. 
We use exponential form factors of the type (\ref{off-shell_form_factors}).
To exclude resonance regions we "correct" the Regge parametrization 
(\ref{Regge_amplitude})
by multiplying by a factor $f_{cont}^{\pi N}(W_{34})$ 
(as in (\ref{Correction_factor})).

In the case of central production of pion pairs not far from the threshold 
rather large transverse momenta squared $t_1$ and $t_2$ are involved
and one has to include non-point-like and off-shellness nature of the particles 
involved in corresponding vertices. This is incorporated via $F_{\pi NN}(t_1)$
or $F_{\pi NN}(t_2)$ vertex form factors. 
In the meson exchange approach \cite{MHE87} they 
are parameterized in the monopole form as
\begin{equation}
F_{\pi N N}(t_{1,2}) = \frac{\Lambda^2 - m_{\pi}^2}{\Lambda^2 - t_{1,2}} \;.
\label{F_piNN_formfactor}
\end{equation} 
%
Typical values of the form factor parameters are $\Lambda$ = 1.2--1.4 GeV \cite{MHE87},
however the Gottfried Sum Rule violation prefers smaller 
$\Lambda \approx$ 0.8 GeV \cite{GSR}.

\section{The differential cross section}

The differential cross section for the $2 \to 4$ reaction
is given as
\begin{eqnarray}
d \sigma = \frac{1}{2s} \overline{ |{\cal M}|^2} (2 \pi)^4 
\delta^4 (p_a + p_b - p_1 - p_2 - p_3 - p_4)
\frac{d^3 p_1}{(2 \pi)^3 2 E_1} 
\frac{d^3 p_2}{(2 \pi)^3 2 E_2} 
\frac{d^3 p_3}{(2 \pi)^3 2 E_3}
\frac{d^3 p_4}{(2 \pi)^3 2 E_4} \;. 
\label{dsigma_for_2to4}
\end{eqnarray}
This can be written in a useful form:
\begin{eqnarray}
d \sigma &=& \frac{1}{2s} \overline{ |{\cal M}|^2} 
\delta^4 (p_a + p_b - p_1 - p_2 - p_3 - p_4) 
\frac{1}{(2 \pi)^{8}}\frac{1}{2^{4}}\nonumber \\
&\times& 
(dy_{1}p_{1t}dp_{1t}d\phi_{1})(dy_{2}p_{2t}dp_{2t}d\phi_{2})
(dy_{3}d^{2}p_{3t})(dy_{4}d^{2}p_{4t}) \;. 
\label{dsigma_for_2to4}
\end{eqnarray}
This can be further simplified:
\begin{eqnarray}
d \sigma &=&  \frac{1}{2s} \overline{ |{\cal M}|^2} 
\delta (E_a + E_b - E_1 - E_2 - E_3 - E_4)
\delta^3 (p_{1z} + p_{2z} + p_{3z} + p_{4z})
\frac{1}{(2 \pi)^{8}}\frac{1}{2^{4}}
\nonumber \\
&\times& 
(dy_{1}p_{1t}dp_{1t}d\phi_{1}) 
(dy_{2}p_{2t}dp_{2t}d\phi_{2}) 
dy_{3} dy_{4} d^{2}p_{m} \;.
\label{dsigma_for_2to4}
\end{eqnarray}
Above we have introduced an auxiliary quantity:
\begin{equation}
\textbf{p}_{m} = \textbf{p}_{3t} - \textbf{p}_{4t}  \;. 
\label{pmt}
\end{equation}
We choose transverse momenta of the outgoing nucleons ($p_{1t},p_{2t},p_{3t},p_{4t}$),
azimuthal angles of outgoing nucleons ($\phi_{1},\phi_{2}$) and rapidity of the
pions ($y_{3},y_{4}$) as independent kinematically complete variables.
Then the cross section can be calculated as:
\begin{eqnarray}
d \sigma &=&\sum_{k} {\cal J}^{-1}(p_{1t},\phi_{1},p_{2t},\phi_{2},y_{3},y_{4},p_{m},\phi_{m})|_{k} 
\frac{\overline{ |{\cal M}(p_{1t},\phi_{1},p_{2t},\phi_{2},y_{3},y_{4},p_{m},\phi_{m})|^2 }  }
{2 \sqrt{s (s-4 m^2)}} 
\frac{1}{(2 \pi)^{8}}\frac{1}{2^{4}}
\nonumber \\
&\times& 
p_{1t}dp_{1t}d\phi_{1}
p_{2t}dp_{2t}d\phi_{2}
\frac{1}{4} dy_{3} dy_{4} d^{2}p_{m} \;, 
\label{dsigma_for_2to4_end}
\end{eqnarray}
where the $\delta$ functions have been totally eliminated 
and $k$ denotes symbolically discrete solutions of the set
of equations for energy and momentum conservation:
%
\begin{eqnarray}
\left\{ \begin{array}{rcl}
\sqrt{s} - E_3 - E_4  &=&
\sqrt{m_{1t}^2+p_{1z}^{2}} + \sqrt{m_{2t}^2+p_{2z}^{2}} \; ,  \\
-p_{3z} - p_{4z} &=& p_{1z} + p_{2z}   \; ,
\end{array} \right.
\label{energy_momentum_conservation}
\end{eqnarray}
where $m_{1t}$ and $m_{2t}$ are transverse masses of outgoing nucleons.
The solutions of Eq.(\ref{energy_momentum_conservation})
depend on the values of integration variables:
$p_{1z} = p_{1z}(p_{1t},p_{2t},p_{3t},p_{4t},\phi_{1},\phi_{2},y_{3},y_{4})$ and
$p_{2z} = p_{2z}(p_{1t},p_{2t},p_{3t},p_{4t},\phi_{1},\phi_{2},y_{3},y_{4})$.

In Eq. (\ref{dsigma_for_2to4_end}) an extra Jacobian of the transformation $(y_{1},y_{2})\rightarrow(p_{1z},p_{2z})$ has appeared:
\begin{equation}
{\cal J}_k =
\left| \frac{p_{1z}(k)}{\sqrt{m_{1t}^2+p_{1z}(k)^2}} -
  \frac{p_{2z}(k)}{\sqrt{m_{2t}^2+p_{2z}(k)^2}} \right| \; .
\end{equation}
In the limit of high energies and central production, i.e. $p_{1z} \gg$ 0 (very forward nucleon1), 
$-p_{2z} \gg$ 0 (very backward nucleon2) the Jacobian becomes a constant
${\cal J} \to \tfrac{1}{2}$.

To calculate the total cross section one has to calculate 
the 8-dimensional integral (see Eq.(\ref{dsigma_for_2to4_end})) numerically. 
This requires some care.

In the next section we shall show our predictions
for several differential distributions in different
variables.

\section{Results}
\begin{figure}[!h]  
\includegraphics[width=0.32\textwidth]{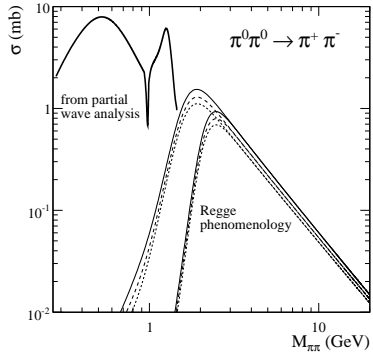}
   \caption{\label{fig:partial_vs_Regge}
   \small 
The angle-integrated cross section for the reaction $\pi^0 \pi^0 \to \pi^+ \pi^-$.
We present contributions obtained from partial wave analysis 
\cite{LSK09} and Regge phenomenology \cite{SNS02}
for corrected ($W_0$ = 1.5, 2 GeV and $a$ = 0.2 GeV 
in Eq.(\ref{Correction_factor}))
extrapolations to low energies.
}
\end{figure}

Before we go to our four-body reaction 
let us focus for a moment on $\pi^0 \pi^0 \to \pi^+ \pi^-$
on-shell scattering. In Fig.\ref{fig:partial_vs_Regge} we show 
the total (angle-integrated) cross section 
for the $\pi^0 \pi^0 \to \pi^+ \pi^-$ process.
We include both the pion-pion rescattering contribution obtained
from partial wave analysis \cite{LSK09} as well as contribution from 
the Regge phenomenology at higher energies.
The parameters of the Regge amplitude for the $\pi \pi \to \pi \pi$
scattering were obtained in Ref.\cite{SNS02} from different
isospin combinations of nucleon-(anti)nucleon, and pion-nucleon
scattering assuming Regge factorization. For our case of
$\pi^0 \pi^0 \to \pi^+ \pi^-$ reaction only the $\rho$-reggeon exchange
is relevant.
We show predictions for the Regge contribution
for corrected ($W_0$ = 1.5, 2 GeV and $a$ = 0.2 GeV in Eq.(\ref{Correction_factor}))
extrapolations to low energies and
for different values of the slope parameter $B_{\pi\pi}$ = 4 GeV$^{-2}$ (dotted lines),
$B_{\pi\pi}$ = 5 GeV$^{-2}$ (dashed line) and $B_{\pi\pi}$ = 6 GeV$^{-2}$ (solid lines).
A relatively good matching is achieved without extra fitting
the model parameters. In the following we shall focus on the 
higher-$M_{\pi\pi}$ Regge component which dominates at higher
energies (see a next figure).

\begin{figure}[!h]  
\includegraphics[width=0.6\textwidth]{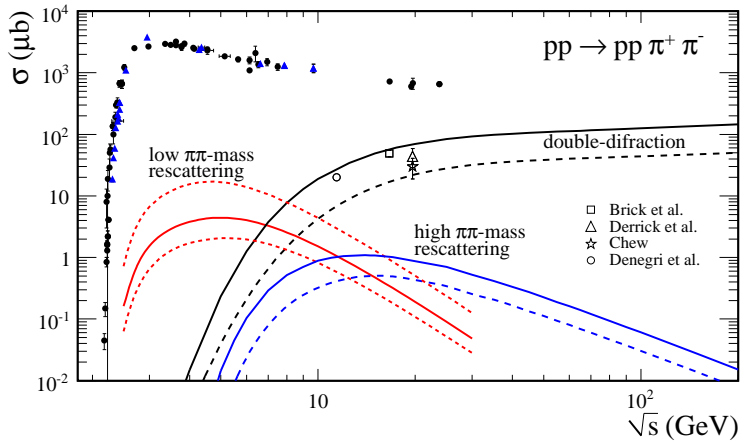}
   \caption{\label{fig:pipi_vs_pom}
   \small 
Cross section for the $p p \to p p \pi^+ \pi^-$ reaction
integrated over phase space as a function of the center-of-mass energy.
We compare the pion-pion rescattering 
and double-diffractive contributions with the
experimental data (open symbols represent DPE contribution 
from Refs.\cite{Denegri75,Brick83,Derrick,Chew}
and filled symbols show the cross sections 
for the $ p p \to p p \pi^+ \pi^-$ reaction (black circles) 
from Ref.\cite{bib_pp_pppipi}
and the $p \bar{p} \to p \bar{p} \pi^+ \pi^-$ reaction (blue triangles)
from Ref.\cite{bib_ppbar_ppbarpipi} ).
The theoretical uncertainties for these contributions are shown in addition.
}
\end{figure}

In Fig.\ref{fig:pipi_vs_pom} we present the total cross section for 
the $pp \to pp\pi^+ \pi^-$ reaction,
i.e. the cross section integrated over full phase space,
as a function of the center-of-mass energy.
We show theoretical predictions from the models
calculations with $\Lambda = 0.8$ GeV
and $\Lambda^{2}_{off, E} = 1$ GeV$^{2}$ (solid lines) and $\Lambda^{2}_{off, E} = 0.5$ GeV$^{2}$ (dashed lines).
The bottom dotted line was obtained with $\Lambda = 0.8$ GeV and $\Lambda^{2}_{off, E} = 0.5$ GeV$^2$
while the top dotted line with $\Lambda = 1.4$ GeV and 
$\Lambda^{2}_{off, E} = 2$ GeV$^2$.
Details of the low-$M_{\pi \pi}$ rescattering contribution
can be found in Ref.\cite{LSK09}.
The search for a double pomeron exchange mechanism contribution
leads to an upper limits of 
$\simeq 20 \mu$b (for $M_{\pi\pi}\leqslant$ 0.7 GeV) \cite{Denegri75}, 
$(49 \pm 5.5)\mu$b \cite{Brick83}, $(30 \pm 11)\mu$b \cite{Derrick} 
and $(44 \pm 15)\mu$b \cite{Chew}.
The experimental value of the cross section taken from \cite{Derrick} 
was obtained for $M_{p\pi}$ $>$ 2 GeV and no limitation
on $M_{\pi\pi}$; reduces however to 9 $\mu$b 
for $M_{\pi\pi}\leqslant$ 0.6 GeV \cite{Derrick}.
For comparison we show the full cross sections 
for the $pp \to pp\pi^+\pi^-$ reaction (filled black circles) 
from Ref.\cite{bib_pp_pppipi}
and for the $p \bar{p} \to p \bar{p} \pi^+ \pi^-$ reaction (filled blue triangles)
from Ref.\cite{bib_ppbar_ppbarpipi}
which are more than 1 mb for ($2.5<\sqrt{s}<10$) GeV
\footnote{This is a significant contribution to the total $pp$
cross section.}.
Clearly for low energies ($\sqrt{s} <$ 20 GeV) neither
exclusive double diffraction nor pion-pion rescattering
constitute the dominant mechanism. Here the production
of single and double resonances is the dominant mechanism 
(see e.g. \cite{LSK09}). The mechanism of the resonant production
is rather complicated and will not be discussed in the present
analysis.

The results depend on the value of the nonperturbative, 
a priori unknown parameter of the form factor responsible for 
off-shell effects.
In Table \ref{tab:sig_tot_ff} we have collected integrated cross sections
for selected energies and different values of model parameters.
We show how the uncertainties of the form factor parameters 
affect our final results.

\begin{table}
\caption{Full-phase-space integrated cross section (in $\mu$b) 
for exclusive double diffractive $\pi^+ \pi^-$ production at selected 
center-of-mass energies
and different values of the off-shell-form factor parameters.
Here $W_0$ = 2 GeV and $a$ = 0.2 GeV in Eq.(\ref{Correction_factor}).
No absorption effects were included explicitly.}
\label{tab:sig_tot_ff}
\begin{center}
\begin{tabular}{|c|c|c|c|c|c|}
\hline
$F(t_{1,2})$ & $\Lambda^{2}_{off}$ (GeV$^{2}$) & W = 5.5 GeV & W = 200 GeV & W = 1960 GeV & W = 14 TeV \\ 
\hline
$\exp\left( (t_{1,2}-m_{\pi}^{2})/\Lambda^{2}_{off, E}\right) $ 
    &   0.5   &   0.1   &   50.3   &    96.4   &   179.1 \\
	&   1     &   0.6   &  146.2   &   287.2   &   535.2 \\
\hline
$(\Lambda^{2}_{off, M} - m_{\pi}^2)/(\Lambda^{2}_{off, M} - t_{1,2})$
	&   0.5   &   0.02  &   18.9   &    35.6   &    66   \\
 	&   1     &   0.18  &   64.6   &   125.2   &   232.8 \\
\hline
$\left( (\Lambda^{2}_{off, D} - m_{\pi}^2)/(\Lambda^{2}_{off, D} - t_{1,2})\right) ^2$
	&   0.5   &   0.31  &   83.6   &   164.2   &   306.5 \\
	&   1     &   1.15  &  217.5   &   437.9   &   822.4 \\
\hline
\end{tabular}
\end{center}
\end{table}

\begin{figure}[!h]  
\includegraphics[width=0.45\textwidth]{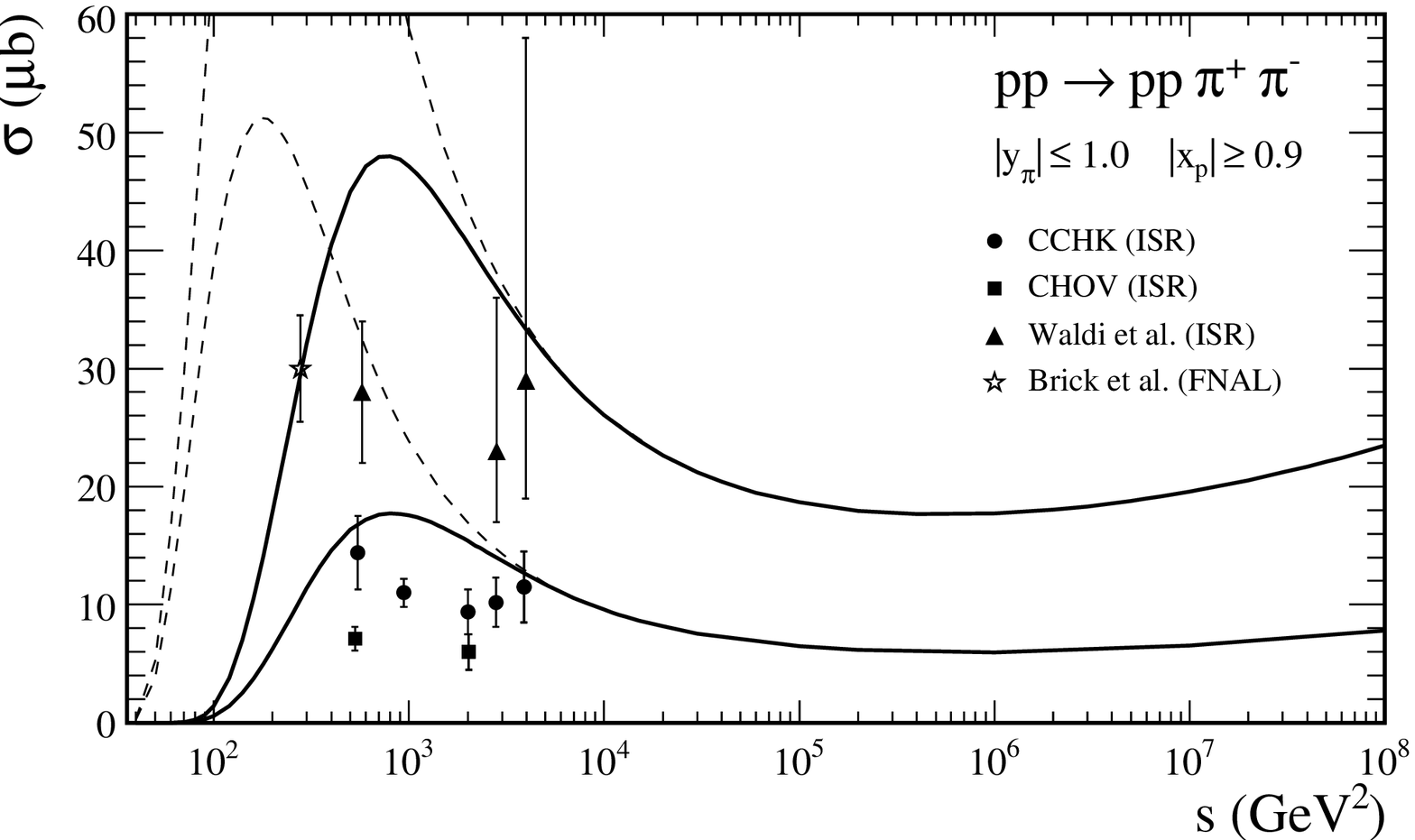}\qquad
\includegraphics[width=0.45\textwidth]{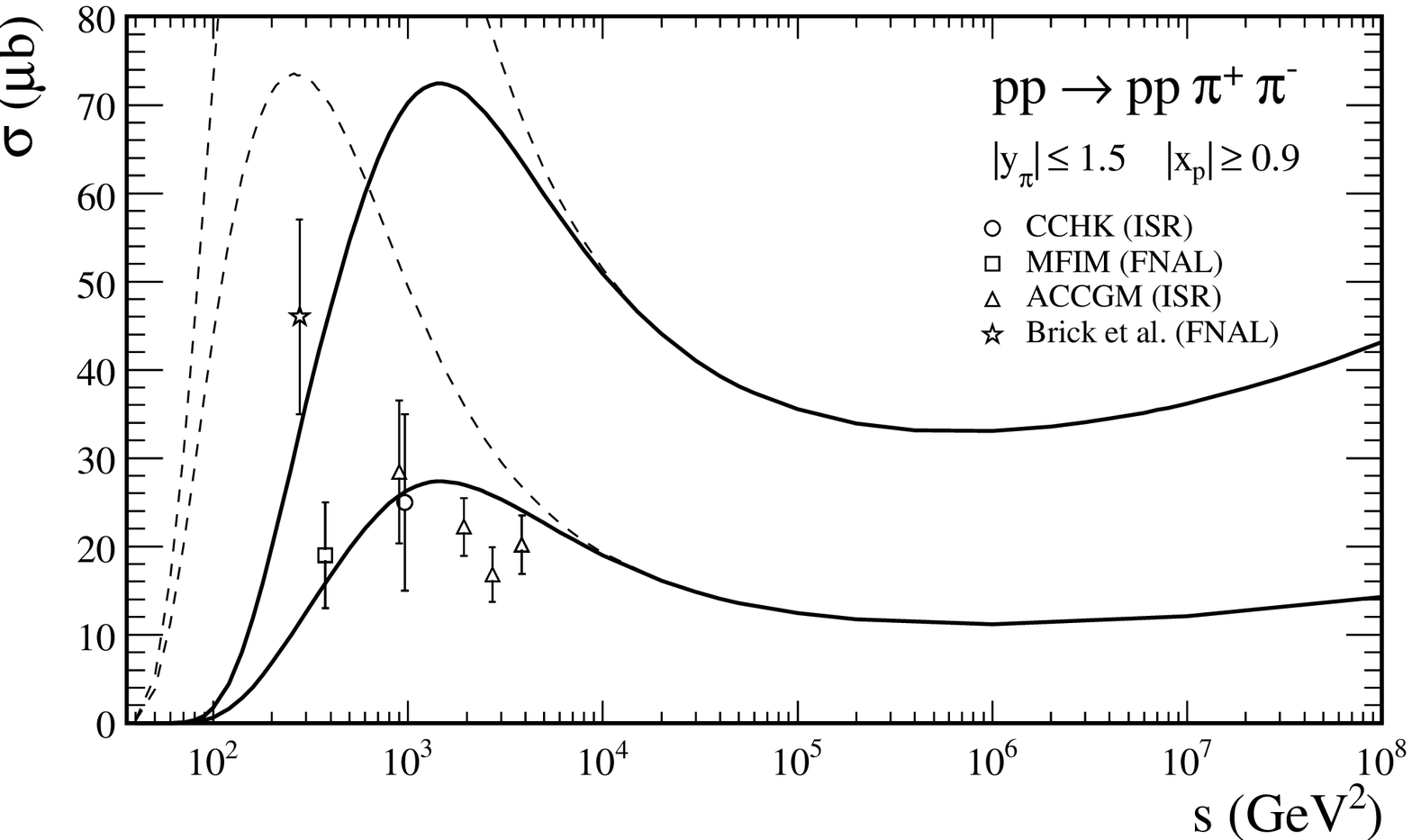}
   \caption{\label{fig:total_W_cuts}
   \small 
Cross section for the $p p \to p p \pi^+ \pi^-$ reaction
integrated over phase space with cuts relevant for a given
experiments \cite{Brick83,BDD,Waldi83,DellaNegra}.
The experimental value from \cite{DellaNegra} 
was obtained for the different cut $\Delta y = |y_{p}-y_{\pi}|$ $>$ 2.
We show results for different values of the parameter 
$\Lambda^{2}_{off, E}$ = 0.5 GeV$^{2}$ (lower lines),
$\Lambda^{2}_{off, E}$ = 1 GeV$^2$ (upper lines)
and for the naive (dashed lines) 
and corrected (solid lines with $W_0$ = 2 GeV and $a$ = 0.2 GeV) 
extrapolations to low energies.
}
\end{figure}

In Fig.\ref{fig:total_W_cuts} we show predictions
for different values of the parameter $\Lambda^{2}_{off, E}$ = 0.5 GeV$^{2}$ (lower lines),
$\Lambda^{2}_{off, E}$ = 1 GeV$^2$ (upper lines) and for naive (dashed lines) 
and corrected (solid lines with $W_0$ = 2 GeV and $a$ = 0.2 GeV)
extrapolations to low energies.
The experimental cuts on the rapidity of the pions 
are included when comparing our results with
existing experimental data.
Although not all the data are in good agreement with the predictions, 
their general trend follows the theoretical expectations.
No absorption effects were included in this calculation.
In general, the higher energy the higher absorption effects.
The bare cross section rises with energy. The absorption corrections are
expected to lower or even stop the rise.
Consistent inclusion of absorption effects is rather difficult
and will not be studied here.

\begin{figure}[!h] 
\includegraphics[width = 0.24\textwidth]{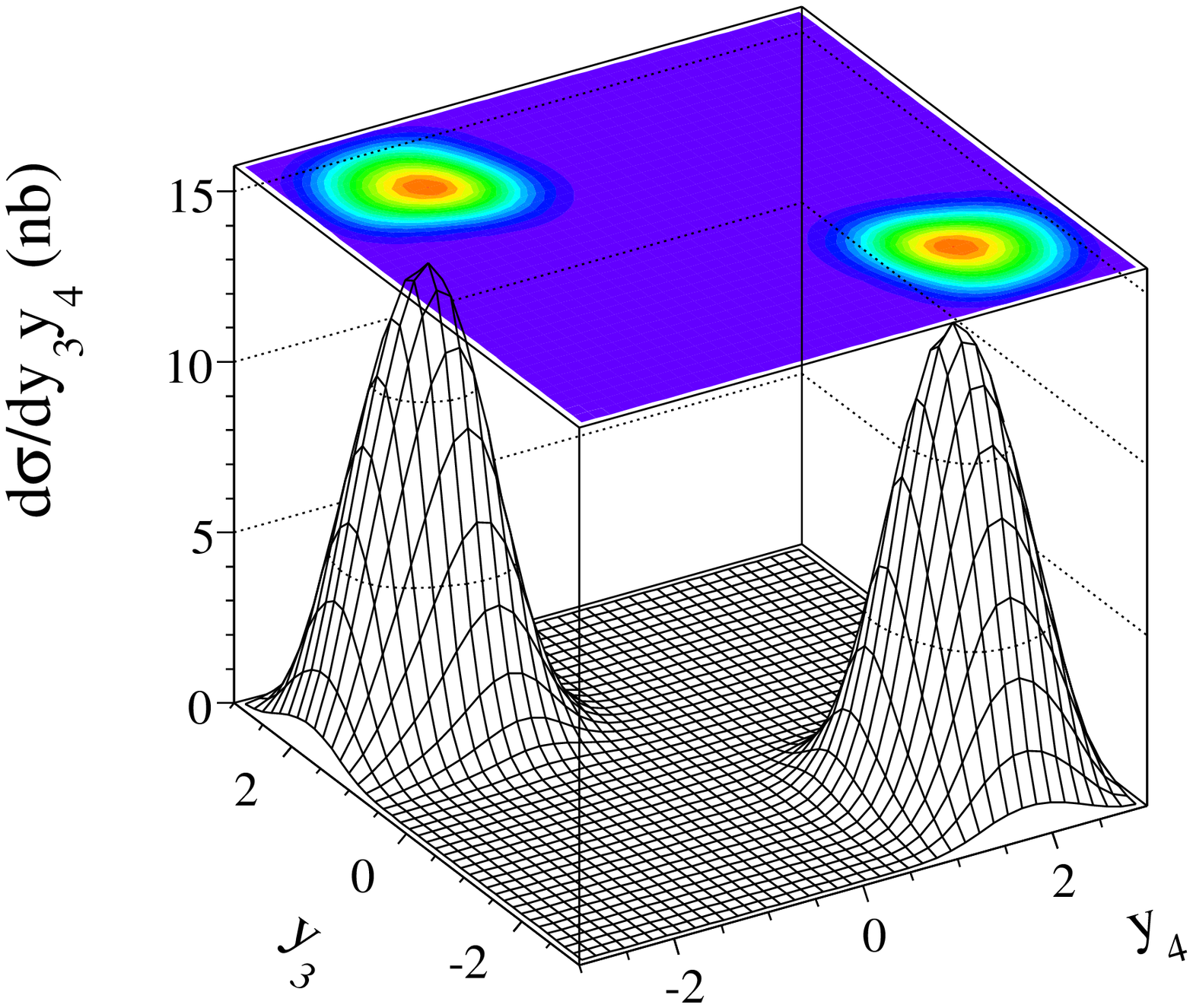}
\includegraphics[width = 0.24\textwidth]{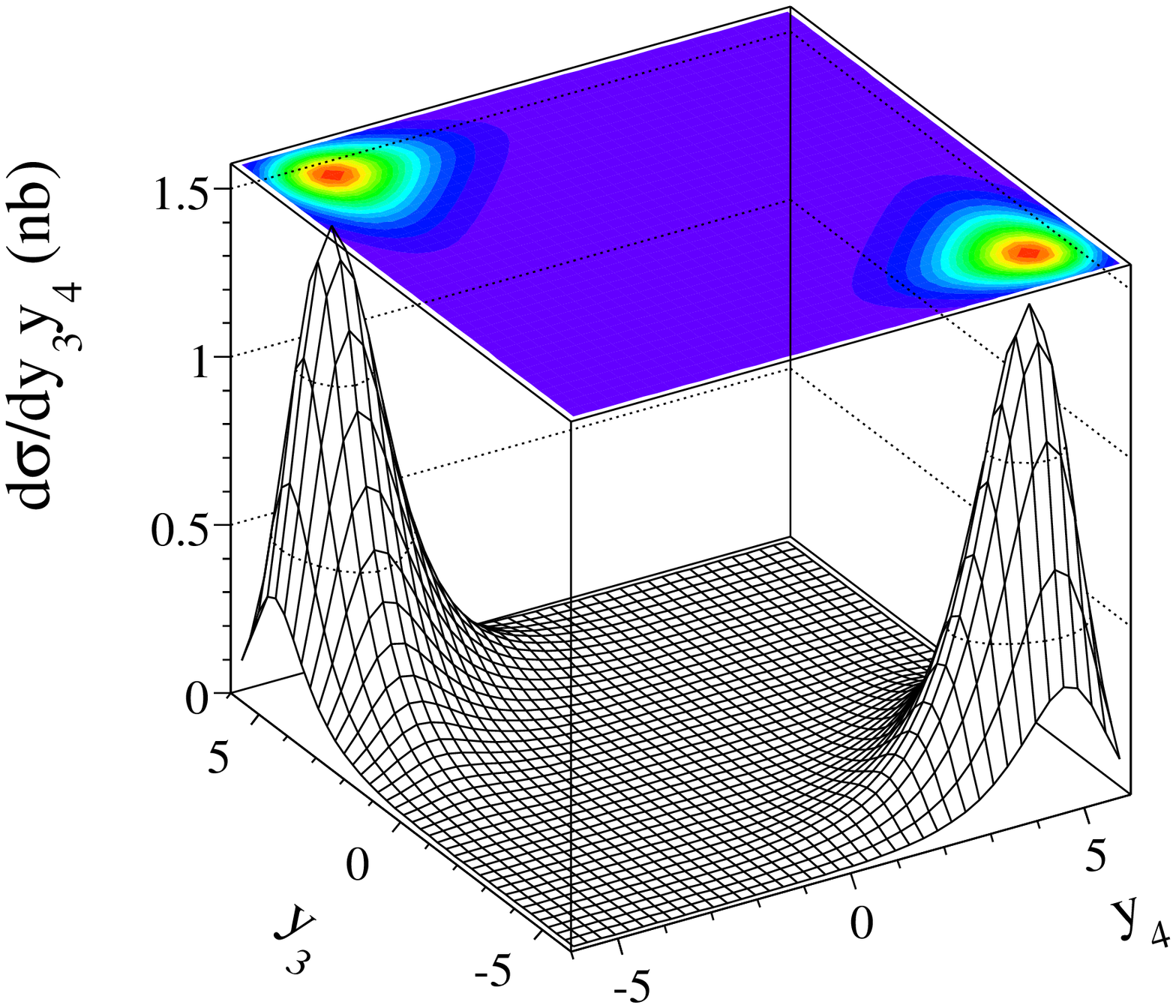}
\includegraphics[width = 0.24\textwidth]{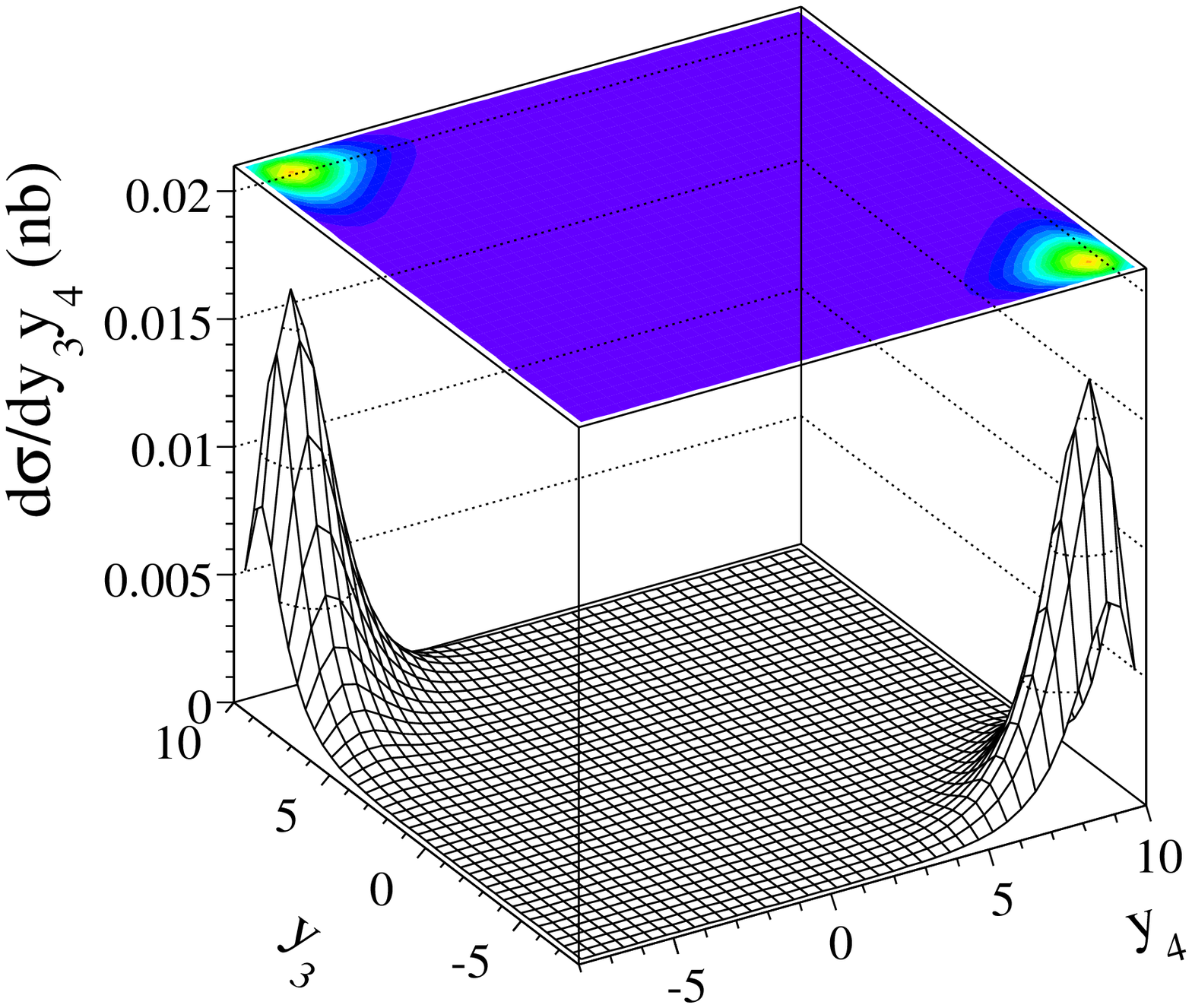}
\includegraphics[width = 0.24\textwidth]{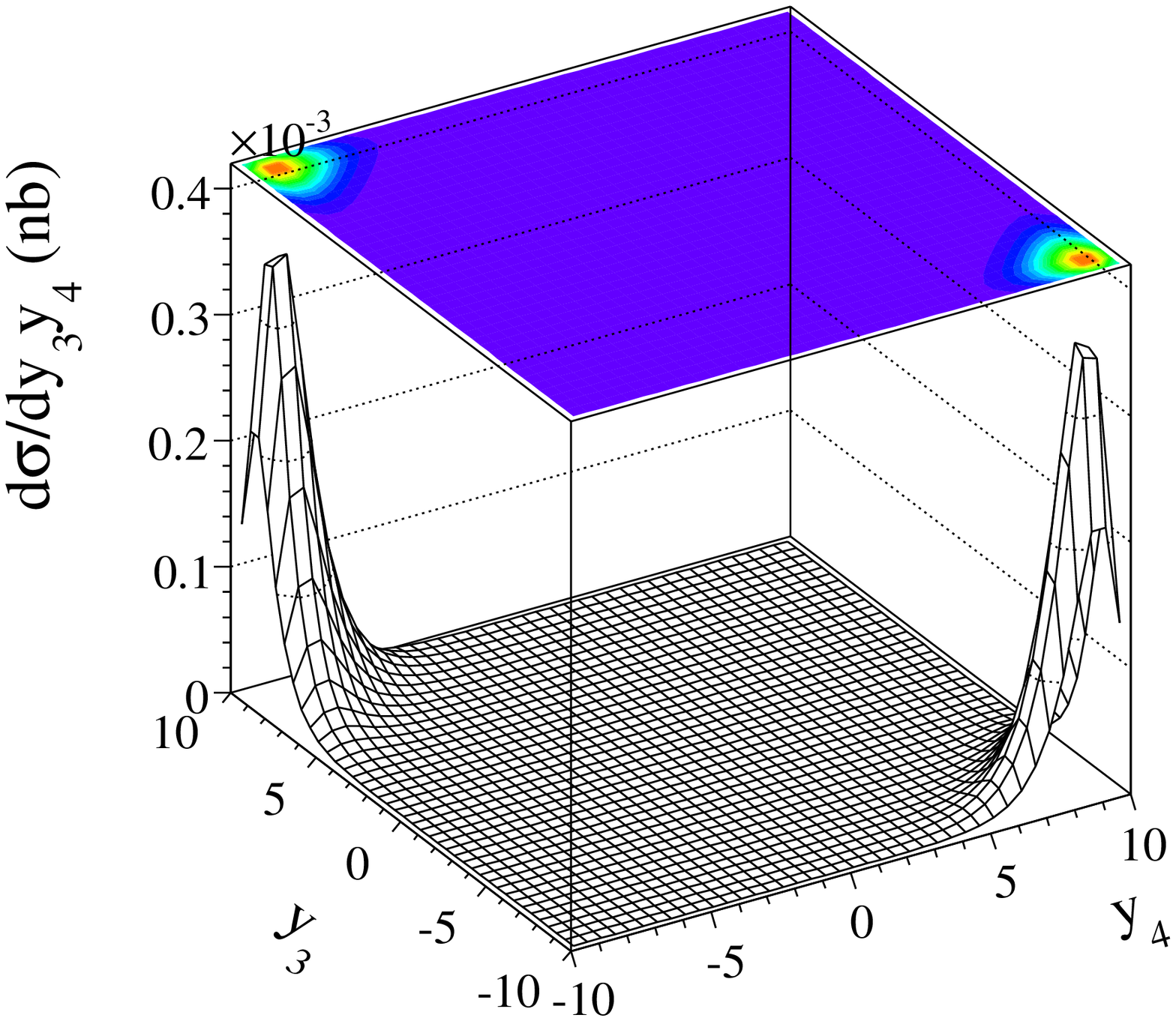}
   \caption{\label{fig:y3y4_pipi_rescattering}
   \small 
Differential cross section in $(y_3,y_4)$ for the pion-pion rescattering 
contribution for different incident energies: 
$W$ = 5.5 (PANDA), 200 (RHIC), 1960 (Tevatron), 14000 (LHC) GeV.
}
\end{figure}

\begin{figure}[!h]  
\includegraphics[width = 0.24\textwidth]{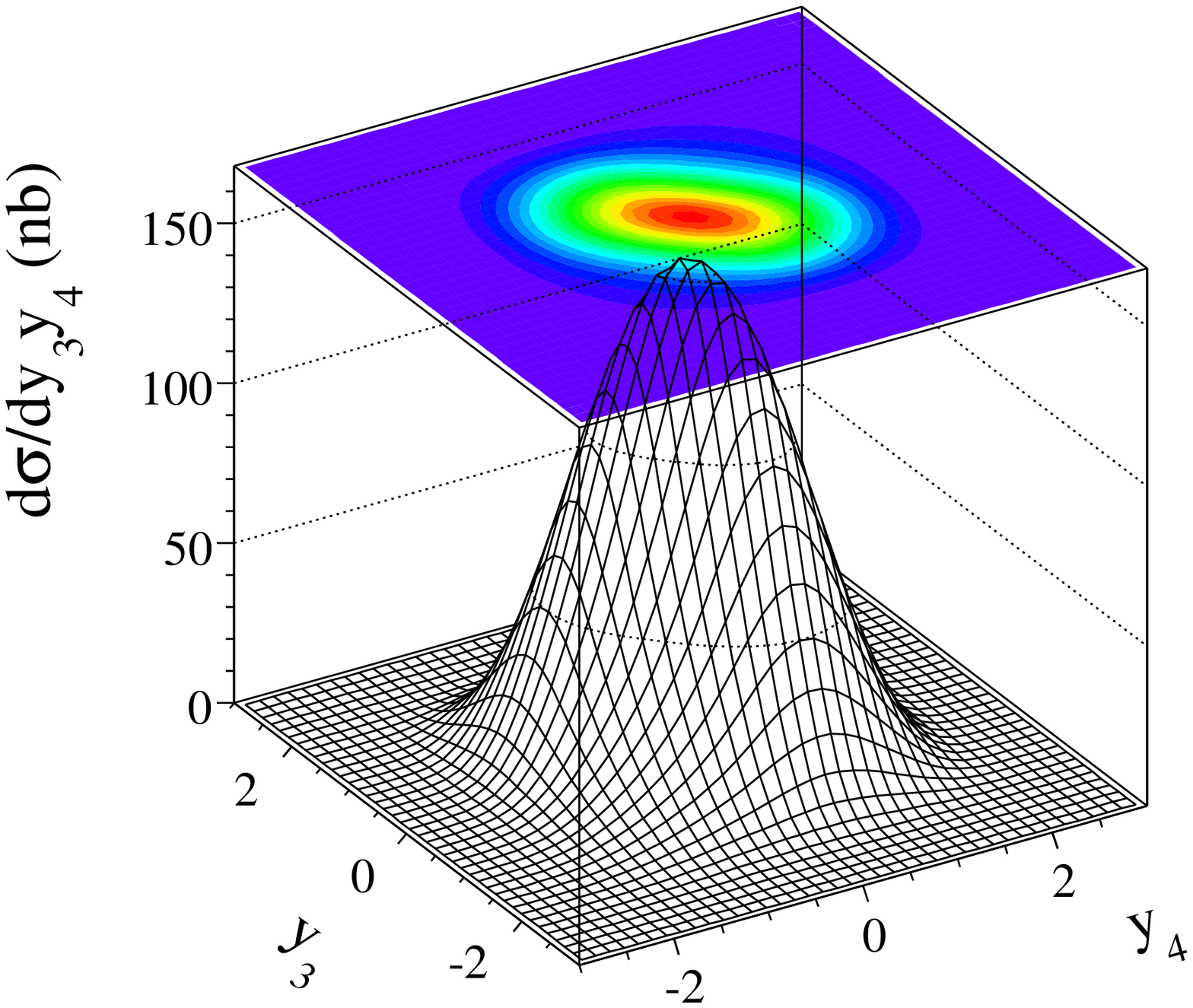}
\includegraphics[width = 0.24\textwidth]{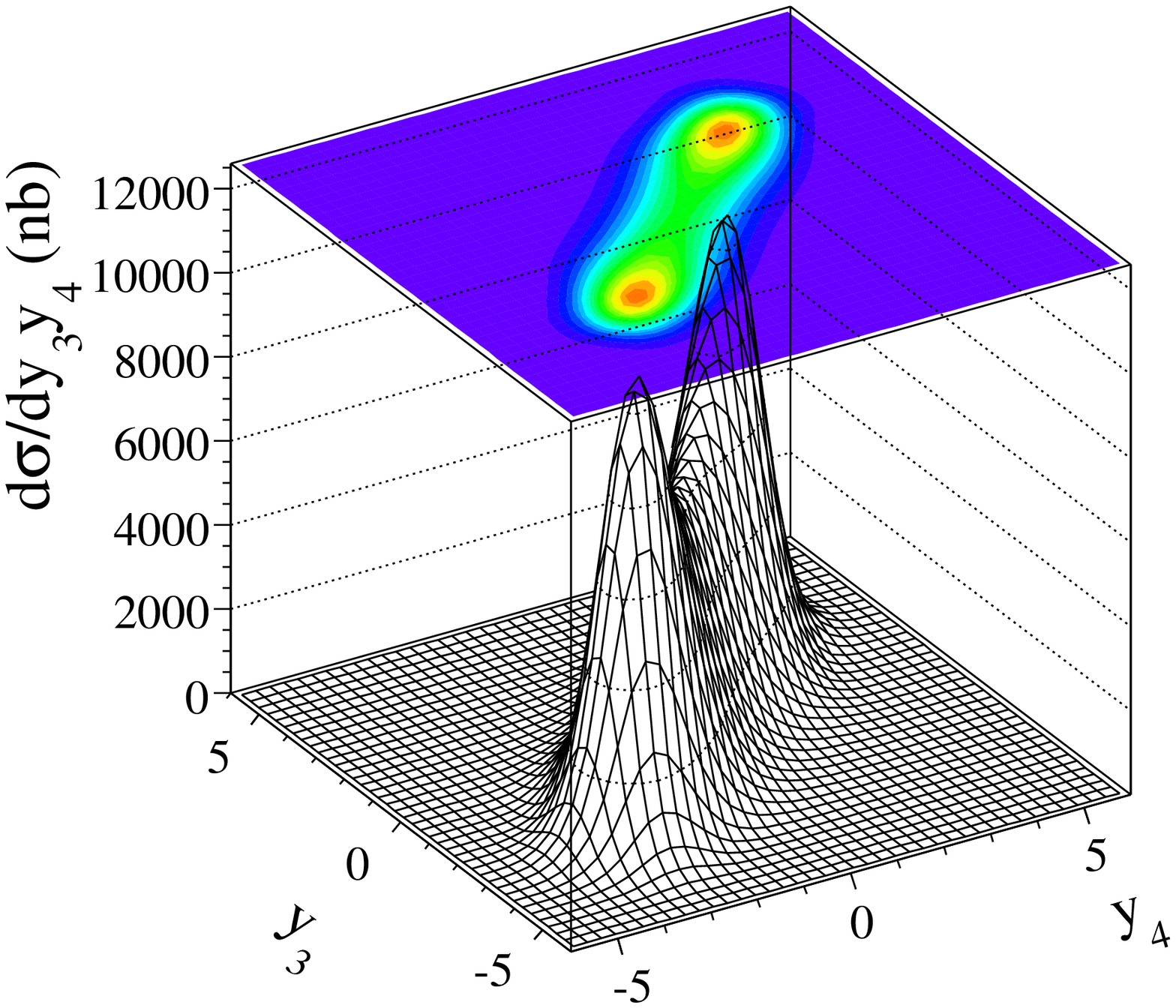}
\includegraphics[width = 0.24\textwidth]{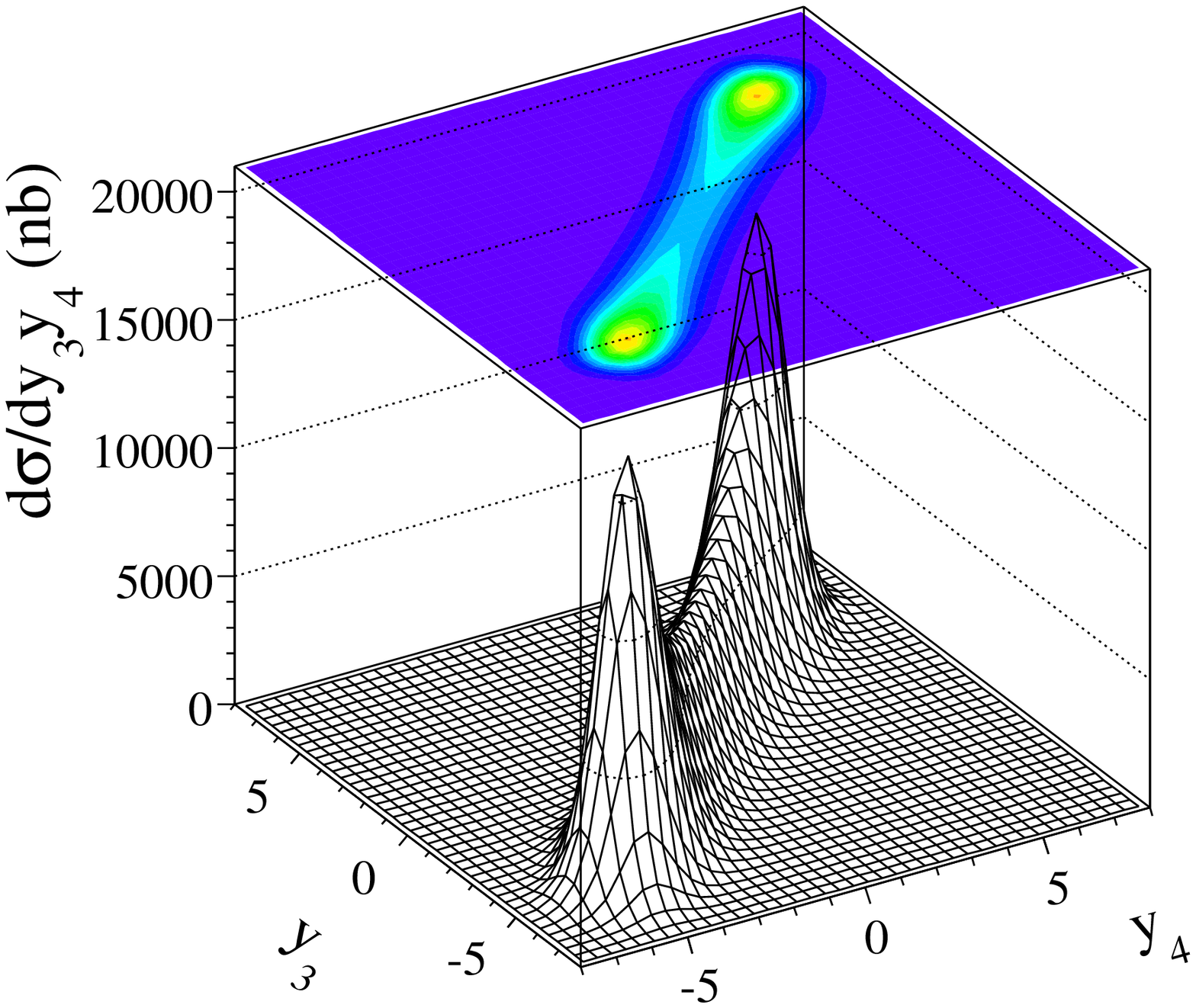}
\includegraphics[width = 0.24\textwidth]{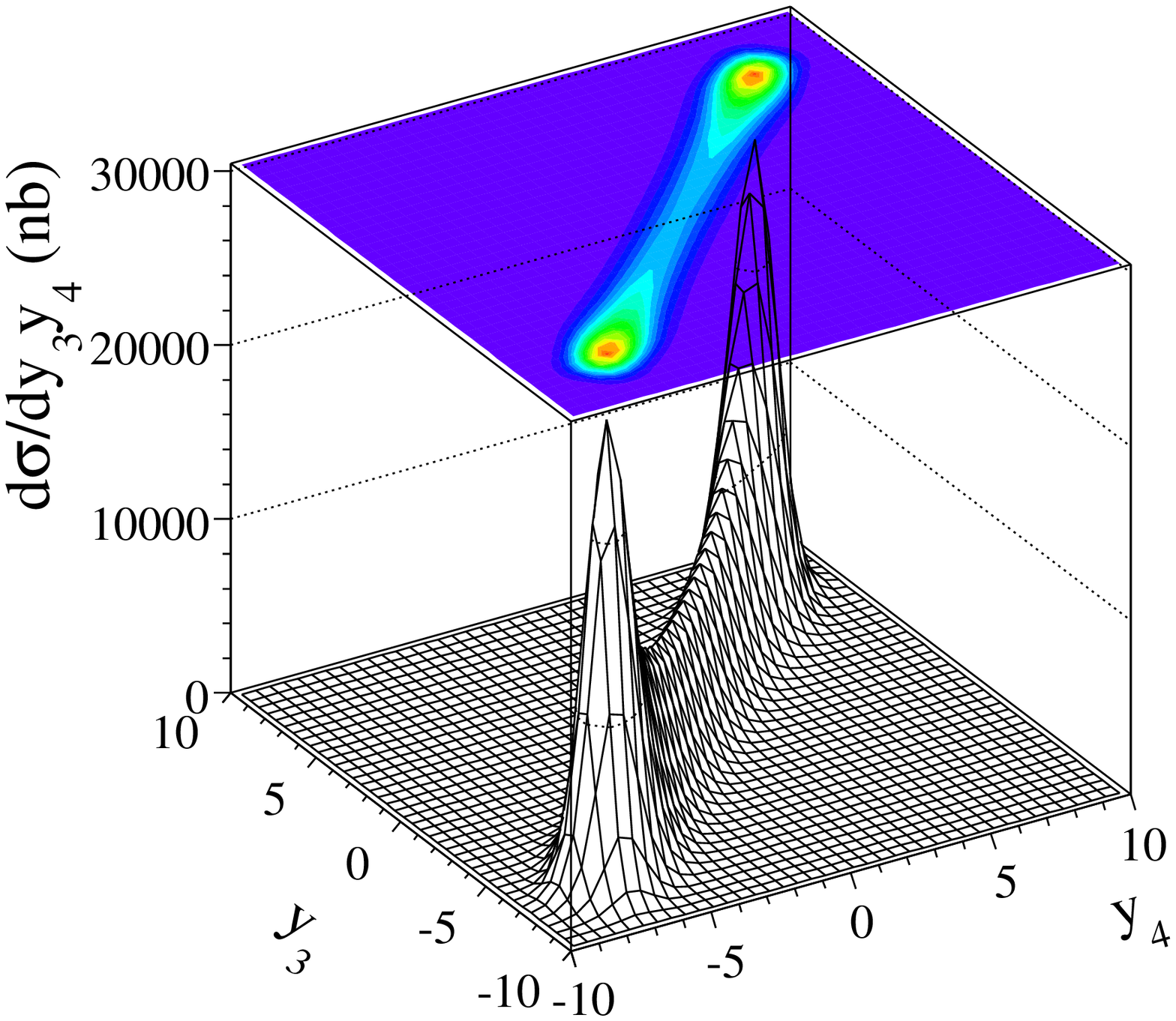}
   \caption{\label{fig:y3y4_central_double_diffraction}
   \small 
Differential cross section in $(y_3,y_4)$ for the double-diffractive
contribution for different incident energies: 
$W$ = 5.5 (PANDA), 200 (RHIC), 1960 (Tevatron), 14000 (LHC) GeV.
}
\end{figure}

The distribution in the $(y_3, y_4)$ space is particularly 
interesting.
In Fig.\ref{fig:y3y4_pipi_rescattering} and
Fig.\ref{fig:y3y4_central_double_diffraction} we show distributions
for the pion-pion rescattering and double-diffractive contributions,
respectively.
In this calculation the cut-off parameter $\Lambda^{2}_{off, E}$ = 1 GeV$^{2}$.
The cross section for the pion-pion rescattering drops quickly with
the center-of-mass energy. The rescattered pions are emitted preferentially 
in different hemispheres, $\pi^+$ at positive $y_3$ and $\pi^-$
at negative $y_4$ or $\pi^+$ at negative $y_3$ and $\pi^-$ at positive 
$y_4$.
The bare (without absorption effects) cross section for 
the double-diffractive contribution grows with energy. 
At high energies the pions are emitted preferentially
in the same hemispheres, i.e. $y_3, y_4 >$ 0 or $y_3, y_4 <$ 0.
While at low energies (PANDA) both contributions (exclusive double diffraction
and pion-pion rescattering) 
overlap, at high energies (RHIC, Tevatron, LHC) they are well separated, i.e. can,
at least in principle, be measured.


\begin{figure}[!h]  
\includegraphics[width = 0.24\textwidth]{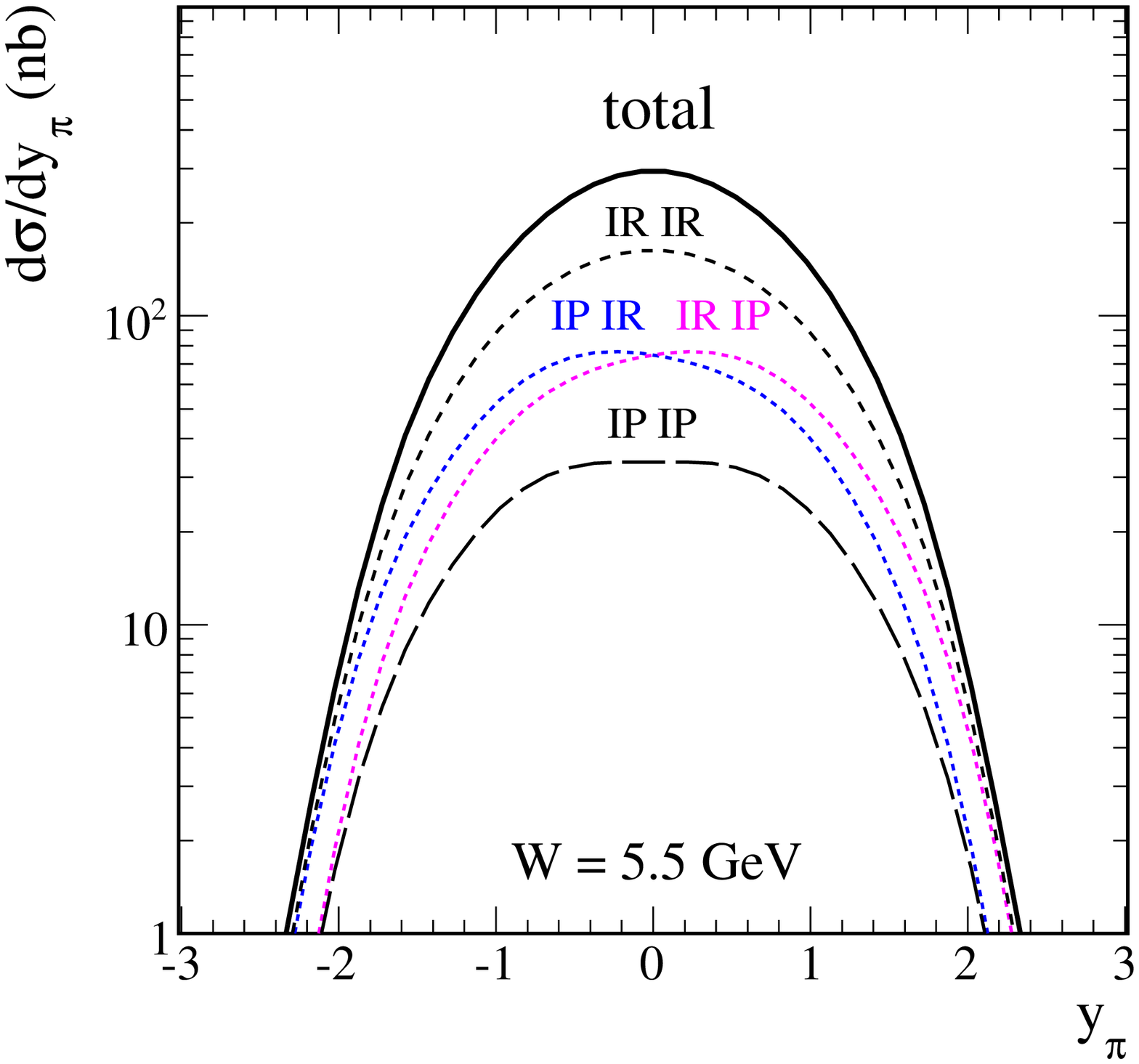}
\includegraphics[width = 0.24\textwidth]{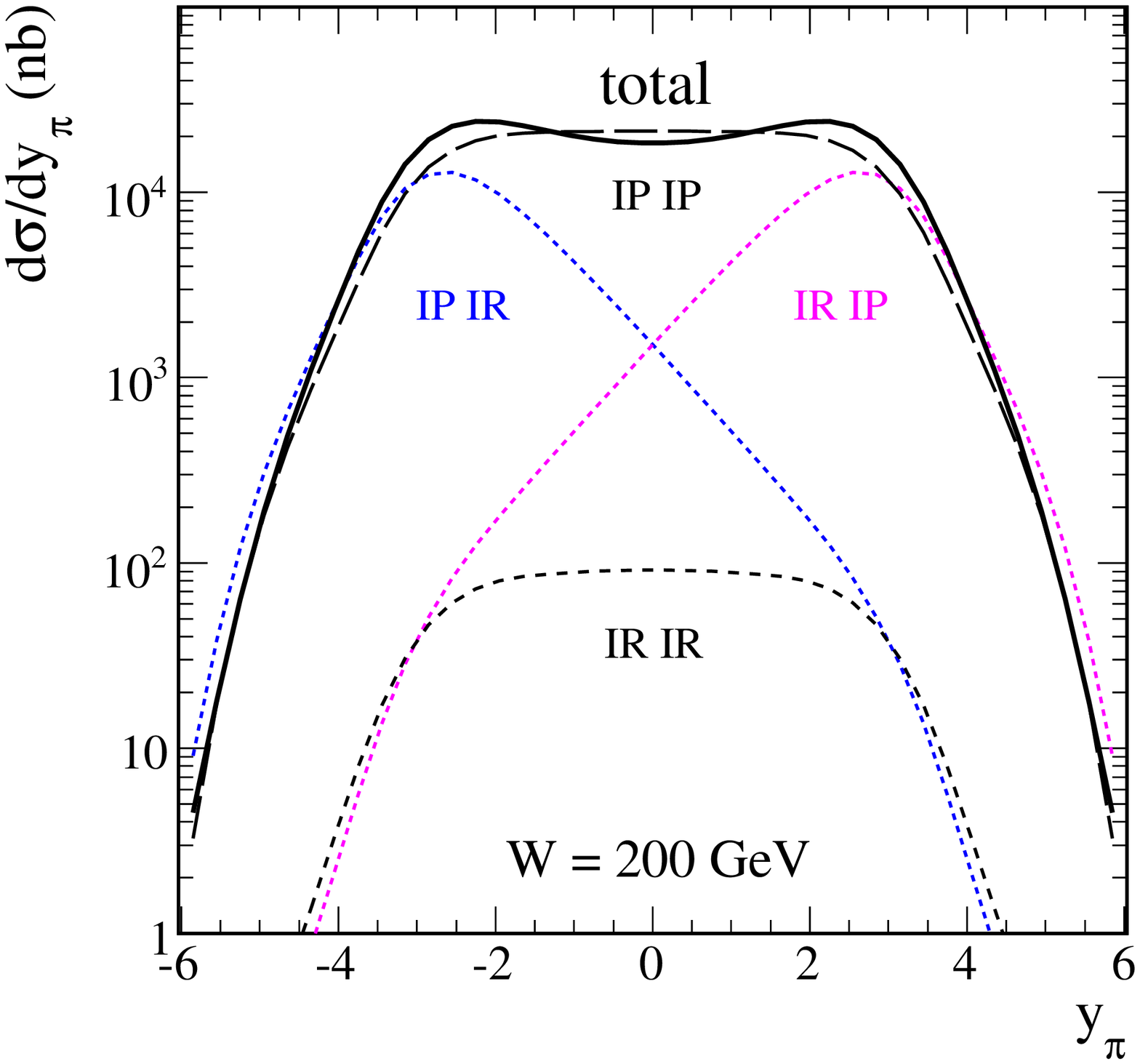}
\includegraphics[width = 0.24\textwidth]{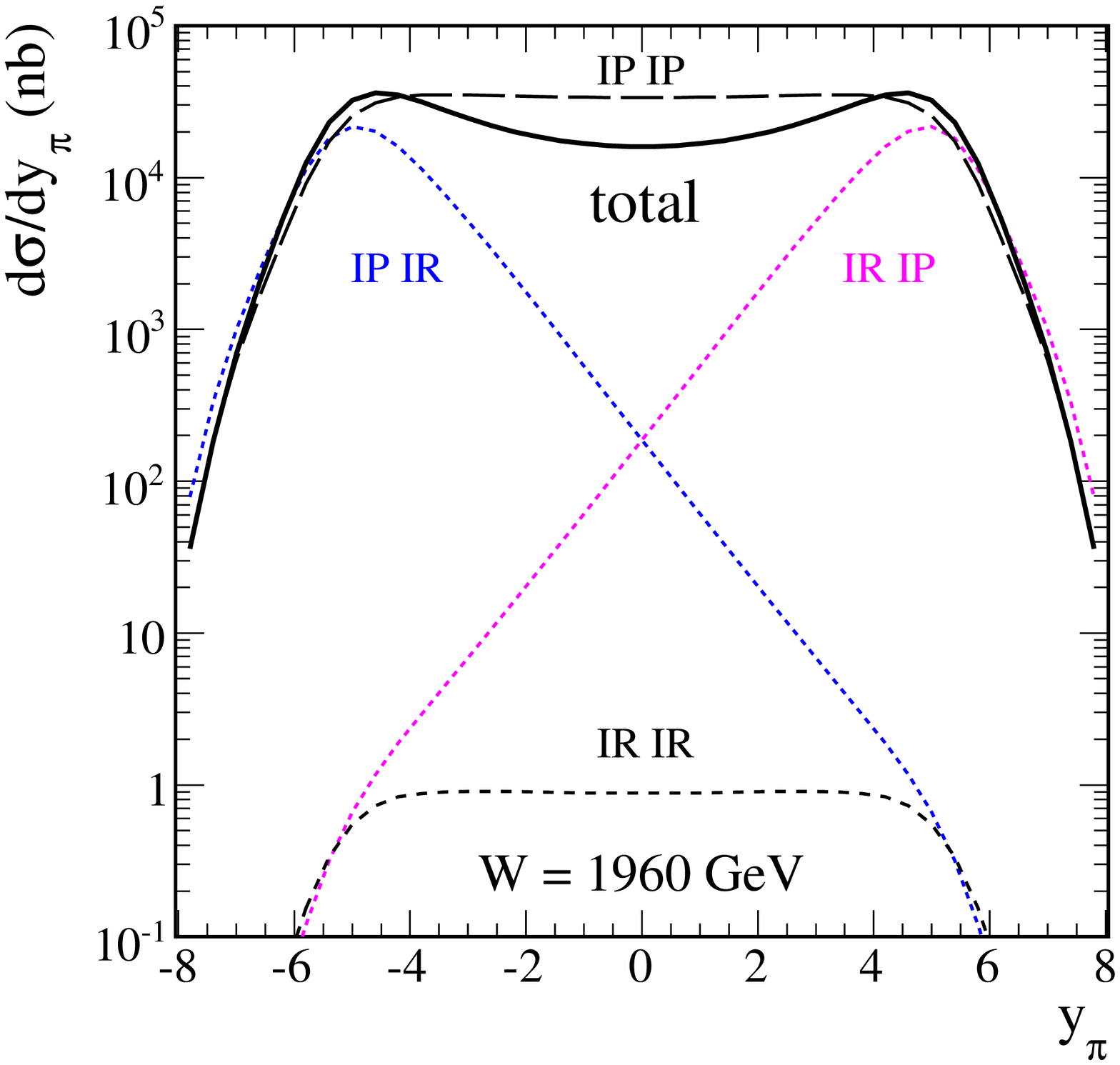}
\includegraphics[width = 0.24\textwidth]{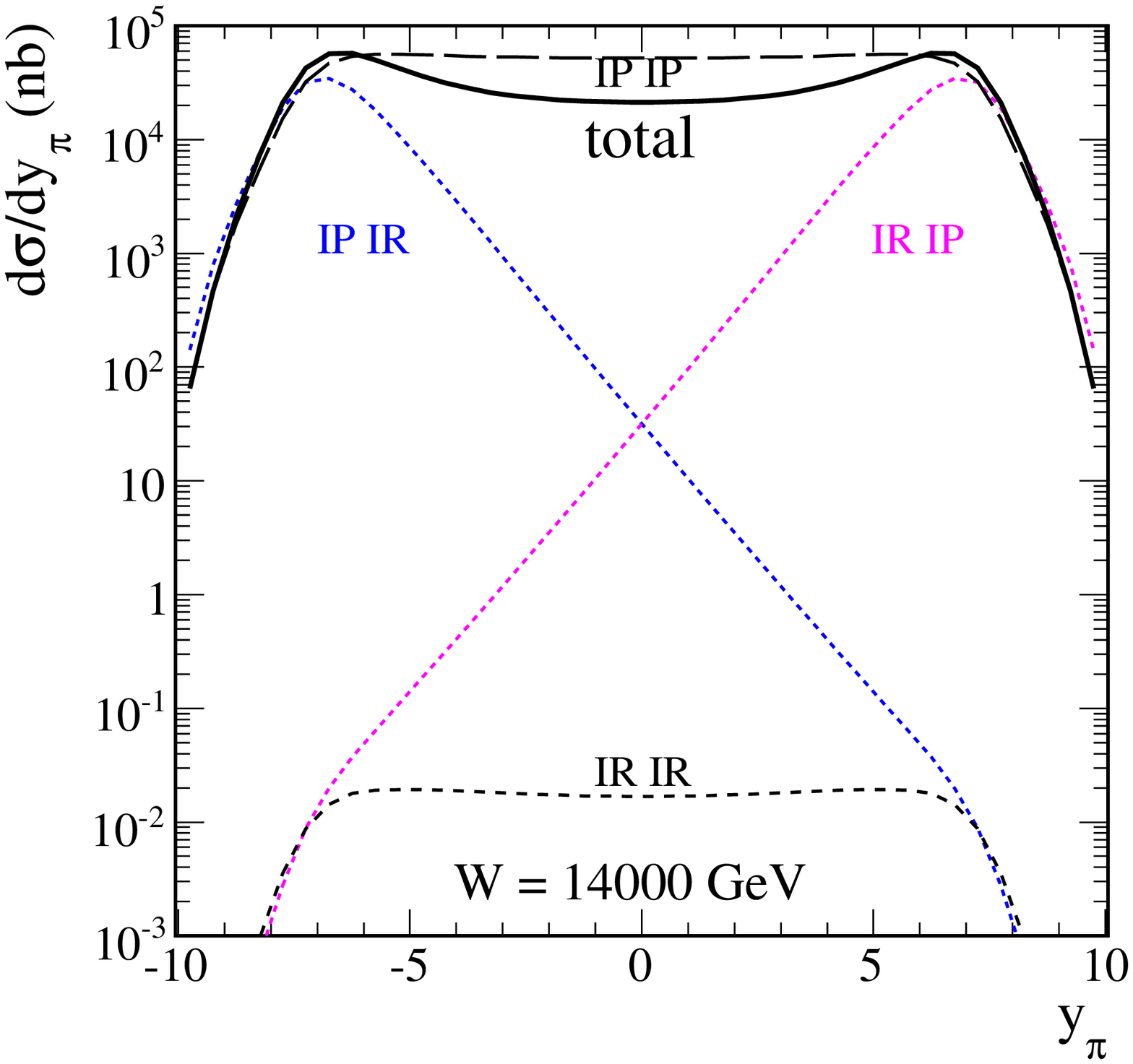}
   \caption{\label{fig:dsig_dy}
   \small 
Rapidity distribution of pions ($\pi^+$ or $\pi^-$)
for different center-of-mass energies. The different lines
corresponds to the situation when only some components in the amplitude
are included. The details are explained in the main text.
}
\end{figure}

At high energies the diffractive contribution seems more interesting.
The camel-like shape of the $y_3, y_4$ distribution requires
a separate discussion.
In our calculation we include both pomeron and reggeon exchanges.
In Fig.\ref{fig:dsig_dy} we show the cross section in $y_{\pi} = y_3 = y_4$
for all ingredients included (thick solid line) and when only
pomeron exchanges are included (long dashed line),
separately for pomeron-reggeon 
and reggeon-pomeron exchanges (dotted lines) and when only
reggeon exchanges are included (dashed line).
In this calculation the cut-off parameter $\Lambda^{2}_{off, E}$ = 1 GeV$^{2}$.
At low energies all individual cross sections when isolated are comparable.
They strongly interfere leading to increase of the cross section.
At higher energies each of the "isolated" cross section peak in different
region of $y_3$ or $y_4$. The $I\!\!P \otimes I\!\!P$ cross section peaks at
midrapidities of pions, while $I\!\!P \otimes I\!\!R$ and $I\!\!R \otimes I\!\!P$
at backward and forward pion rapidities, respectively. When interfering the three 
components in the amplitude produce significant (camel-like) enhancements 
of the cross section at forward/backward rapidities. It would be desirable
to identify the camel-like structure experimentally
\footnote{The ALICE experiment seems
to be able to study the dependence because of the much lower threshold
on pion transverse momenta.}.
At even more forward/backward rapidities 
one may expect single-diffraction contributions 
(e.g. diffractive production of nucleon resonances and their decays) 
not included in the present analysis. This will be discussed elsewhere
\cite{LS2010}.

\begin{figure}[!h]
\includegraphics[width = 0.24\textwidth]{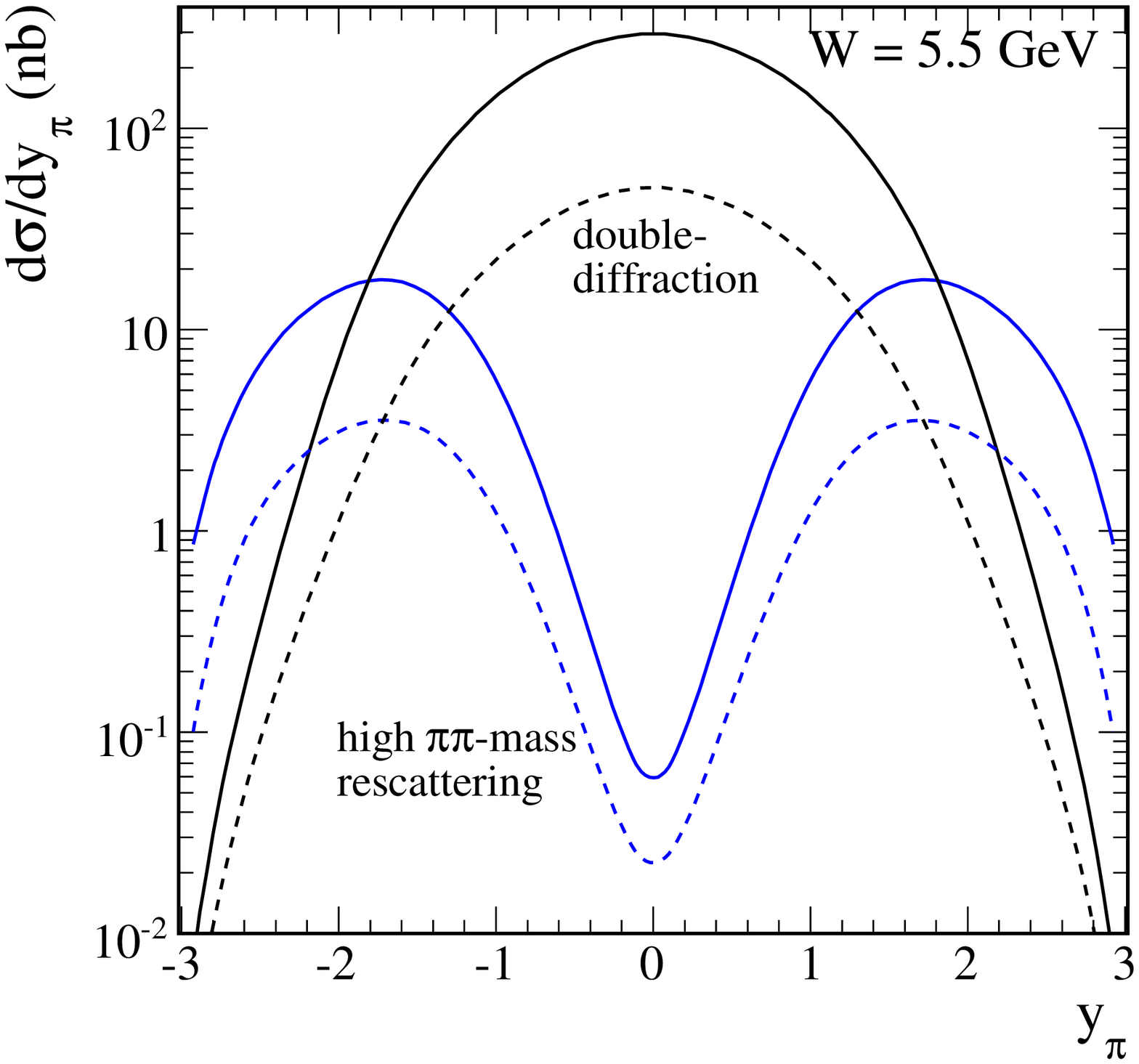}
\includegraphics[width = 0.24\textwidth]{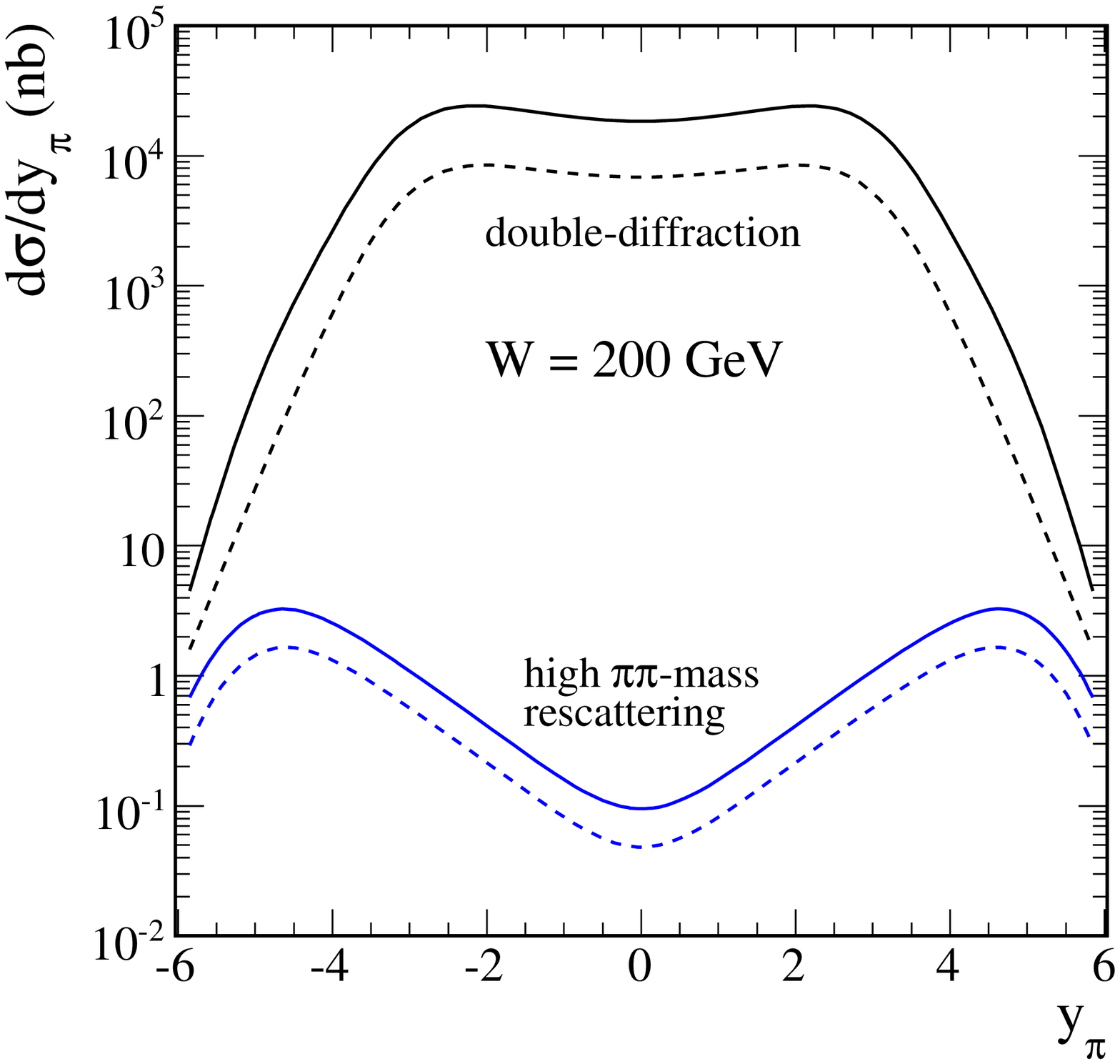}
\includegraphics[width = 0.24\textwidth]{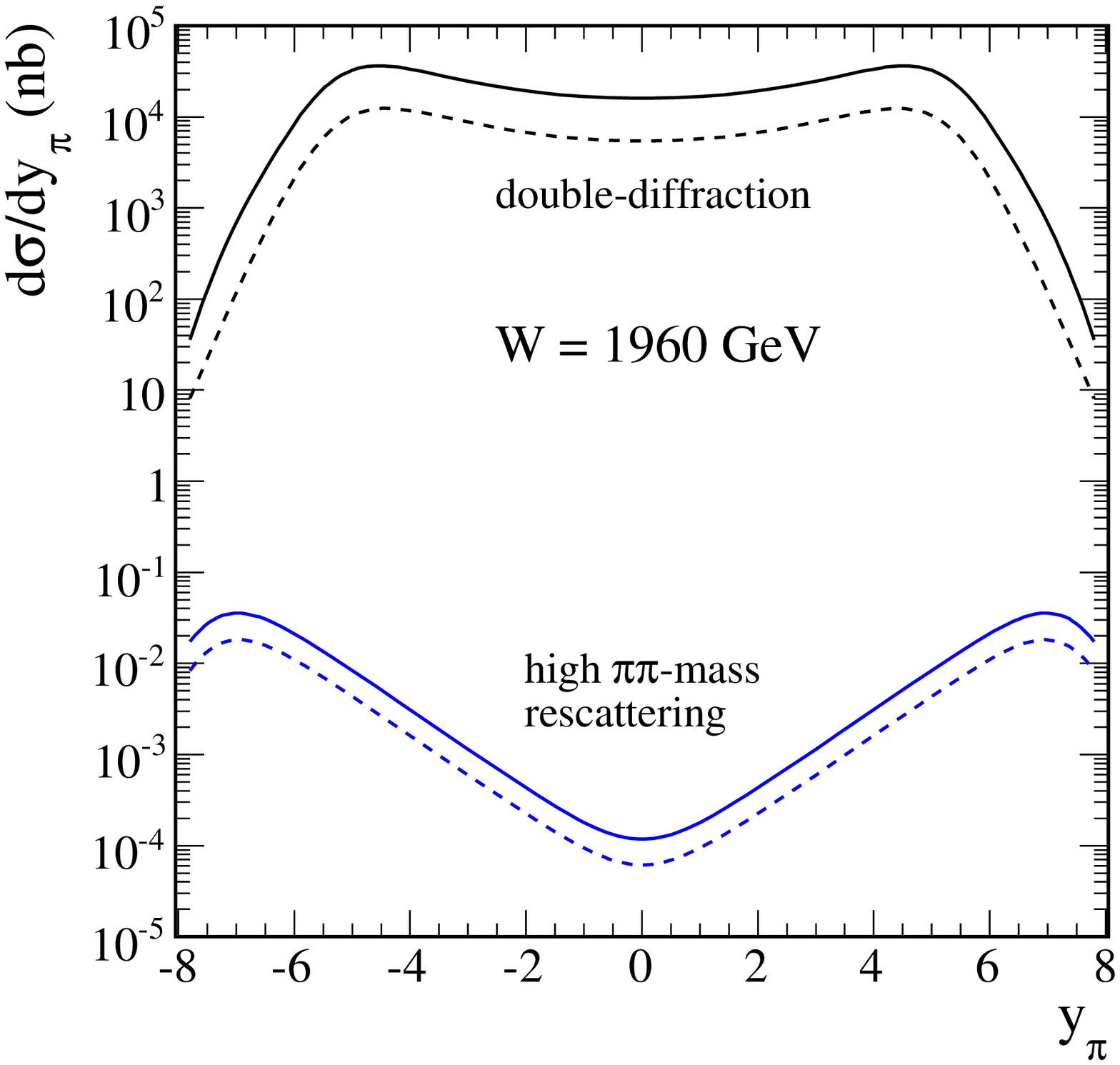}
\includegraphics[width = 0.24\textwidth]{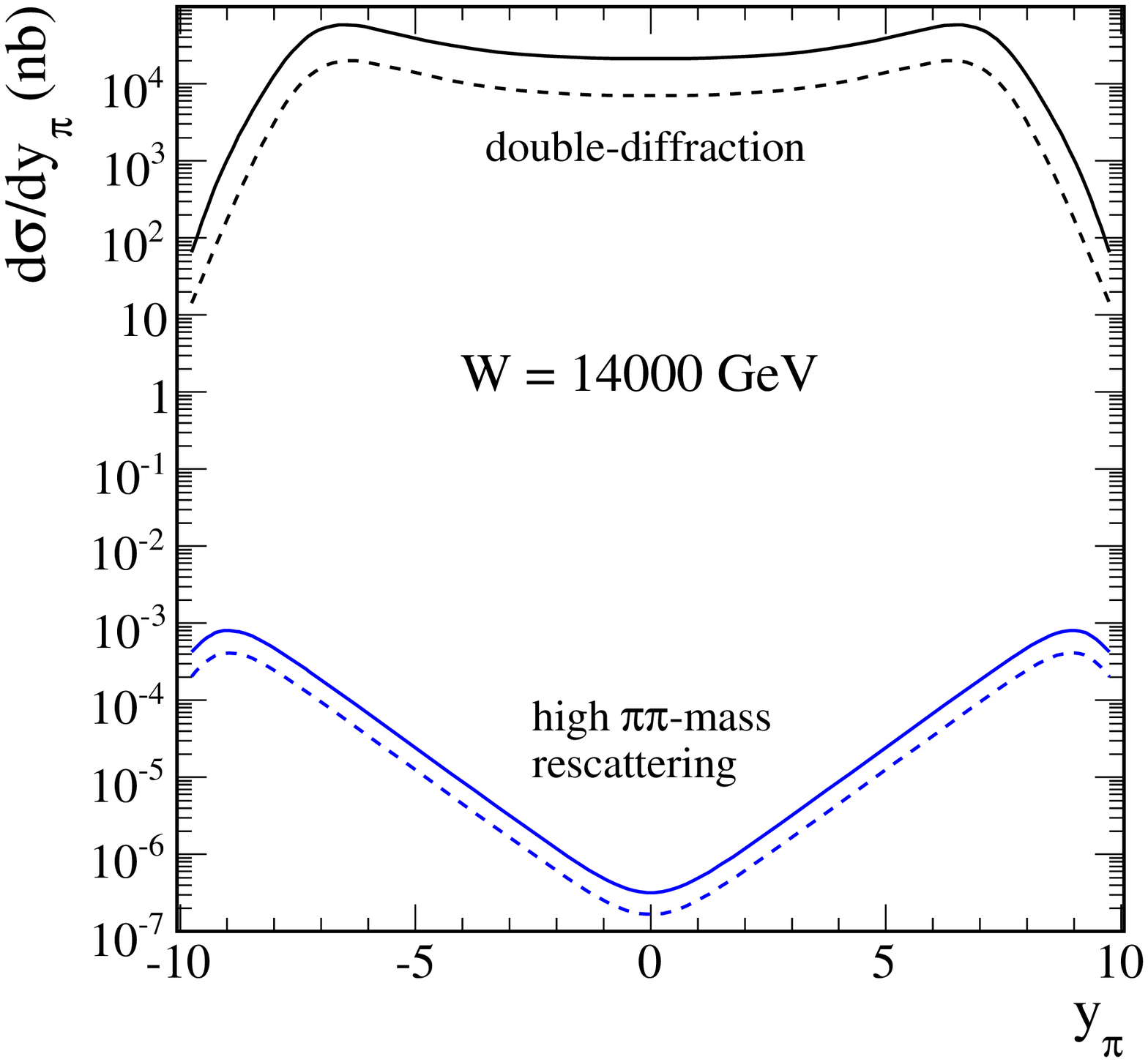}
   \caption{\label{fig:dsig_dy_comparison}
   \small 
Differential cross section $d\sigma/dy_{\pi}$
for the double diffractive and high-$M_{\pi\pi}$ pion-pion 
rescattering contributions at the PANDA, RHIC, Tevatron and LHC energies.
The solid lines was obtained with 
$\Lambda^{2}_{off, E} = 1$ GeV$^2$ 
and dashed lines with $\Lambda^{2}_{off, E} = 0.5$ GeV$^2$.
}
\end{figure}

In Fig.\ref{fig:dsig_dy_comparison} we compare distributions of pion
rapidities $y_{\pi}$ for exclusive double diffraction and high-$M_{\pi\pi}$
pion-pion rescattering at the PANDA, RHIC, Tevatron and LHC energies.

\begin{figure}[!h]   
\includegraphics[width = 0.24\textwidth]{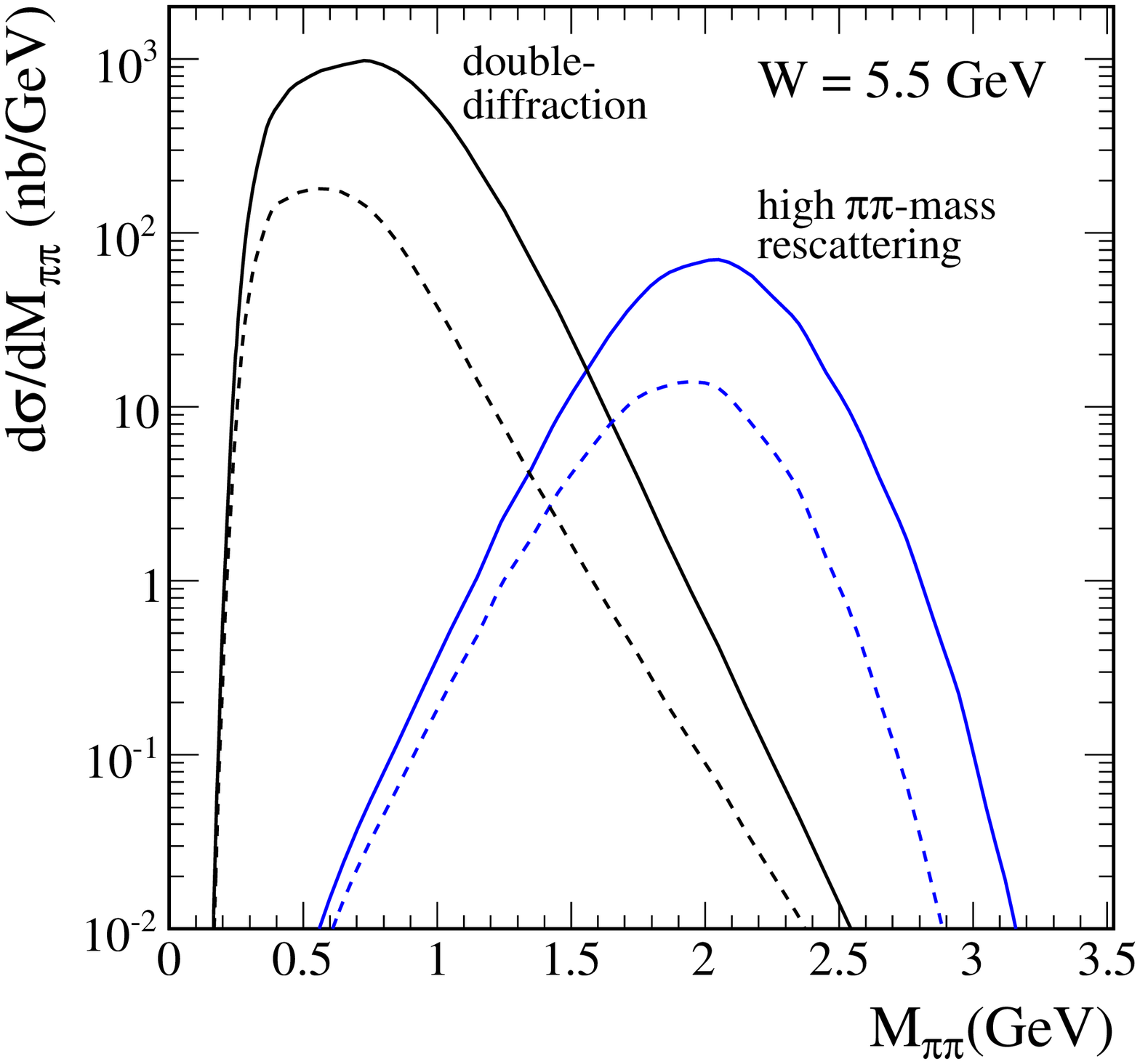}
\includegraphics[width = 0.24\textwidth]{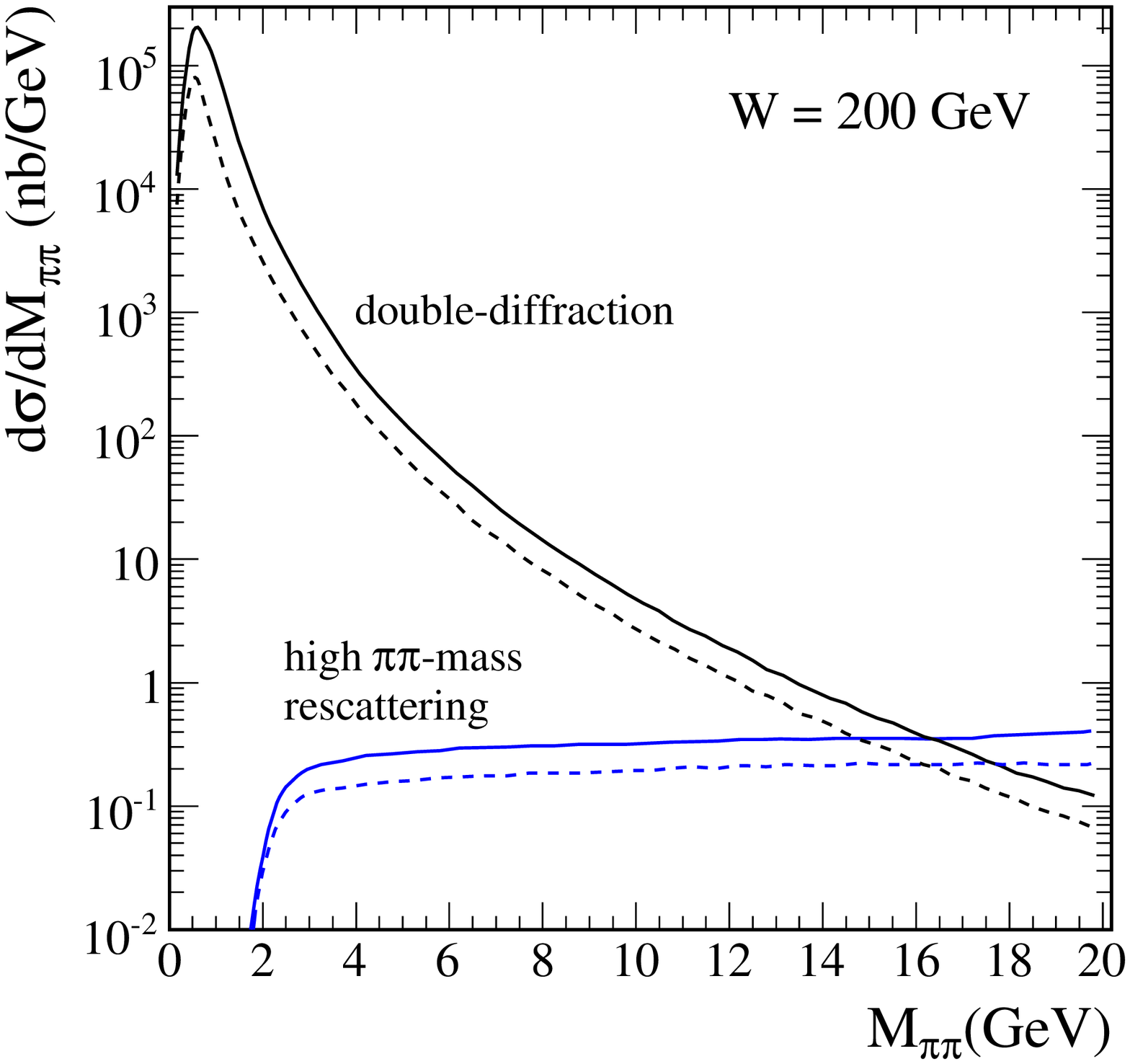}
\includegraphics[width = 0.24\textwidth]{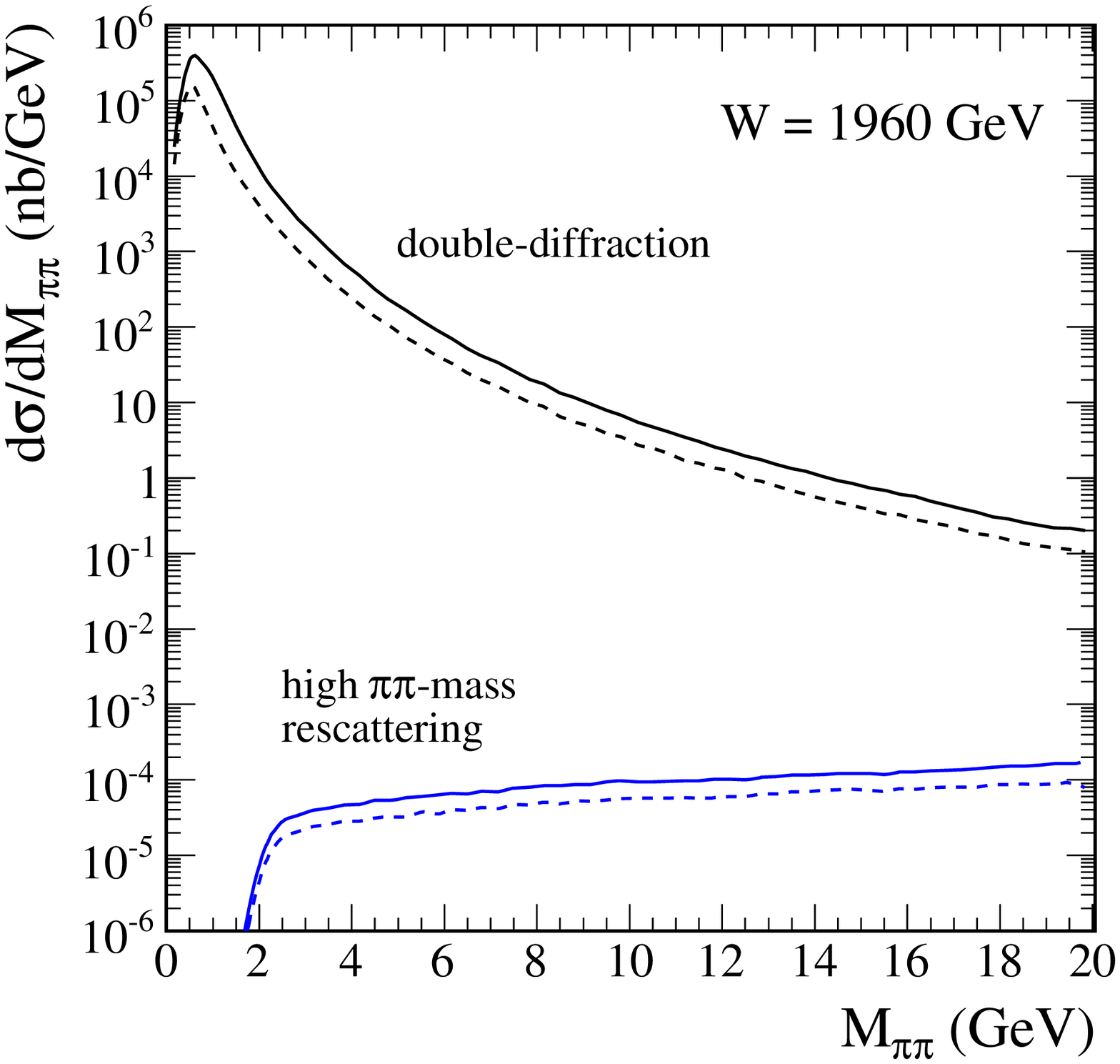}
\includegraphics[width = 0.24\textwidth]{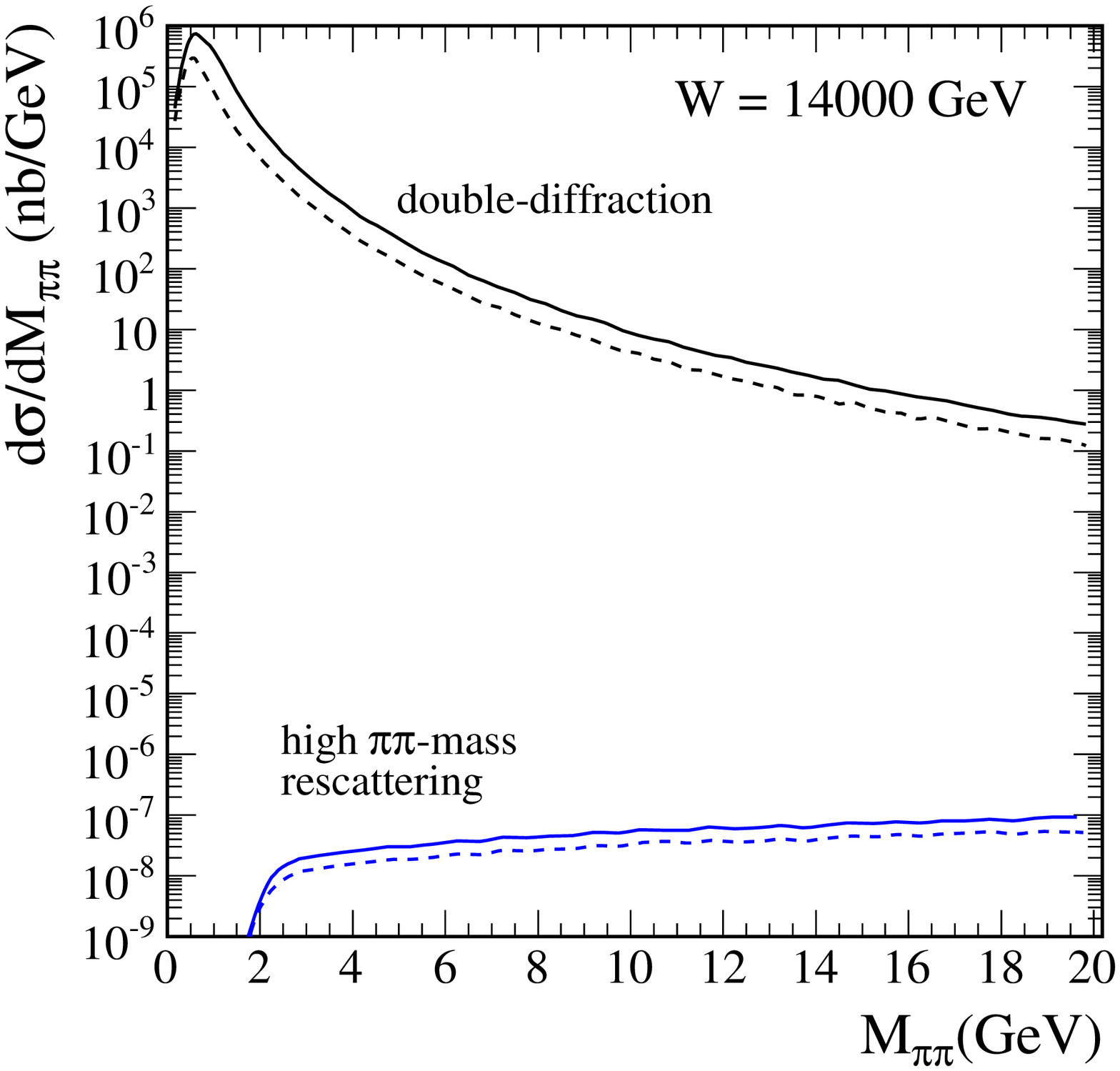}
   \caption{\label{fig:dsig_dm34}
   \small 
Differential cross section $d\sigma/dM_{\pi\pi}$
for diffractive and high-$M_{\pi\pi}$ pion-pion rescattering contributions
at the PANDA, RHIC, Tevatron and LHC energies.
The solid lines was obtained with 
$\Lambda^{2}_{off, E} = 1$ GeV$^{2}$ 
and dashed lines with $\Lambda^{2}_{off, E} = 0.5$ GeV$^{2}$.
}
\end{figure}

In Fig.\ref{fig:dsig_dm34} we show the two-pion invariant-mass
distribution at the PANDA, RHIC, Tevatron and LHC energies.
At the lowest energy the pion-pion rescattering and
the double-diffractive components strongly overlap.
While the double-diffractive component dominates at low two-pion
invariant masses, the pion-pion rescattering components dominates
at large invariant masses. This dependence can be used to
improve purity of one of the two components by imposing extra cuts.
At the Tevatron and LHC energies the double-diffractive component dominates over the pion-pion rescattering in the whole range of $M_{\pi\pi}$.

Finally we show distributions in transverse momentum of the pions.
The double-diffractive component dominates
over the pion-pion rescattering component.
The pions are produced preferentially back-to-back. The smearing
in $p_{t \pi}$ around zero as well as with respect
to $\phi$ = $\pi$ (relative azimuthal angle between charged pions)
is caused by the disbalance of transverse
momenta of exchanged pomerons and/or reggeons from
both proton/antiproton lines.
Whether the distributions can be measured at the LHC requires
a Monte Carlo studies of the ALICE detector.

\begin{figure}[!h]  
\includegraphics[width = 0.24\textwidth]{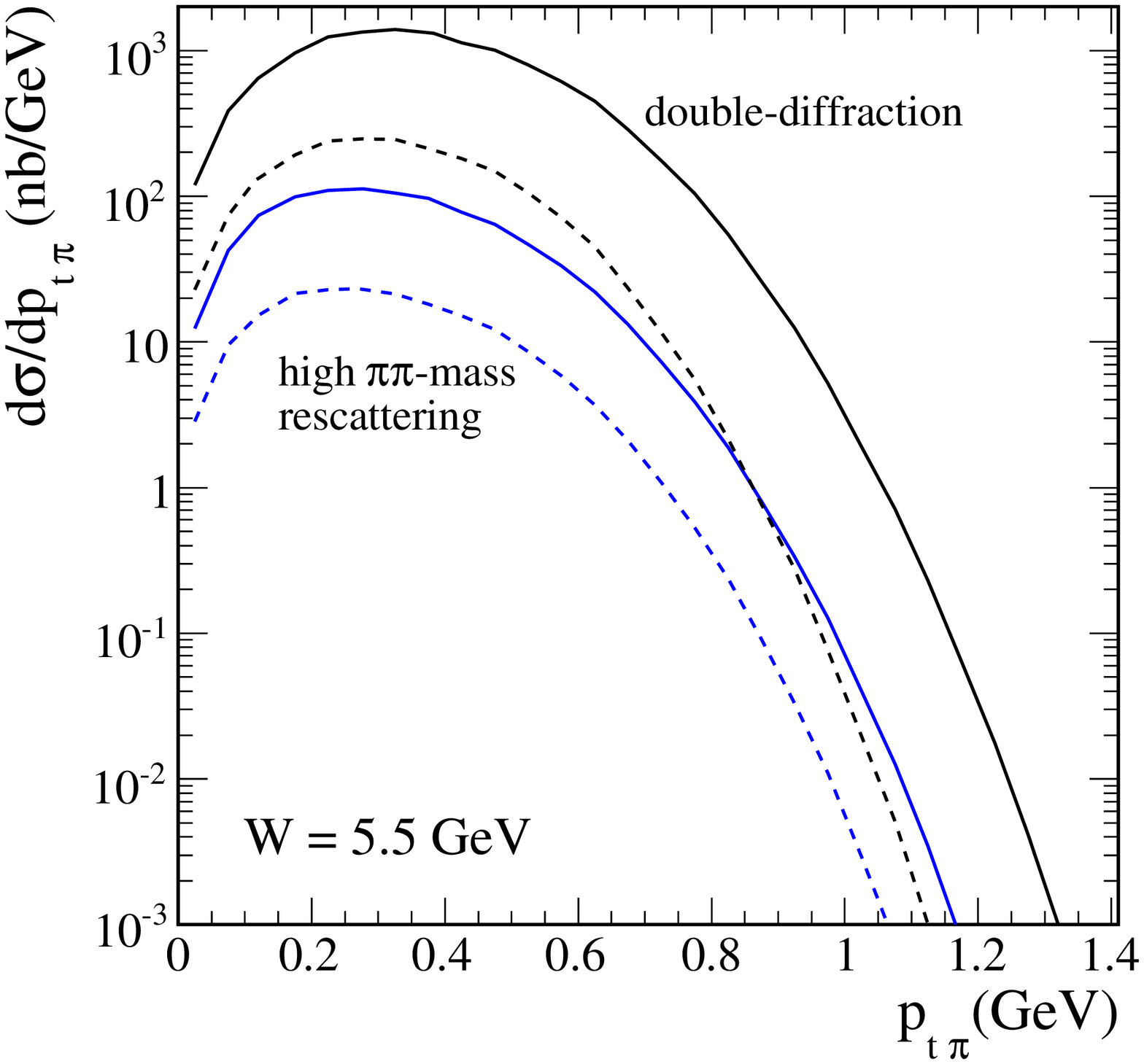}
\includegraphics[width = 0.24\textwidth]{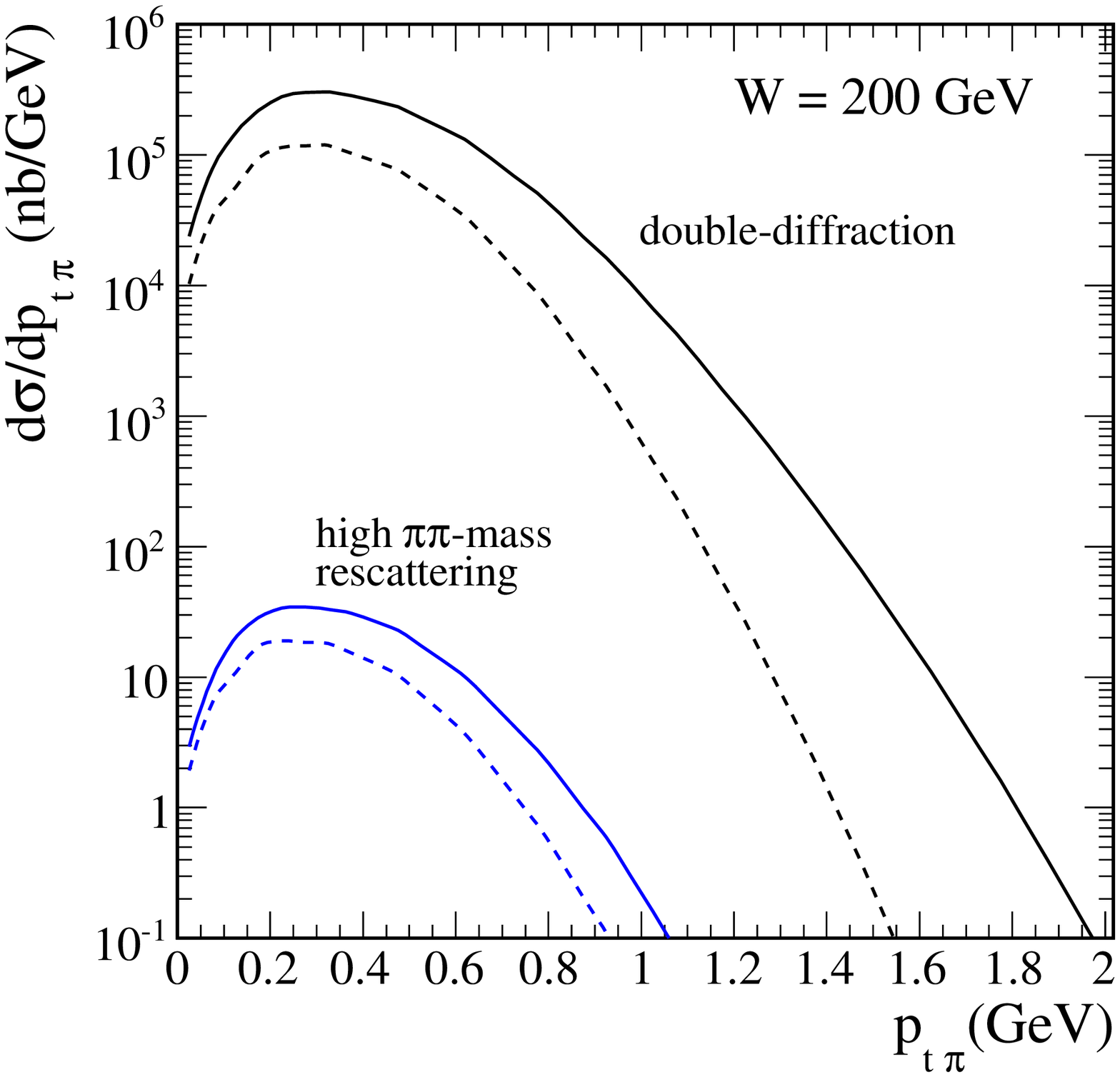}
\includegraphics[width = 0.24\textwidth]{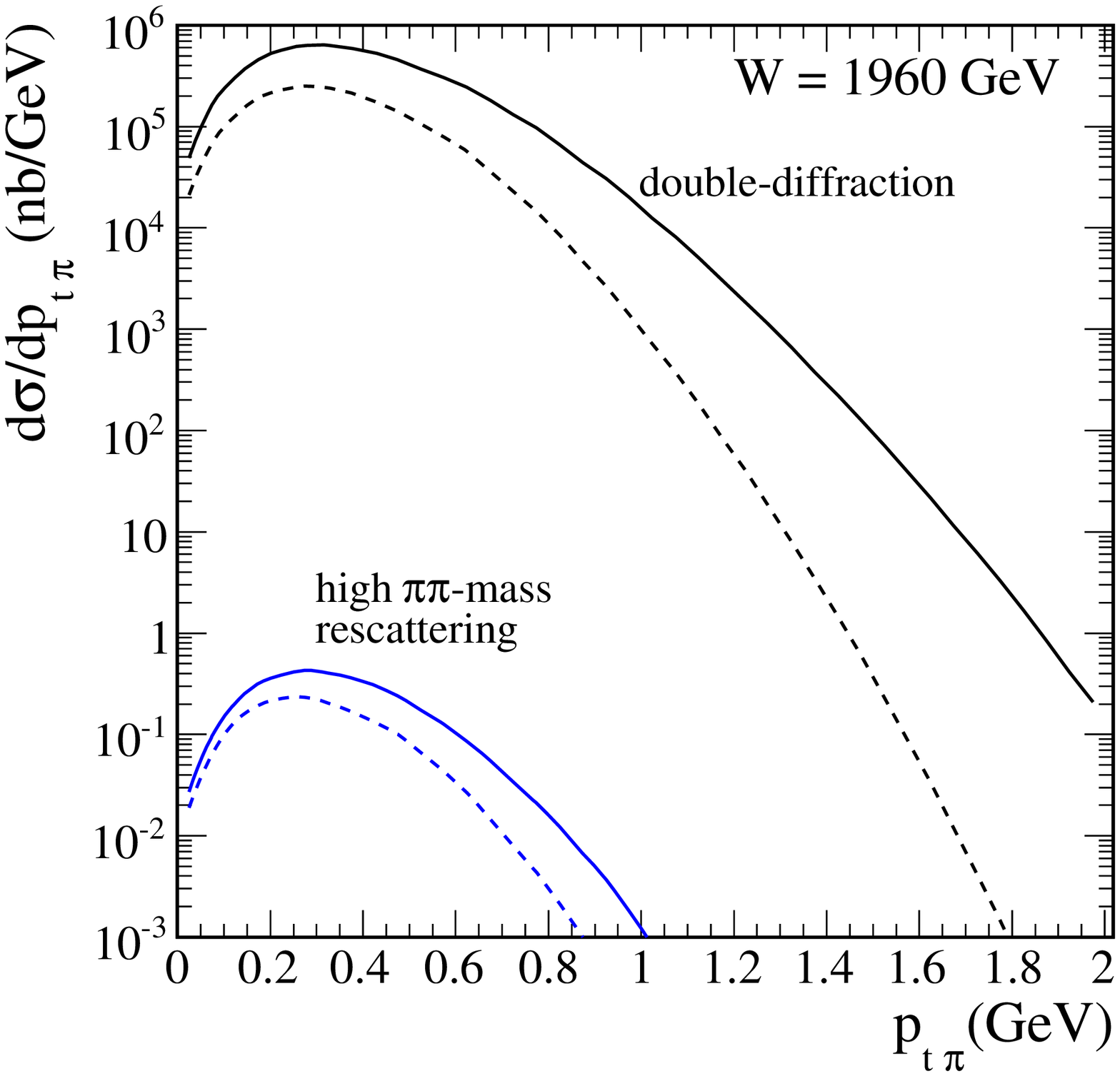}
\includegraphics[width = 0.24\textwidth]{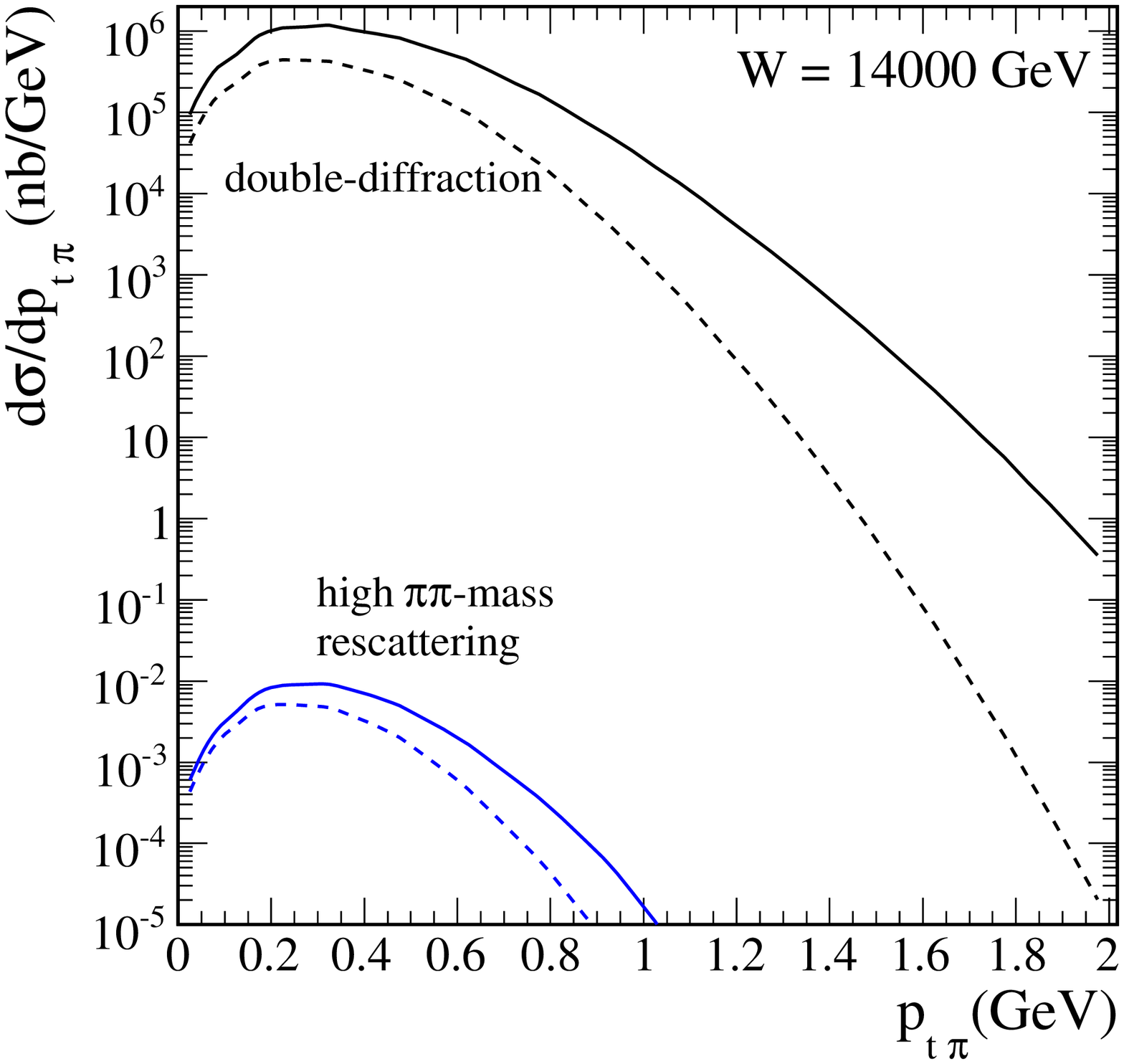}
   \caption{\label{fig:diff}
   \small 
Differential cross section $d\sigma/dp_{t \pi}$
for diffractive and pion-pion rescattering contributions
at the PANDA, RHIC, Tevatron and LHC energies.
The solid lines was obtained with 
$\Lambda^{2}_{off, E} = 1$ GeV$^{2}$ 
and the dashed lines with $\Lambda^{2}_{off, E} = 0.5$ GeV$^2$.
}
\end{figure}

\section{Outlook}

\subsection{Beyond the Born approximation}

\begin{figure}[!h]  
\includegraphics[width=0.32\textwidth]{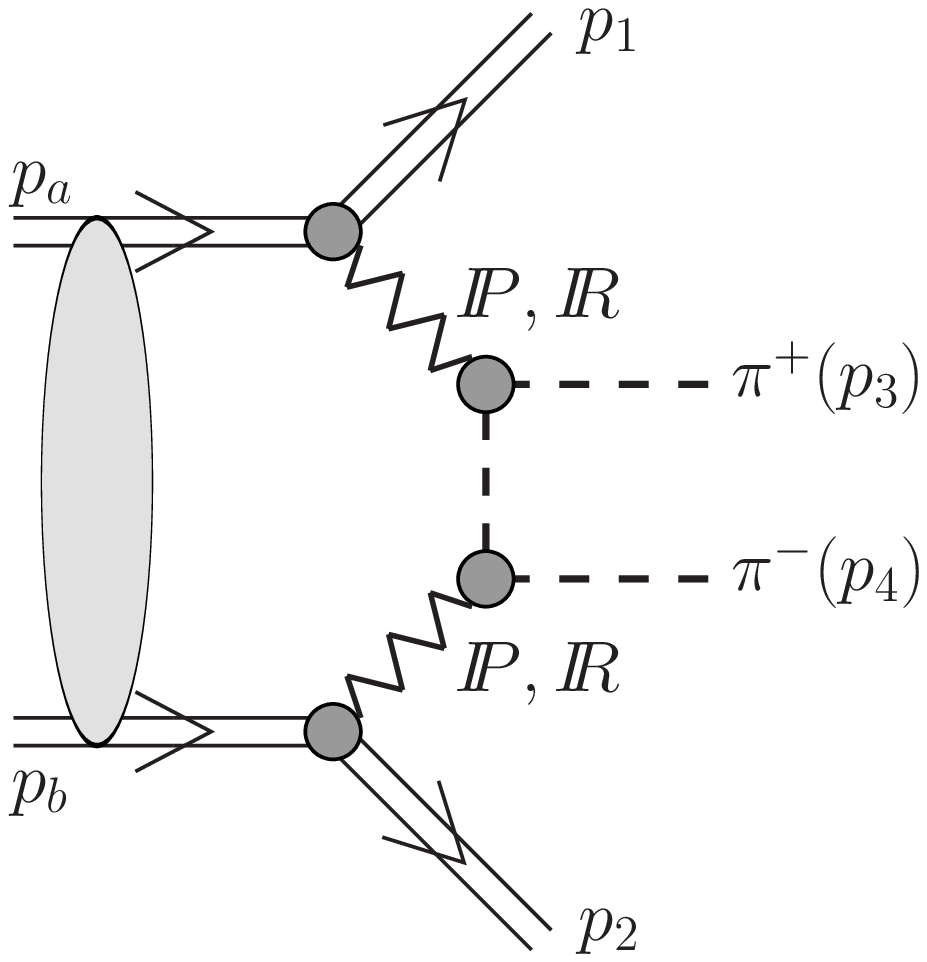}
\includegraphics[width=0.32\textwidth]{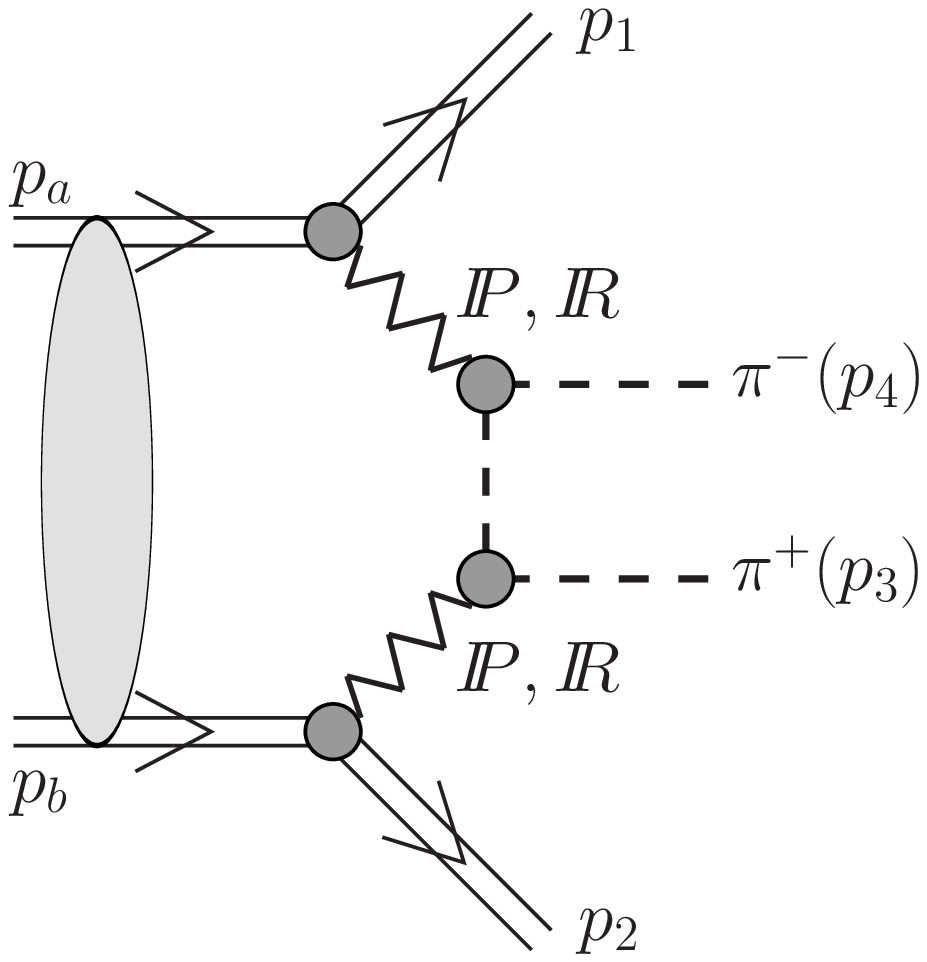}
   \caption{\label{fig:absorption_diagrams}
   \small
Diagrams representing the absorption effects due to proton-proton
interaction.
}
\end{figure}

\begin{figure}[!h]  
\includegraphics[width=0.32\textwidth]{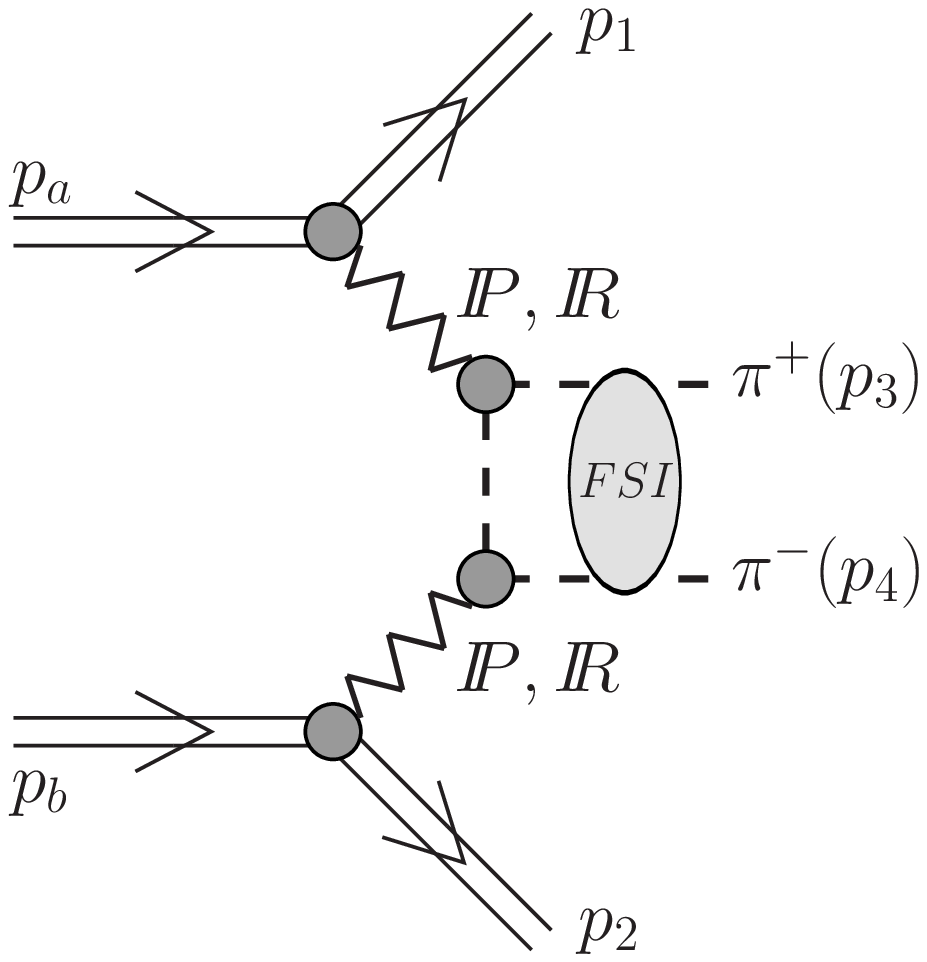}
\includegraphics[width=0.32\textwidth]{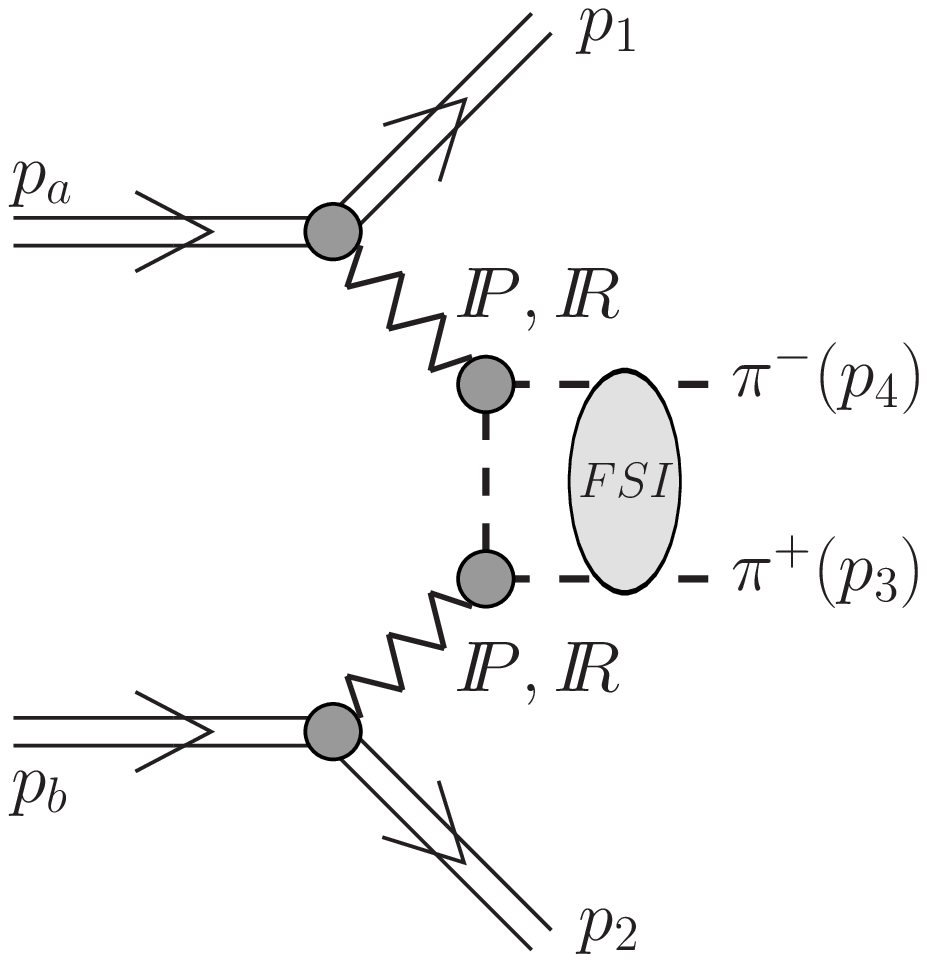}
   \caption{\label{fig:pipifsi_diagrams}
   \small
Diagrams representing pion-pion final state interaction.
}
\end{figure}

In the present, intentionally simplified, analysis
we have performed calculation in the Born approximation
with the form factor parameter roughly adjusted to existing "low-energy" 
experimental data.
In a more microscopic approach one has to include
higher-order diagrams shown in Fig.\ref{fig:absorption_diagrams}
and in Fig.\ref{fig:pipifsi_diagrams}.

The first type of the interaction was studied e.g. for three-body
reactions. For the four-body reaction discussed here a similar effect 
is expected, i.e. large energy-dependent damping of the cross section
which is often embodied in the soft gap survival probability.

When going from the Born (Fig.\ref{fig:central_double_diffraction_diagrams}) 
to the diagrams with the pion-pion FSI
(Fig.\ref{fig:pipifsi_diagrams}) the following replacement is formally 
required:
\begin{equation}
\frac{F_{off}^A(k) F_{off}^B(k)}{k^2 - m_{\pi}^2}  \to
\int \frac{d^4 k}{(2 \pi)^4} \frac{1}{k^2 - m_{\pi}^2}
\frac{F_{off}^A(k,k_3)}{k_3^2 - m_{\pi}^2}
\frac{F_{off}^B(k,k_4)}{k_4^2 - m_{\pi}^2}
\sum_{ij} 
{\cal M}_{\pi_i \pi_j \to \pi^{\pm} \pi^{\mp}}^{off-shell}(k_3 k_4 \to p_3 p_4)
\; ,
\label{pipi_FSI}
\end{equation}
where the sum runs over different isospin combinations of pions.
In general the integral above is complicated (singularities, unknown elements), 
the vertex form factors (A and B) with two pions being off-mass-shell 
are not well known, 
and even the off-shell matrix element is not fully under control.
Usually a serious simplifications are done to make the calculation
useful on a practical level.
Limiting to the $S$-wave ($L=0$) one can correct the Born amplitude
by a phenomenological function which causes an enhancement close 
to the two-pion threshold and damping at $M_{\pi \pi} \sim$ 0.8 GeV.
Dealing with higher partial waves is more complicated.
At even larger $M_{\pi\pi}$ the interaction becomes absorptive
and was not much studied. Some work can be found in Ref. \cite{SNS02}.
Clearly much more theoretical afford is required.

The second type of diagrams leads approximately 
to a redistribution of the strenght but seems to modify the pion-pion
integrated cross section very little \cite{PH76}. 
The effect of pion-pion FSI must be, however, included if
the spectrum of invariant mass is studied.
At high invariant masses one may expect also a strong damping
due to absorption in the pion-pion subsystem.
Only low-invariant-mass spectra were studied in the past experiments 
\cite{Akesson86}.
The experiments at LHC could study the potential damping of
large-mass dipion production and therefore could shed more light on 
the not fully understood problem of absorption effects in a few-body 
hadronic systems, so important in understanding e.g. the exclusive 
production of the Higgs boson discussed recently in the literature.

\subsection{Other not included processes at high energies}

\begin{figure}[!h]  
\includegraphics[width=0.32\textwidth]{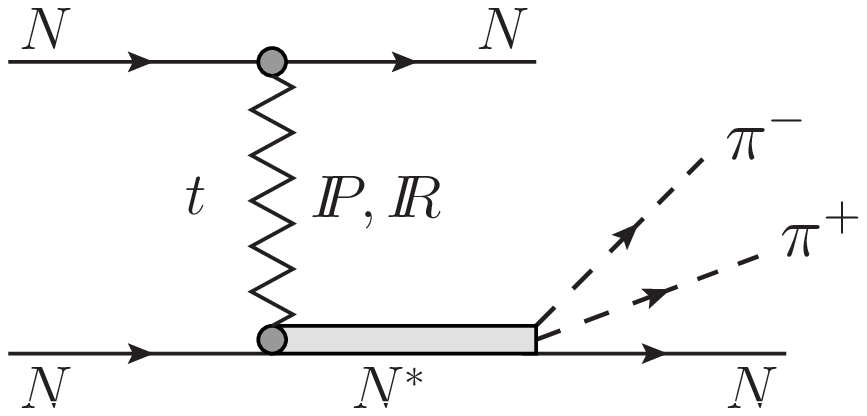}
\includegraphics[width=0.32\textwidth]{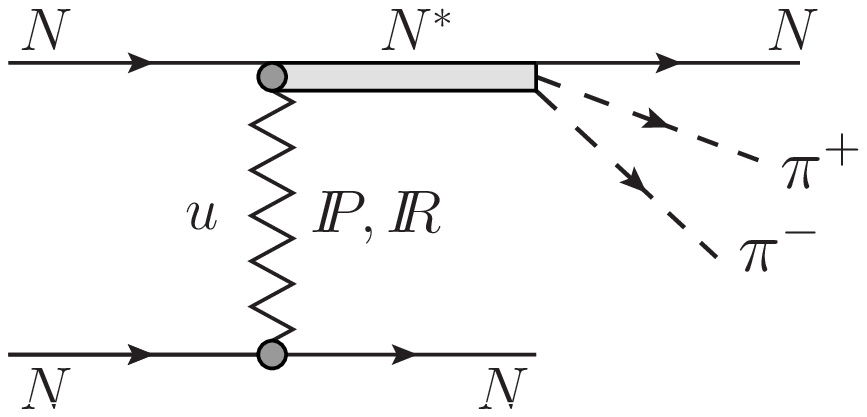}
   \caption{\label{fig:DSRE_diagram}
   \small
Resonance contributions leading to the $p p \to p p \pi^+ \pi^-$
channel through diffractive single resonance excitation (DSRE).
}
\end{figure}

\begin{figure}[!h]  
\includegraphics[width=0.32\textwidth]{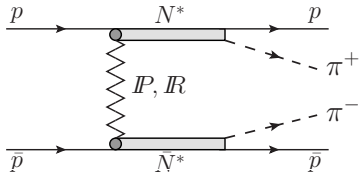}
   \caption{\label{fig:DDRE_diagram}
   \small
Resonance contributions leading to the $p \bar p \to p \bar p \pi^+ \pi^-$
channel through diffractive double resonance excitation (DDRE). 
}
\end{figure}

Up to now we have discussed only central double-diffractive (CDD) 
contribution to the $pp \to pp\pi^+\pi^-$ reaction.
In general, there are also contributions with
diffractive single or double proton/antiproton excitations
followed by the resonance decays shown in
Figs.\ref{fig:DSRE_diagram} and \ref{fig:DDRE_diagram}.
The first mechanism contribute both to the $p p \to p p \pi^+\pi^-$
and $p \bar p \to p \bar p \pi^+ \pi^-$ reaction while the second
mechanism only to the $p \bar{p} \to p \bar{p} \pi^+ \pi^-$ reaction
at high energy
\footnote{At low energy the double $\Delta$ isobar excitation contribute
to $p p \to p p \pi^+ \pi^-$.}.

\begin{figure}[!h]  
\includegraphics[width=0.4\textwidth]{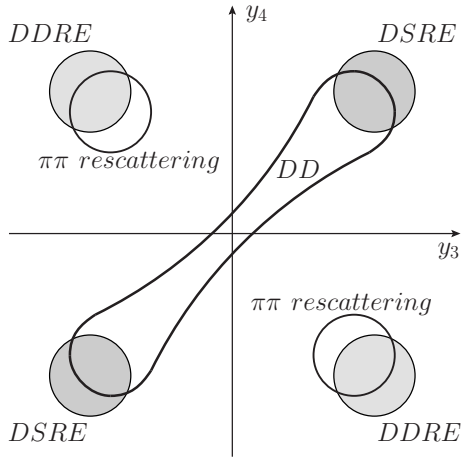}
   \caption{\label{fig:y3y4_mechanisms_localization}
   \small
A schematic localization of different mechanisms for  
the $p p \to p p \pi^+ \pi^-$ or $p \bar p \to p \bar p \pi^+ \pi^-$ 
reactions at high energies. The acronyms used in the figure are explained 
in the main text.
}
\end{figure}

Can these processes be separated from the CDD contribution.
The general situation at high energy is sketched
in Fig.\ref{fig:y3y4_mechanisms_localization}.
The discussed in this paper CDD contributions lays along 
the diagonal $y_3 = y_4$ and the classical DPE in the center
$y_3 \approx y_4$.
The diffractive single resonance excitation (DSRE) contribution
\footnote{The Roper resonance excitation is a good example.} is expected
at $y_3,y_4 \sim y_{beam}$ or $y_3,y_4 \sim y_{target}$, i.e.
situated at the end points of the CDD contribution.
The diffractive double resonance excitation (DDRE) contribution
is expected at ($y_3 \sim y_{beam}$ and $y_4 \sim y_{target}$)
or ($y_3 \sim y_{target}$ and $y_4 \sim y_{beam}$), i.e.
well separated from the CDD contribution discussed in 
the present paper. The Tevatron 
is the only place where one could
look at the DDRE contribution, never studied so far at high energies, 
when it is clearly separated from other mechanisms (CDD, DSRE).

\section{Conclusions}

We have calculated several differential observables 
for the exclusive $p p \to p p \pi^+ \pi^-$ and
$p \bar p \to p \bar p \pi^+ \pi^-$ reactions. 
Both double diffractive and pion-pion rescattering proceses were considered.
The full amplitude was parametrized in terms of subsystem
amplitudes. Only continuum processes (classical DPE) 
were included in the present analysis.

In the first case the energy dependence of the amplitudes of $\pi N$
subsystems was parametrized in the Regge form
which describes total and elastic cross section for $\pi N$
scattering. This parametrization includes both 
leading pomeron trajectory as well as subleading
reggeon exchanges. Even at relatively high energies
the inclusion of reggeon exchanges is crucial
as amplitudes with different combination
of exchanges interfere or/and $\pi N$ subsystem
energies can be relatively small $W_{\pi N}<10$ GeV.
The latter happens when $y_{\pi^{+}},y_{\pi^{-}}\gg$ 0 or
$y_{\pi^{+}},y_{\pi^{-}}\ll$ 0. In this region
of the phase space one can expect a competition 
of single diffractive mechanism.
In the literature mainly total single diffractive 
was calculated. We leave the estimation of the SD mechanism 
contributions to the $pp \pi^{+} \pi^{-}$ channel 
for a separate study.

The integrated cross section of the central
double-diffractive component grows slowly with incident energy
if absorption effects are ignored. In principle, the absorption effects 
may even reverse the trend.

In the second case the pion-pion amplitude was parametrized 
using a recent phase shift analysis at the low pion-pion energies 
and a Regge form of the continuum obtained by assumption
of Regge factorization. The factorization assumption 
is made to estimate the process contribution.

The two contributions occupy slightly different parts of the phase space, 
have different energy dependence and in principle can
be resolved experimentally. The interference of amplitudes
of the both processes is almost negligible.

The energy dependence of the "diffractive" 
central production of two-pions is quite different
than the one for elastic scattering, single- or double-diffraction.
This is due to the specificity of the reaction, 
where rather the  subsystem energies dictate the energy 
dependence of the process.

At high energies we find a preference for the same hemisphere
(same-sign rapidity) emission of $\pi^{+}$ and $\pi^{-}$.
At ISR energies the same size emission is about 50 $\%$
while at LHC energies the same hemisphere emission constitutes about
90 $\%$ of all cases.

In the present analysis we have excluded several resonance 
contributions. 
Formally they belong to the category (c) and not (d) which we 
consider in the present analysis.
But the distinction between the different categories
is a bit arbitrary and may be quite involved experimentally.
Further work is required to estimate contribution
of such a process. This clearly goes beyond the scope 
of the present analysis but will be done in the future.

\vspace{0.5cm}

{\bf Acknowledgments}

We are indebted to Mike Albrow and Valeri Khoze for exchange
of information and Wolfgang Sch\"afer for a discussion. 
This study was partially supported by the Polish grant 
of MNiSW No. N202 249235.




\begin{thebibliography}{1000}
\bibitem{AG81} 
G. Alberi and G. Goggi, Phys. Rep. {\bf 74} (1981) 1.


\bibitem{K79} 
A.B. Kaidalov, Phys. Rep. {\bf 50} (1979) 157.

\bibitem{Good-Walker} 
M.L. Good and W.D. Walker, Phys.Rev. {\bf 120} (1960) 1857.

\bibitem{PH76}
J. Pumplin and F.S. Henyey, Nucl. Phys. {\bf B117} (1976) 377.

\bibitem{AKLR75}
Y.I. Azimov, V.A. Khoze, E.M. Levin and M.G. Ryskin, Sov. J. Nucl. Phys. {\bf 21} (1975) 21.

\bibitem{LSK09}
P. Lebiedowicz, A. Szczurek and R. Kami\'nski, Phys. Lett. {\bf B680} (2009) 459.

\bibitem{Alde97}
D. Alde \textit{et al}. [GAMS Collaboration], Phys. Lett. {\bf 397} (1997) 350.

\bibitem{dsig_dt}
A. Eide \textit{et al.}, Nucl. Phys. {\bf B60} (1973) 173;
I. Ambats \textit{et al.}, Phys. Rev. {\bf D9} (1974) 1179;
C.W. Akerlof \textit{et al.}, Phys. Rev. {\bf D14} (1976) 2864;
D.S. Ayres \textit{et al.}, Phys. Rev. {\bf D15} (1977) 3105;
A. Schiz \textit{et al.}, Phys. Rev. {\bf D24} (1981) 26. 

\bibitem{DL92}
A. Donnachie and P.V. Landshoff, Phys. Lett. {\bf B296} (1992) 227.

\bibitem{SNS02}
A. Szczurek, N.N. Nikolaev and J. Speth, Phys. Rev. {\bf C66} (2002) 055206.

\bibitem{KMR}
V.A. Khoze, A.D. Martin and M.G. Ryskin, Phys. Lett. B {\bf 401},
330 (1997);\\
V.A. Khoze, A.D. Martin and M.G. Ryskin, Eur. Phys. J. C {\bf 23},
311 (2002).

\bibitem{PST08}
R.S. Pasechnik, A. Szczurek and O.V. Teryaev, Phys. Rev. {\bf D78} (2008) 014007.

\bibitem{SL09} 
A. Szczurek and P. Lebiedowicz, Nucl. Phys. {\bf A826} (2009) 101.

\bibitem{ELT02}
T.E.O. Ericson, B. Loiseau and A.W. Thomas, Phys. Rev. {\bf C66} (2002) 014005.

\bibitem{MHE87}
R. Machleidt, K. Holinde and Ch. Elster, Phys. Rep. {\bf 149} (1987) 1;\\
D. V. Bugg, R. Machleidt, Phys. Rev. {\bf C52} (1995) 1203.

\bibitem{GSR}
A. Szczurek and J. Speth, Nucl. Phys. {\bf A555} (1993) 249;\\
B.C. Pearce, J. Speth and A. Szczurek, Phys. Rep. {\bf 242} (1994) 193;\\
J. Speth and A.W. Thomas, Adv. Nucl. Phys. {\bf 24} (1997) 83.


\bibitem{Amsler_PDG08}
C. Amsler \textit{et al.}, (Particle Data Group), Phys. Lett. {\bf B667} (2008) 1,

[{\tt http://pdg.lbl.gov/2009/hadronic-xsections/}].

\bibitem{Denegri75}
D. Denegri \textit{et al.}, [France-Soviet-Union Collaboration], Nucl. Phys. {\bf B98} (1975) 189.

\bibitem{Brick83} 
D.H. Brick \textit{et al.}, Z. Phys. {\bf C19} (1983) 1.

\bibitem{Derrick}
M. Derrick \textit{et al.}, Phys. Rev. Lett. {\bf 32} (1974) 80.

\bibitem{Chew} 
D.M. Chew, Nucl. Phys. {\bf B82} (1974) 422. 

\bibitem{bib_pp_pppipi}
E. Pickup \textit{et al.},
Phys. Rev. {\bf 125} (1962) 2091;
E.L. Hart \textit{et al.},
Phys. Rev. {\bf 126} (1962) 742;
A.M Eisner  \textit{et al.},
Phys. Rev. {\bf 138} (1965) B670;
E. Gellert  \textit{et al.},
Phys. Rev. Lett. {\bf 17} (1966) 884;
G. Alexander \textit{et al.},
Phys. Rev. {\bf 154} (1967) 1284;
A.P. Colleraine and U. Nauenberg,
Phys. Rev. {\bf 161} (1967) 1387;
W. Chinowsky  \textit{et al.},
Phys. Rev. {\bf 171} (1968) 1421;
S.P. Almeida  \textit{et al.},
Phys. Rev. {\bf 174} (1968) 1638;
R. Ehrlich \textit{et al.},
Phys. Rev. Lett. {\bf 21} (1968) 1839;
G. Kayas \textit{et al.}, Nucl. Phys. {\bf B5} (1968) 169;
C. Caso \textit{et al.},
Nuovo Cim. {\bf A55} (1968) 66;
C. Caso \textit{et al.}, Nuovo Cim. {\bf A33} (1976) 671;
C.D. Brunt \textit{et al}., Phys. Rev. {\bf 187} (1969) 1856;
G. Yekutieli \textit{et al.}, Nucl. Phys. {\bf B18} (1970) 301;
E. Colton \textit{et al.},
Phys. Rev. {\bf D3} (1971) 1063;
J.G. Rushbrooke \textit{et al.},
Phys. Rev. {\bf D4} (1971) 3273;
H. Boggild \textit{et al.}, Nucl. Phys. {\bf B27} (1971) 285;
J. Le Guyader \textit{et al.}, Nucl. Phys. {\bf B35} (1971) 573;
D.R.F. Cochran \textit{et al}., Phys. Rev. {\bf D6} (1972) 3085;
W. Burdett \textit{et al.}, Nucl. Phys. {\bf B48} (1972) 13;
B.Y. Oh \textit{et al.} [MFIM Collaboration], FERMILAB-PUB-77-114-E (1977);
M. Derrick \textit{et al.}
Phys. Rev. {\bf D9} (1974) 1215;
Zh.S. Takibaev \textit{et al.},
Yad. Fiz. {\bf 21} (1975) 1015;
F.H. Cverna \textit{et al}., Phys. Rev. {\bf C23} (1981) 1698;
S.A. Azimov  \textit{et al.},
Yad. Fiz. {\bf 34} (1981) 77;
F. Shimizu \textit{et al}., Nucl. Phys. {\bf A386} (1982) 571;
L.G. Dakhno \textit{et al}., Sov. J. Nucl. Phys. {\bf 37} (1983) 540;
D.H. Brick \textit{et al.}, Z. Phys. {\bf C19} (1983) 1;
J. Johanson \textit{et al.} [PROMICE/WASA Collaboration], Nucl. Phys. {\bf A712} (2002) 75; 
S. Abd El-Bary \textit{et al.} [COSY-TOF Collaboration], Eur. Phys. J. {\bf A37} (2008) 267.







\bibitem{bib_ppbar_ppbarpipi}
H. C. Dehne \textit{et al.},
Phys. Rev. {\bf 136} (1964) B843-B851;
C. Walck \textit{et al.},
Nucl. Phys. {\bf B100} (1975) 61; 
M.A. Jabiol \textit{et al.}, Nucl. Phys. {\bf B127} (1977) 365;
C.K. Chen \textit{et al.},
Phys. Rev. {\bf D17} (1978) 42;
D.R. Ward \textit{et al.}, Nucl. Phys. {\bf B172} (1980) 302;
D.E. Zissa \textit{et al.},
Phys. Rev. {\bf D22} (1980) 2642;
M.Yu. Bogolyubsky \textit{et al.}, Yad. Fiz. {\bf 43} (1986) 350;
B.V. Batyunya \textit{et al.}, Sov. J. Nucl. Phys. {\bf 46} (1987) 650, Yad. Fiz. {\bf 46} (1987) 1117; 
L. Bertolotto \textit{et al}. [JETSET Collaboration], Phys. Lett. {\bf B345} (1995) 325;
A. Buzzo \textit{et al}. [JETSET Collaboration], Z. Phys. {\bf C76} (1997) 475.

\bibitem{Waldi83} 
R. Waldi, K.R. Schubert and K. Winter, Z. Phys. {\bf C18} (1983) 301.

\bibitem{BDD} 
L. Baksay \textit{et al.}, [ACCGM Collaboration], Phys. Lett. {\bf B61} (1976) 89;\\
H. De Kerret \textit{et al.}, [CHOV Collaboration], Phys. Lett. {\bf B68} (1977) 385;\\
D. Drijard \textit{et al.}, [CCHK Collaboration], Nucl. Phys. {\bf B143} (1978) 61;\\
B.Y. Oh \textit{et al.}, [MFIM Collaboration], FERMILAB-PUB-77-114-E (1977).

\bibitem{DellaNegra}
M. Della Negra \textit{et al.}, [CCHK Collaboration], Phys. Lett. {\bf B65} (1976) 394.



\bibitem{Akesson86}
T. Akesson \textit{et al.}, Nucl. Phys. {\bf B264} (1986) 154.

\bibitem{LS2010}
P. Lebiedowicz and A. Szczurek, to be presented in the future.

\end{thebibliography}
\end{document}